\documentclass[aps,prd,reprint,amsmath,amssymb,nofootinbib,superscriptaddress,longbibliography,floatfix]{revtex4-2}

\usepackage{graphicx}
\usepackage{booktabs}
\usepackage{array}
\usepackage{xcolor}
\usepackage{tikz}
\newcommand{\pcell}[2]{\parbox[t]{#1\columnwidth}{\raggedright #2\strut}}
\usetikzlibrary{positioning,arrows.meta}

\graphicspath{{fig/}}

\newcommand{\nubar}{\bar\nu}
\newcommand{\GeV}{\,\mathrm{GeV}}
\newcommand{\MeV}{\,\mathrm{MeV}}
\newcommand{\cm}{\,\mathrm{cm}}
\newcommand{\fm}{\,\mathrm{fm}}
\newcommand{\DS}{\Delta S}
\newcommand{\sqrts}{\sqrt{s}}
\newcommand{\Msun}{M_\odot}
\newcommand{\rhoz}{\rho_0}
\newcommand{\Mmax}{M_{\max}}

\begin{document}

\title{Neutrino-induced hyperon final-state interactions as a probe of the
in-medium hyperon potential}

\author{J.\,A.\ Nowak}
\thanks{ORCID: \href{https://orcid.org/0000-0001-8637-5433}{0000-0001-8637-5433}}
\email[Corresponding author: ]{j.nowak@lancaster.ac.uk}
\affiliation{School of Physics and Astronomy, Lancaster University,
Lancaster LA1 4YB, United Kingdom}

\date{\today}

\begin{abstract}
Hyperon single-particle potentials $U_Y(\rho)$ govern hyperon propagation in
nuclei and the appearance of hyperons in dense matter, where they soften the
neutron-star equation of state (EOS) and lower the maximum mass $\Mmax$, the
hyperon puzzle. Charged-current (anti)neutrino interactions on
$^{40}$Ar create $\Lambda$ and $\Sigma$ hyperons inside the nucleus. The
reconstructed $\Lambda$ yield and momentum and a kaon-vetoed final-state-interaction
(FSI) $\Sigma^+$ tag respond to $U_\Lambda$ and $U_\Sigma$ at SBND and DUNE energies.
We write the potential at low density as $U_Y=U_Y(\rhoz)\,(\rho/\rhoz)^\gamma$, plus a
correction that vanishes at saturation. A detector-level Fisher forecast, with
production cross sections, fluxes and efficiencies as nuisance parameters and the
muon kinematics calibrating the production model, determines $U_\Lambda$ best at
$\rho^*\simeq0.82\rhoz$, independently of $\gamma$. The statistical precision is
$1.5\MeV$; with the production model profiled and the remaining systematic
uncertainties added in quadrature, the total is $\pm31\MeV$, of which the largest item, $18\MeV$, is
the convention by which the cascade applies the potential. The measurement is a check,
with different systematics, of the hypernuclear constraint, $\pm3.3\MeV$ at $\rho^*$.
$U_\Sigma$ is not constrained. In relativistic
mean-field EOSs anchored to hypernuclear and $\Sigma$-atom depths, $\Mmax$ for
$np\Lambda\Sigma^-$ matter is $1.94\,\Msun$ for GM1 and $2.03\,\Msun$ for FSU2H with
the $\phi$ meson; it depends on a high-density $\Lambda NN$ term $c_\Lambda$ that the
neutrino data do not constrain. The calculations use the StrangeMC
Monte Carlo generator.
\end{abstract}

\maketitle

\section{Introduction}
\label{sec:intro}
The appearance of hyperons in the dense interior of a neutron star relieves the
nucleon Fermi pressure and softens the equation of state, lowering the maximum mass
below the $2\,\Msun$ of the heaviest observed pulsars---the \emph{hyperon puzzle}.
Whether hyperons appear, and how far they soften the star, is set by their
in-medium single-particle potentials $U_Y(\rho)$, which are only loosely fixed by
hypernuclei and $\Sigma$-atoms at and below saturation density and are essentially
unmeasured above it.

Accelerator (anti)neutrinos can provide a terrestrial handle on $U_Y$. Charged-current
interactions on $^{40}$Ar create $\Lambda$ and $\Sigma$ hyperons inside the
nucleus---through the Cabibbo-suppressed quasi-elastic reaction
$\bar\nu_\mu N\to\mu^+ Y$ on antineutrino beams and Cabibbo-favoured associated
production on neutrino beams---and the hyperons propagate out through nuclear
matter, feeling $U_Y(\rho)$ on the way. The reconstructed $\Lambda$ yield and
momentum and the $\Sigma\!\to\!\Lambda$ conversion respond monotonically to
$U_\Lambda$ and $U_\Sigma$ at SBND and DUNE, complemented by a
final-state-interaction (FSI) $\Sigma^+$ tag whose momentum tracks $U_\Sigma$. Slow
$\Lambda$'s that cannot leave the well are captured, which we treat with a
transport-level energy criterion (Sec.~\ref{sec:smc}). Through a specified hyperonic
equation of state and an assumed high-density continuation, the potential these
observables constrain enters the neutron-star maximum mass.

The calculations use StrangeMC, an internal multi-channel Monte Carlo of
strange-particle production and intranuclear transport on argon~\cite{nowakphd}. Its production
layer is calibrated to the published $\DS=0,1$ cross
sections~\cite{singh2006,alam2010,alam2013,fatima2025} and compared with the
inclusive MicroBooNE CC-$K^+$-on-argon measurement~\cite{microboone2025}; the nuclear
initial state and the intranuclear cascade are inherited, with their existing
validation, from the LUNAR proton-decay package~\cite{Nowak:2026czc}
(Sec.~\ref{sec:methods}). A short account of the neutrino handle is given in
Ref.~\cite{Nowak:2026hyperonPRL}.

\emph{Directly constrained} by the projected data, through transport-model templates, is
the in-medium $\Lambda$
potential at the sub-saturation densities at which the hyperons are produced. The
observables respond to $U_Y$ at the production density, so the anchor $U_Y(\rhoz)$ and
the low-density exponent $\gamma$ of Eq.~\eqref{eq:potform} enter through a
near-degenerate combination. The combination is best determined at the decorrelation
density $\rho^*\simeq0.82\rhoz$, where the fitted $U_\Lambda$ does not depend on
$\gamma$. \emph{Constrained with an assumption} on $\gamma$ and on the functional form
is $U_\Lambda(\rhoz)$ (Sec.~\ref{sec:gamma}). \emph{Not determined} are $U_\Sigma$,
$U_Y$ above saturation and the high-density turn-over of the potential. The fitted
$U_\Sigma$ is a parameter of the transport prescription, and with present $YN$ input
the neutrino data do not determine it. The production model, the transport
prescription by which the potential acts, the $YN$ cross sections used in the cascade
and the nucleon initial state are inputs of the templates. With the production model
calibrated by the muon kinematics of the same events, they limit $U_\Lambda(\rho^*)$
to $\pm31\MeV$, against a statistical $1.5\MeV$ and the hypernuclear $\pm3.3\MeV$ at
the same density, and leave $U_\Sigma(\rhoz)$ uncertain by about $110\MeV$, mainly
through the $YN$ cross sections (Sec.~\ref{sec:robust}).
\emph{Inferred through an EOS model} are the onset density, the maximum mass $\Mmax$
and the tidal deformability, which depend on the choice of RMF parametrisation (GM1 or
FSU2H), the baryon content and the high-density continuation of $U_Y$. The neutrino
data do not constrain the high-density sector, and
the stellar numbers are inferences conditional on a stated EOS framework and an
external high-density prior.

\section{The neutron-star hyperon puzzle}
\label{sec:puzzle}
Pulsar timing has established the existence of neutron stars with masses close to
or above $2\,\Msun$ (PSR~J1614$-$2230~\cite{demorest2010},
J0348$+$0432~\cite{antoniadis2013}, J0740$+$6620~\cite{fonseca2021}). Such masses
require a stiff equation of state
(EOS) at several times nuclear saturation density $\rhoz\!\simeq\!0.16\fm^{-3}$.
At the same time, microscopic calculations of dense matter find that once the
baryon density exceeds $\sim 2\rhoz$ it becomes energetically favourable to
convert the most energetic neutrons into hyperons ($\Lambda$ and $\Sigma^-$). The
new degrees of freedom relieve the Fermi pressure and \emph{soften} the
EOS~\cite{chatterjee2016,tolos2020,burgio2021}.

Whether, and at what density, each hyperon appears is set by its in-medium
single-particle potential $U_Y(\rho)$, and these differ sharply by species:
$U_\Lambda(\rhoz)\approx-28\MeV$ (attractive, fixed by $\Lambda$ hypernuclei
\cite{gal2016}) while $U_\Sigma(\rhoz)\approx+30\MeV$ (repulsive, from
$\Sigma^-$ atoms and $(\pi^-,K^+)$ data \cite{saha2004}). A sufficiently repulsive
$\Sigma$ potential removes $\Sigma^-$ from neutron-star matter
\cite{bednarek2012}, and the high-density slope of $U_Y$ governs how much the
remaining hyperons soften the EOS. The strange sector is correspondingly the
least-constrained part of the dense-matter EOS, and the three established
terrestrial handles---$\Lambda$ hypernuclear spectroscopy for $U_\Lambda$,
$\Sigma^-$ atoms and $(\pi^-,K^+)$ data for $U_\Sigma$, and heavy-ion $\Lambda$
directed flow~\cite{ohnishi2022,nara2022,Jinno:2025lfu}---each carry their own model
dependence.
An independent terrestrial constraint enters the astrophysical prediction as the
low-density input of a hyperonic EOS.

\subsection{Proposed resolutions and their terrestrial inputs}
\label{sec:routes}
With hyperons included, most relativistic mean-field (RMF) models give
$\Mmax=1.4$--$1.8\,\Msun$ \cite{Weissenborn:2011ut,Vidana:2018bdi}, and
Brueckner--Hartree--Fock (BHF) calculations with the Nijmegen ESC08 potentials give
less than $1.4\,\Msun$ \cite{Schulze:2011zza}. The proposed resolutions
\cite{chatterjee2016,Vidana:2018bdi,tolos2020,burgio2021,Sedrakian:2022ata,Oertel:2016bki}
fall into four groups, which draw on different terrestrial inputs.

In RMF models, relaxing the SU(6) vector couplings to SU(3), adding the
hidden-strangeness $\phi$ meson or quartic vector terms, or weakening the
hyperon--scalar couplings stiffens the hyperonic EOS enough for $2\,\Msun$
\cite{Weissenborn:2011ut,Weissenborn:2011kb,bednarek2012,Colucci:2013pya,Oertel:2014qza,Tolos:2016hhl}.
The scalar couplings are fixed by $U_Y(\rhoz)$, as in Eq.~\eqref{eq:anchor}
\cite{glendenning1991}. $\Lambda$ hypernuclei calibrate the $\Lambda$
\cite{Millener:1988hp,Hashimoto:2006aw,gal2016,Fortin:2020qin}, and with SU(6)
vector couplings the calibrated $U_\Lambda(\rhoz)$ is $-30$ to $-32\MeV$
\cite{Fortin:2017cvt}. $\Sigma^-$ atoms and $(\pi^-,K^+)$ spectra require a
$\Sigma$--nucleus potential that is repulsive inside the nucleus
\cite{Batty:1997zp,Friedman:2007zza,Noumi:2001tx,saha2004,Harada:2005hs}. The
results depend strongly on the hyperon--hyperon channels \cite{Oertel:2014qza},
which only double-$\Lambda$ hypernuclei \cite{gal2016} and $\Lambda\Lambda$
femtoscopy \cite{ALICE:2019eol} constrain.

A repulsive hyperon--nucleon--nucleon (YNN) force that grows with density can delay
or prevent the hyperon onset, with a model-dependent effect on $\Mmax$
\cite{Vidana:2010ip,Yamamoto:2014jga}. In auxiliary-field diffusion Monte Carlo
calculations, two $\Lambda NN$ forces that both describe medium-mass hypernuclei
give $\Mmax=1.36\,\Msun$ and $\Mmax\ge2.09\,\Msun$ \cite{Lonardoni:2014bwa}. In SU(3)
chiral effective field theory ($\chi$EFT) the $\Lambda$ potential turns strongly
repulsive at $2$--$3\rhoz$, more so with chiral $\Lambda NN$ forces
\cite{Haidenbauer:2016vfq,Kohno:2018gby,Gerstung:2020ktv}, which give $2\,\Msun$
in BHF \cite{Logoteta:2019utx}. $\Lambda$ binding energies across the periodic table attribute about
$14\MeV$ of repulsion at $\rhoz$ to $\Lambda NN$ terms \cite{Friedman:2022bpw}.
J-PARC E40 measured $\Sigma^\pm p$ and $\Sigma^-p\to\Lambda n$ scattering
\cite{miwa2021,miwa2022,nanamura2022}, and CLAS $\Lambda p$ scattering
\cite{CLAS:2021gur}. Femtoscopy \cite{alice2020nature,Fabbietti:2020bfg} resolves
the $N\Sigma\leftrightarrow N\Lambda$ cusp in $p\Lambda$ correlations
\cite{alice2022coupled} and gives a first $p$--$p$--$\Lambda$ cumulant, consistent
with zero \cite{ALICE:2022boj}. HAL QCD lattice potentials
\cite{Aoki:2012tk,HALQCD:2019wsz} give $U_\Lambda=-28\MeV$ and $U_\Sigma=+15\MeV$ at
$\rhoz$ in BHF \cite{Inoue:2018axd}. For NLO $\chi$EFT interactions
\cite{Haidenbauer:2013oca,Petschauer:2015nea,petschauer2020} the scattering data
leave the $\Lambda N$--$\Sigma N$ conversion strength open, and it matters in the
medium \cite{Haidenbauer:2019boi,alice2022coupled}. $\Lambda$
directed flow in heavy-ion collisions probes the potential above saturation; it is
reproduced with $\chi$EFT potentials of weak momentum dependence and is more
sensitive to the momentum dependence than to the density dependence
\cite{nara2022,ohnishi2022}, and the $\Lambda+\Sigma^0$ flow also depends on
$U_\Sigma$ \cite{Jinno:2025lfu}.

In the third group the high-density pressure comes from a crossover to strongly
interacting quark matter near $3\rhoz$ \cite{Masuda:2012kf,Baym:2017whm}, from quark
cores \cite{Annala:2019puf} or from quarkyonic matter
\cite{McLerran:2018hbz,Fujimoto:2023mzy,Margueron:2021dtx}, in which quark
saturation moves the hyperon threshold from $2$--$3\rhoz$ to $5$--$6\rhoz$
\cite{Fujimoto:2024doc}; the crossover result is largely insensitive to the
hyperonic EOS below it \cite{Masuda:2012kf}. In the fourth, the $\Delta(1232)$ can
appear before the hyperons and delay them, but it softens the EOS as well
\cite{chatterjee2016,Vidana:2018bdi}; it reduces the radii at nearly unchanged
$\Mmax$ \cite{Sedrakian:2022ata} and stabilises the compact hadronic branch of the
two-family scenario \cite{Drago:2013fsa}. Its input, the $\Delta$--nucleus
potential, is placed by electron, pion and photoabsorption data between the
nucleon potential and $30\MeV$ below it \cite{Sedrakian:2022ata}.

Outside the third group the hyperon--nucleus potential near saturation is a
calibration input: it fixes the RMF scalar couplings, the two-body baseline against
which a $\Lambda NN$ repulsion is measured, and the hyperon onset relative to the
$\Delta$ onset. The neutrino observables constrain $U_\Lambda$ at
$\rho^*=0.82\rhoz$ (Sec.~\ref{sec:gamma}), through hyperons produced
inside argon and observed after they leave the nucleus. The systematics come from
the hyperon--nucleon cross sections, the production model, the transport, the nucleon
initial state and the detector response (Sec.~\ref{sec:robust}); hypernuclear
structure calculations do not enter. The
groups differ mainly above $2\rhoz$, and this measurement does not discriminate
between them: at fixed $U_\Lambda(\rhoz)$ the $\rho^2$ term of Eq.~\eqref{eq:potform}
($\gamma=1$, $\beta=2$) shifts $U_\Lambda(\rho^*)$ by
$-c_\Lambda x^*(1-x^*)\approx-0.15\,c_\Lambda$, with $x^*=\rho^*/\rhoz$,
about $-2\MeV$ for the $\Lambda NN$ repulsion of Ref.~\cite{Friedman:2022bpw}.
This is comparable to the statistical precision on $U_\Lambda(\rho^*)$, $1.5\MeV$, and
an order of magnitude below its total uncertainty, $\pm31\MeV$
(Sec.~\ref{sec:robust}).
For the $\Sigma$ the neutrino observables are limited by the hyperon--nucleon
conversion cross sections (Sec.~\ref{sec:robust}), and $\Sigma^-$ atoms and
$(\pi^-,K^+)$ data remain the input.

GM1 with SU(6) vector couplings, scalar couplings anchored to $U_\Lambda=-28\MeV$
and $U_\Sigma=+30\MeV$, and no hidden-strangeness mesons (Sec.~\ref{sec:rmf}) is the
minimal form of the first group. Its octet maximum mass, $1.82\,\Msun$ at
$U_\Xi=-14\MeV$, lies in the range $1.69$--$1.92\,\Msun$ found for GM1 with the
octet, SU(6) couplings, $U_\Lambda\approx-30\MeV$, $U_\Sigma=+30\MeV$ and $U_\Xi$ from
$-40$ to $+40\MeV$ \cite{Weissenborn:2011kb}; the $np\Lambda\Sigma^-$ benchmark,
$1.94\,\Msun$, is higher because the $\Xi^-$ is omitted. The $\phi$ meson raises GM1
to $1.96\,\Msun$ \cite{Weissenborn:2011kb} ($2.00\,\Msun$ for the octet in our
implementation), and FSU2H with the $\phi$ meson gives $2.02$--$2.03\,\Msun$
\cite{Tolos:2016hhl,Tolos:2017lgv}, which we reproduce (Sec.~\ref{sec:eossys}).
Without the $\phi$ meson, GM1 at the anchored depths exceeds $2\,\Msun$ only through
a positive turn-over $c_\Lambda$ [Eq.~\eqref{eq:vextra}], a phenomenological form of
the second group. The GM1 maximum masses above $2\,\Msun$ in Sec.~\ref{sec:bayes} come
from $c_\Lambda$ and follow its prior.

\section{In-medium potentials}
\label{sec:potentials}
Produced hyperons and strange mesons feel a density-dependent single-particle
potential, written as a power law anchored at saturation plus a correction that
vanishes there,
\begin{equation}
  U_Y(\rho)=U_Y(\rhoz)\,(\rho/\rhoz)^\gamma
            + c_Y\big[(\rho/\rhoz)^\beta-(\rho/\rhoz)^\gamma\big] ,
  \label{eq:potform}
\end{equation}
with $\rhoz=0.16\fm^{-3}$, applied either as an energy shift as the particle leaves
the nucleus or, optionally, as continuous gradient-force transport or, for hyperons,
from the start of the cascade as in NuWro (Sec.~\ref{sec:methods}). Setting $c_Y=0$ recovers the pure power law.

The form follows the density expansion of the single-particle potential. At low
density the potential is linear in $\rho$, $U_Y\simeq t_{YN}\,\rho$, with $t_{YN}$
fixed by the forward $YN$ scattering amplitude~\cite{Friedman:2007zza}; $\gamma=1$ is
this limit, and $\gamma\neq1$ parametrises the leading departure from it (Pauli
blocking and short-range correlations) without a specific microscopic model. The
term $c_Y[(\rho/\rhoz)^\beta-(\rho/\rhoz)^\gamma]$ with $\beta=2$ adds the leading
$\rho^2$ contribution of three-body $YNN$ forces. Chiral effective field theory and
quantum Monte Carlo calculations find that this repulsion dominates $U_\Lambda$ at
$2$--$3\rhoz$~\cite{Lonardoni:2014bwa,Haidenbauer:2016vfq,Gerstung:2020ktv}. For
$\gamma=1$, $\beta=2$ and $x=\rho/\rhoz$, Eq.~\eqref{eq:potform} reads
$U_Y=[U_Y(\rhoz)-c_Y]\,x+c_Y\,x^2$, the linear-plus-quadratic form of hypernuclear
potential fits. The local part of the Millener--Dover--Gal potential fitted to
$\Lambda$ single-particle levels, $U_\Lambda=-340\rho+1087.5\rho^2\MeV$ with $\rho$ in
$\fm^{-3}$~\cite{Millener:1988hp}, is Eq.~\eqref{eq:potform} with
$U_\Lambda(\rhoz)=-26.6\MeV$ and $c_\Lambda=27.8\MeV$. A recent fit that includes
Pauli correlations in the linear term finds a $\Lambda NN$ ($\rho^2$) depth of
$11.3\pm1.4\MeV$ at $0.17\fm^{-3}$~\cite{Friedman:2023ucs}. Writing the $\rho^2$ term
so that it vanishes at $\rhoz$ keeps $U_Y(\rhoz)$, the quantity that hypernuclear and
$\Sigma$-atom analyses quote, as an explicit parameter. The term also acts below
saturation: for $\gamma=1$ it equals $-c_Y\,x(1-x)$, deepening the potential by up to
$c_Y/4$ at $\rhoz/2$ and by $0.15\,c_Y$ at $0.82\rhoz$, the density at which the
neutrino observables constrain $U_\Lambda$ (Sec.~\ref{sec:gamma}). The $\Sigma$ depth
can be set separately for each charge ($\Sigma^+,\Sigma^0,\Sigma^-$), since the two
beam polarities populate different charges and it is the $\Sigma^-$ that matters for
the neutron star (Sec.~\ref{sec:puzzle}); the default, used throughout, is one
charge-independent depth. The $\Lambda$ well,
$U_\Lambda(\rhoz)\approx-28\MeV$, traps slow $\Lambda$'s (hypernucleus capture);
the $\Sigma$ potential is mildly repulsive. The $K^+$ feels a weak repulsion
($\approx+25\MeV$) while the $K^-$ is deeply attractive ($\approx-90\MeV$),
leading to kaonic capture.

\begin{figure}[t]
\centering
\includegraphics[width=\columnwidth]{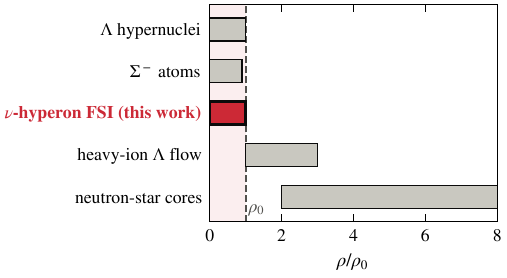}
\caption{Density ``lever arm'': the approximate range of $\rho/\rhoz$ over which each
terrestrial and astrophysical probe constrains the in-medium hyperon potential
$U_Y(\rho)$. Neutrino-induced hyperon final-state interactions (this work,
highlighted) probe the same range as
hypernuclear spectroscopy and $\Sigma^-$ atoms, below the
supra-saturation regime ($\gtrsim2\rhoz$) probed by heavy-ion flow and realised in
neutron-star cores. The maximum mass is governed by the high-density end, which the
neutrino data do not directly constrain.}
\label{fig:leverarm}
\end{figure}

Figure~\ref{fig:leverarm} shows the density range that each probe constrains.
Hypernuclear spectroscopy and $\Sigma^-$ atoms constrain $U_Y$ at and below
saturation; the neutrino-induced hyperon FSI probes the same
$\rho\lesssim\rhoz$ regime; heavy-ion $\Lambda$ flow and neutron-star cores probe the
supra-saturation densities that dominate the maximum mass.

\section{(Anti)neutrinos as a probe of the hyperon potential}
\label{sec:nuhandle}
Charged-current (anti)neutrino interactions on nuclei produce hyperons through
several channels. The cleanest is the Cabibbo-suppressed ($\DS=1$) quasi-elastic
reaction on a bound nucleon,
\begin{equation}
  \bar\nu_\mu + N \;\to\; \mu^+ + Y , \qquad Y\in\{\Lambda,\Sigma^0,\Sigma^-\},
  \label{eq:qe}
\end{equation}
which supplies $\sim\!77\%$ of the strange final states for the SBND antineutrino
beam (reverse horn current, RHC) and $49\%$ for DUNE RHC, where associated production supplies $45\%$ (Table~\ref{tab:comp}).
Neutrino (forward horn current, FHC) beams add the Cabibbo-favoured ($\DS=0$) associated production
$\nu N\to\ell\,Y K$ and single-kaon production. In all cases the hyperon is born
inside the nucleus and must propagate out through nuclear matter before
detection, feeling the in-medium potential $U_Y(\rho)$. Four
consequences follow:
\begin{enumerate}
  \item \textbf{Energy/momentum shift.} As the hyperon leaves the well its kinetic
  energy is shifted by $U_Y$ at the production density; an attractive $\Lambda$
  well lowers the escaping momentum, a repulsive $\Sigma$ well raises it.
  \item \textbf{Trapping and the detector threshold.} A slow $\Lambda$ that cannot
  climb out of the attractive well is captured, which we model with a transport-level
  energy criterion (Sec.~\ref{sec:smc}). The escaping $\Lambda$'s are slowed by the
  same well, and the fraction whose decay products pass the reconstruction thresholds
  falls as the well deepens.
  \item \textbf{$\Sigma\!\to\!\Lambda$ conversion.} The strong reaction
  $\Sigma N\to\Lambda N$ in the medium, and the $\Sigma/\Lambda$ yield, respond to
  how the two species are transported.
  \item \textbf{The FSI $\Sigma^+$ tag.} In antineutrino quasi-elastic
  production on a single nucleon the weak vertex obeys $\Delta S=\Delta Q$ and can make
  only $\Lambda,\Sigma^0,\Sigma^-$. A $\Sigma^+$ in a kaon-less
  quasi-elastic event is therefore an FSI product ($\Lambda p\to\Sigma^+ n$,
  $\Sigma^0 p\to\Sigma^+ n$). It is a kaon-vetoed tag of in-medium dynamics,
  potential-sensitive both through the in-medium threshold shift
  $U_\Sigma(\rho)-U_\Lambda(\rho)$ and through the exit-energy shift $U_\Sigma$.
\end{enumerate}
The associated channel at DUNE energies also produces $\Sigma^+$, which the
antineutrino quasi-elastic channel cannot make. A recent calculation of antineutrino-induced hyperon
production off nuclei~\cite{benitez2024} already includes a $\Lambda$-nucleus FSI
potential and $\Sigma\!\to\!\Lambda$ conversion, estimating $\mathcal{O}(10^4)$
hyperon events at SBND, and MicroBooNE has reported the first antineutrino
quasi-elastic $\Lambda$-on-argon measurement~\cite{microboone2023lambda}.

\subsection{Coupling the potential to the cascade}
\label{sec:smc}
The potential of Eq.~\eqref{eq:potform} is applied in the hyperon intranuclear
cascade: a $\Lambda,\Sigma$ propagates out of the Woods--Saxon nucleus, undergoing
elastic $YN$ scattering, Pauli-blocked knockout, and the charge-conserving
channels $\Sigma N\to\Lambda N$, $\Lambda N\to\Sigma N$, $\Sigma N\to\Sigma' N'$
(sampled by their $YN$ cross sections); on escape its energy is shifted by $U_Y$
evaluated at the production density $\rho_v$,
\begin{equation}
  E_{\rm out} = E_{\rm in} + U_Y(\rho_v) .
  \label{eq:exitshift}
\end{equation}
In the alternative gradient mode the hyperon moves under $-\nabla U_Y$ with
$\sqrt{p^2+m_Y^2}+U_Y$ conserved between collisions, so a hyperon that does not
rescatter leaves with the same energy as in Eq.~\eqref{eq:exitshift}. A third mode
follows the convention of the NuWro cascade: $U_Y(\rho_v)$ is subtracted from the
kinetic energy at the start of the cascade and added back on exit, so a hyperon that
does not rescatter leaves with $E_{\rm in}$ [Sec.~\ref{sec:robust}, item~(b)].

\paragraph{Energy bookkeeping at the vertex.} The struck nucleon enters the hard
process off shell, with energy $E_N=M_N-e_{\rm rem}$, where $e_{\rm rem}$ is the
removal energy of the nuclear model (Sec.~\ref{sec:methods}). The hyperon leaves the
vertex on shell, with momentum $\mathbf{k}=\mathbf{p}+\mathbf{q}$ and energy
$E_{\rm in}=\sqrt{k^2+m_Y^2}$; $U_Y$ does not enter the vertex energy balance and acts
only in the cascade and through Eq.~\eqref{eq:exitshift}. Energy conservation at the
vertex fixes the energy transfer,
\begin{equation}
  \omega = \sqrt{k^2+m_Y^2} - M_N + e_{\rm rem} ,
  \label{eq:omega}
\end{equation}
so at fixed momentum transfer $\mathbf{q}$ the removal energy is carried by the
lepton, and a hyperon that does not rescatter reaches the detector with energy
$\sqrt{k^2+m_Y^2}+U_Y(\rho_v)$, independent of $e_{\rm rem}$. In $\DS=1$
quasi-elastic and in associated $YK$ production there is no outgoing nucleon, so the
only nucleon self-energy in the problem is that of the hole. It reaches the hyperon
only through the momentum distribution of $\mathbf{p}$ and through the weight with
which each $(\mathbf{q},\omega)$ is sampled, and a change of $e_{\rm rem}$ does not
mimic a change of $U_Y$ one for one [Sec.~\ref{sec:robust}, item~(f)]. The
energy-conserving treatments of Refs.~\cite{Sobczyk:2019uej,Hernandez:2025snr} place
the hyperon potential in the vertex $\delta$ function,
$\delta(E_h+\omega-E_Y(\mathbf{k})-V)$, with $E_h$ the energy of the hole. A hyperon
that then leaves along a real potential with its energy conserved reaches the
detector with $E_Y(\mathbf{k})+V$, the same asymptotic energy as
Eq.~\eqref{eq:exitshift} at the same $\mathbf{k}$; the two prescriptions differ by $V$
in the energy transfer of Eq.~\eqref{eq:omega}. For $V=-30\,(\rho/\rhoz)\MeV$ that
difference moves the quasi-elastic peak in $\omega$ by about $30\MeV$ and has a small
effect on single-differential and integrated cross
sections~\cite{Sobczyk:2019uej}; the hyperon observables do not involve the lepton
energy, but the muon means of the HM configurations of Sec.~\ref{sec:robust} do.
At fixed energy transfer the energy-conserving hyperon reaches the detector with
$E_h+\omega$, as in the convention of the NuWro cascade, and Eq.~\eqref{eq:exitshift}
gives $E_h+\omega+U_Y(\rho_v)$ [Sec.~\ref{sec:robust}, item~(b)]. In a local Fermi gas whose $\delta$ function
omits the nucleon potential, the produced baryon implicitly feels the same local
potential as the struck nucleon~\cite{Hernandez:2025snr}. StrangeMC keeps the two
separate, with $e_{\rm rem}$ at the vertex and $U_Y$ at the exit.

\paragraph{Capture.} If $E_{\rm out}\le m_Y$ the hyperon is bound (trapped) and is
removed from the event. This energy-threshold criterion is a transport-level proxy for
hypernuclear capture: shell structure, the angular momentum of the captured state, and
photon/nucleon de-excitation are not modelled, and the absolute capture probability is not predicted. The trapped fraction
is a truth-level quantity and is not one of the fitted observables, which are the
reconstructed-$\Lambda$ fraction $f_\Lambda$, its mean momentum
$\langle p_\Lambda\rangle$ and the $\Sigma^+$ fraction (Sec.~\ref{sec:detreach}). A
$\Lambda$ that leaves with less than $5\MeV$ of kinetic energy gives a decay proton
below $19\MeV$, under the $21\MeV$ proton threshold of Table~\ref{tab:detector}, so
moving the capture boundary within this range leaves the reconstructed observables
unchanged [Sec.~\ref{sec:robust}, item~(b)]; their $U_\Lambda$ dependence comes from
the exit shift of the escaping spectrum relative to the detector thresholds.

The conversion channels carry an in-medium threshold shift
$\Delta U = U_{Y'}(\rho)-U_Y(\rho)$, so the endothermic $\Lambda N\to\Sigma N$
---the route to a $\Sigma^+$ in an antineutrino event---is suppressed when the
$\Sigma$ potential is more repulsive than the $\Lambda$. The per-hyperon species,
outcome (escaped/bound/converted), production density and density at the last
scattering are recorded for every event.

\section{Sensitivity of the neutrino observables}
\label{sec:sens}
\begin{figure}[t]
\centering
\includegraphics[width=\columnwidth]{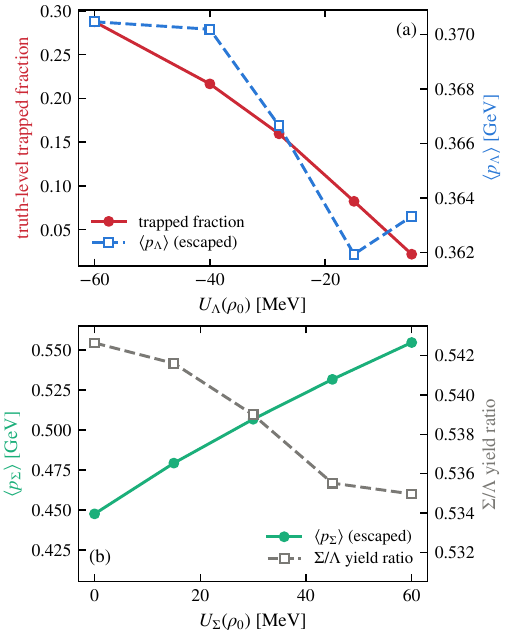}
\caption{Sensitivity of the SBND RHC ($\bar\nu$) hyperon observables to the
in-medium potential. (a)~The truth-level trapped $\Lambda$ fraction (red)
and the mean escaping-$\Lambda$ momentum $\langle p_\Lambda\rangle$ (blue) versus
$U_\Lambda(\rhoz)$. (b)~The mean escaping-$\Sigma$ momentum
$\langle p_\Sigma\rangle$ (green) versus $U_\Sigma(\rhoz)$; the $\Sigma/\Lambda$
yield ratio (grey) is essentially $U_\Sigma$-independent. Each point has
$1.2\times10^4$ hyperons; the statistical uncertainty is $0.002$--$0.005$ on the
trapped fraction and $0.003\GeV$ on the mean momenta.}
\label{fig:potsens}
\end{figure}

We scan the potential parameters around the physical baseline
($U_\Lambda=-28\MeV$, $U_\Sigma=+30\MeV$, $\gamma=1$), one at a time, for the
dominant hyperon channel of each beam (quasi-elastic for $\bar\nu$, associated for
$\nu$) at a representative energy. The potential is an FSI effect and does not
change the hard cross section, so the hyperon kinematics and FSI outcomes are the
relevant observables. Figure~\ref{fig:potsens} shows the response for the SBND
antineutrino beam.

\begin{figure}[b]
\centering
\includegraphics[width=\columnwidth]{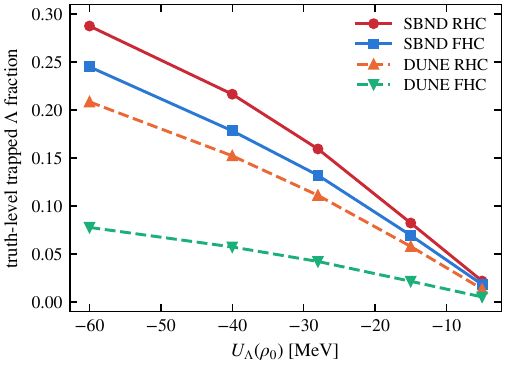}
\caption{Truth-level trapped-$\Lambda$ fraction versus $U_\Lambda(\rhoz)$ for the
four beams. The observable orders by beam energy ($0.160$, $0.132$, $0.111$ and $0.042$
at $U_\Lambda=-28\MeV$ for SBND RHC, SBND FHC, DUNE RHC and DUNE FHC): the
lower-energy SBND beams produce slower, more easily trapped $\Lambda$'s than DUNE.}
\label{fig:potbeams}
\end{figure}

Deepening the attractive $\Lambda$ well
from $U_\Lambda=-5$ to $-60\MeV$ raises the (truth-level) trapped fraction monotonically
from $0.02$ to $0.29$ and hardens the surviving spectrum, $\langle p_\Lambda\rangle:
0.363\to0.370\GeV$ (slow $\Lambda$'s are preferentially captured); between $-5$ and
$-15\MeV$ the change in $\langle p_\Lambda\rangle$ is within the $0.003\GeV$ statistical
uncertainty. Making the
$\Sigma$ well more repulsive from $0$ to $+60\MeV$ raises $\langle p_\Sigma\rangle:
0.447\to0.555\GeV$, while the $\Sigma/\Lambda$ count ratio varies only weakly
($0.54\to0.53$). These are truth-level responses; $\Sigma$ reconstruction is not
modelled, and in the forecast of Sec.~\ref{sec:detreach} the $U_\Sigma$ information
enters through $\langle p_\Lambda\rangle$ and the $\Sigma^+$ fraction.
Figure~\ref{fig:potbeams} compares all four beams: the trapped fraction is largest for
the low-energy SBND beams (where $\Lambda$'s are slow) and smallest for DUNE.

\section{From the potential to the equation of state}
\label{sec:eos}
We build a \emph{toy} EOS that
takes the same $(U_\Lambda,U_\Sigma,\gamma)$ as the simulation. We consider matter of
$n,p,\Lambda,\Sigma^-$ and
leptons $e,\mu$ at zero temperature. Each species is a relativistic Fermi gas with
spin degeneracy $g=2$; the number density and kinetic energy density in terms of
the Fermi momentum $k_F$ are
\begin{widetext}
\begin{align}
  n_i &= \frac{k_{F,i}^3}{3\pi^2}, &
  \varepsilon^{\rm kin}_i &= \frac{1}{8\pi^2}\Big[
    k_F E_F (2k_F^2+m_i^2) - m_i^4\ln\tfrac{k_F+E_F}{m_i}\Big],
    \qquad E_F=\sqrt{k_F^2+m_i^2}.
\end{align}
\end{widetext}
The nucleons carry a Skyrme-like potential energy per baryon,
\begin{equation}
  w_N(u) = a\,u + b\,u^{\sigma} + S_{\rm pot}\,u\,(1-2x_p)^2,\quad u=\frac{n_N}{\rhoz},
  \label{eq:wN}
\end{equation}
with $n_N=n_n+n_p$ and $x_p=n_p/n_N$. The isoscalar coefficients $a,b$ are fixed
analytically by symmetric-matter saturation ($w_N+T_0=-16\MeV$ and
$\partial w_N/\partial n=0$ at $\rhoz$); the exponent $\sigma$ is a stiffness knob
tuned so pure-nucleon matter supports $\Mmax\approx2.45\,\Msun$ ($\sigma=1.8$); the
symmetry term has $S_{\rm pot}=19\MeV$. Each hyperon feels Eq.~\eqref{eq:potform}
at the baryon density, and the total energy density is
\begin{equation}
  \varepsilon = \sum_{i}\varepsilon^{\rm kin}_i
  + n_N\,w_N(u) + \sum_{Y}n_Y\,U_Y(\rho) .
  \label{eq:eps}
\end{equation}
The matter is in $\beta$-equilibrium and electrically neutral,
$\mu_i = \mu_n - q_i\,\mu_e$, $\mu_\mu=\mu_e$, with the Fermi momenta from
$\mu_i = \sqrt{k_{F,i}^2+m_i^2} + U_i(\rho)$. We solve the composition at fixed
$n_B$ by nested bisection ($\mu_e$ for charge neutrality, $\mu_n$ for baryon
number); the pressure follows from the exact relation
\begin{equation}
  P = \sum_i \mu_i n_i - \varepsilon = \mu_n\,n_B - \varepsilon ,
  \label{eq:P}
\end{equation}
which is smooth across a hyperon onset and softens when a new species
appears.

\subsection{TOV equations and the maximum mass}
\label{sec:tov}
A static, spherically symmetric star obeys the Tolman--Oppenheimer--Volkoff (TOV)
equations; with $e=\varepsilon/c^2$,
\begin{align}
  \frac{dm}{dr} &= 4\pi r^2\,e(r), \label{eq:tov1}\\
  \frac{dP}{dr} &= -\,\frac{G\,(e + P/c^2)(m + 4\pi r^3 P/c^2)}
                      {r\,(r - 2Gm/c^2)} . \label{eq:tov2}
\end{align}
For a chosen central pressure we integrate
Eqs.~\eqref{eq:tov1}--\eqref{eq:tov2} outward by fourth-order Runge--Kutta using
$e(P)$ from Eq.~\eqref{eq:P} until $P\to0$; sweeping the central density traces
the mass--radius curve, whose maximum is $\Mmax=\max_{P_c} M(P_c)$. Because the toy
EOS is a functional of $(U_\Lambda,U_\Sigma,\gamma)$, so is its $\Mmax$.

\subsection{Results: maximum mass versus the potential}
\label{sec:eosresults}
\begin{figure}[t]
\centering
\includegraphics[width=\columnwidth]{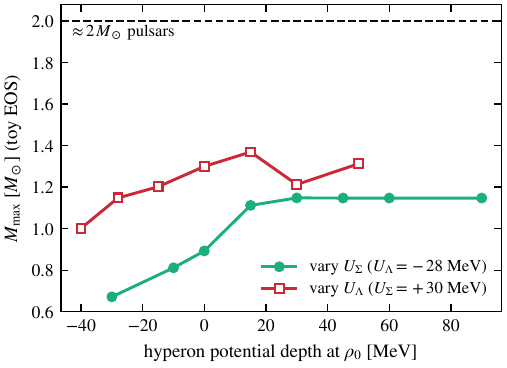}
\caption{Toy neutron-star maximum mass versus the hyperon potential depth at
$\rhoz$, varying $U_\Sigma$ (green, $U_\Lambda=-28\MeV$) and $U_\Lambda$
(red, $U_\Sigma=+30\MeV$). The same parameters drive the simulated observables of
Figs.~\ref{fig:potsens}--\ref{fig:potbeams}; the dashed line marks the
$\approx2\,\Msun$ pulsar constraint.}
\label{fig:eosbridge}
\end{figure}

With hyperons suppressed the toy nucleonic EOS gives $\Mmax=2.45\,\Msun$. Turning
on the physical potentials collapses this to $\Mmax=1.15\,\Msun$, with the
$\Lambda$ appearing at $\rho\approx1.8\rhoz$. Figure~\ref{fig:eosbridge} and
Table~\ref{tab:eos} show the response. (i)~$U_\Sigma$ repulsion
pushes the $\Sigma^-$ onset up until it disappears, raising $\Mmax$ from $0.67$ to
$1.15\,\Msun$. (ii)~Making $U_\Lambda$ less attractive pushes the $\Lambda$ onset up
and lifts $\Mmax$ toward $1.4\,\Msun$, but the composition does not stay fixed along
the scan: at $U_\Lambda\ge0$ the suppressed $\Lambda$ population no longer absorbs
the negative charge, the electron chemical potential rises, and the $\Sigma^-$
returns at $\rho\approx0.67\fm^{-3}$ (Table~\ref{tab:eos}). The lift in
$\Mmax$ is therefore the net of a receding $\Lambda$ and a re-entering $\Sigma^-$.
(iii)~The density slope $\gamma$ acts through the sign of the potential (a
steeper attractive $\Lambda$ softens, a steeper repulsive $\Sigma$ stiffens), and
because the two hyperons pull oppositely the net $\Mmax(\gamma)$ is
non-monotonic---rising slightly from $\gamma=0.5$ to $1.0$ as the steepening
$\Sigma$ repulsion expels the $\Sigma^-$, then falling to $0.81\,\Msun$ at
$\gamma=2.5$ as the deepening $\Lambda$ well takes over. In the toy EOS the density
dependence of $U_Y$ matters as much as its depth at $\rhoz$.

\begin{table}[t]
\centering
\begin{tabular}{@{}lccc@{}}
\toprule
scan & parameter & $\Mmax\,[\Msun]$ & $\Sigma^-$ onset $[\fm^{-3}]$\\
\midrule
vary $U_\Sigma$ & $-30\MeV$ & $0.67$ & $0.23$ \\
($U_\Lambda=-28$) & $0\MeV$ & $0.89$ & $0.27$ \\
                  & $+30\MeV$ & $1.15$ & removed \\
                  & $+60\MeV$ & $1.15$ & removed \\
\midrule
vary $U_\Lambda$ & $-40\MeV$ & $1.00$ & removed \\
($U_\Sigma=+30$)  & $-28\MeV$ & $1.15$ & removed \\
                  & $0\MeV$   & $1.30$ & $0.67$ \\
                  & $+15\MeV$ & $1.37$ & $0.68$ \\
\midrule
vary $\gamma$ & $0.5$ & $1.10$ & $0.64$ \\
($U_\Lambda=-28$, & $1.0$ & $1.15$ & removed \\
$U_\Sigma=+30$) & $1.5$ & $1.00$ & removed \\
                & $2.5$ & $0.81$ & removed \\
\bottomrule
\end{tabular}
\caption{Toy maximum mass and $\Sigma^-$ onset density as the hyperon potential is
varied. ``removed'' means $\Sigma^-$ does not appear up to $8\rhoz$.}
\label{tab:eos}
\end{table}

\section{Two-dimensional mapping}
\label{sec:2d}
\begin{figure}[t]
\centering
\includegraphics[width=\columnwidth]{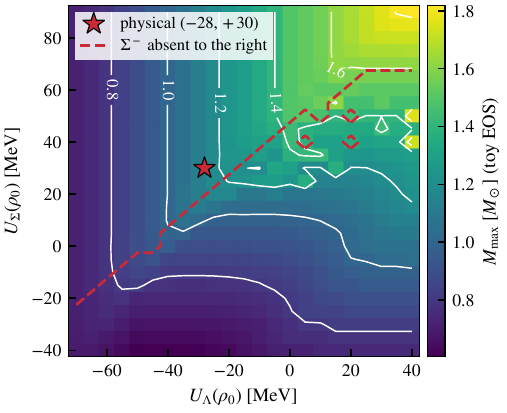}
\caption{Toy neutron-star maximum mass over the $(U_\Lambda,U_\Sigma)$ plane
($\gamma=1$). White contours label $\Mmax$; the red star marks the physical point;
to the right of the red dashed line $\Sigma^-$ is absent from the star.}
\label{fig:eosheatmap}
\end{figure}

\begin{figure}[t]
\centering
\includegraphics[width=\columnwidth]{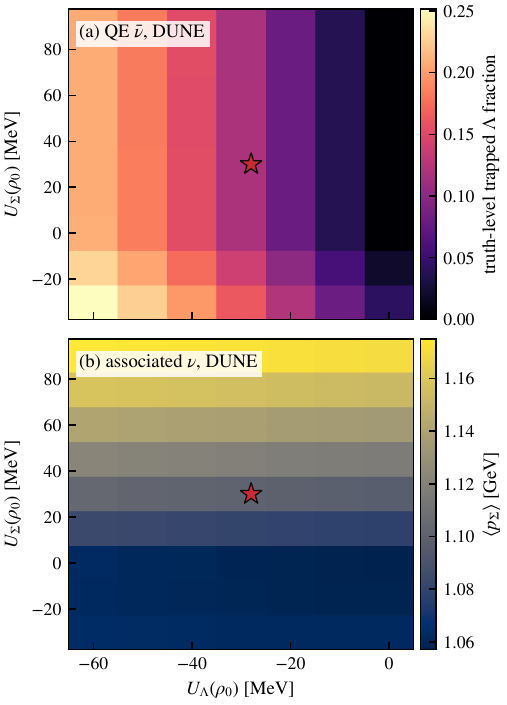}
\caption{Generator observables at truth level over the $(U_\Lambda,U_\Sigma)$ plane
($5\times10^4$ hyperons per node). (a)~The trapped-$\Lambda$ fraction in
quasi-elastic antineutrino events. (b)~The escaping-$\Sigma$ momentum
($\Sigma^+/\Sigma^0$), sensitive to $U_\Sigma$, in associated neutrino (FHC) events at
DUNE energy. The red star is the physical point.}
\label{fig:potheatmap}
\end{figure}

The 1-D scans vary one parameter at a time; the effects of the two potentials on
the observables are correlated and must be mapped jointly. Figure~\ref{fig:eosheatmap} shows the toy
$\Mmax$ over the full $(U_\Lambda,U_\Sigma)$ plane at $\gamma=1$: $\Mmax$ rises
monotonically toward the repulsive corner, the physical point sits near
$1.1$--$1.2\,\Msun$ just inside the $\Sigma^-$-removal boundary, and the strongest
gradient is along $U_\Lambda$. The simulated observables are mapped over the same
plane using both production mechanisms; Fig.~\ref{fig:potheatmap} pairs the
quasi-elastic $\Lambda$-trapping map (truth level) with the associated
$\langle p_\Sigma\rangle$ map at DUNE/FHC (truth level, sensitive to $U_\Sigma$, for
$\Sigma^+$ and $\Sigma^0$).

\subsection{The FSI $\Sigma^+$ tag}
\label{sec:sigmaplus}
\begin{figure}[t]
\centering
\includegraphics[width=\columnwidth]{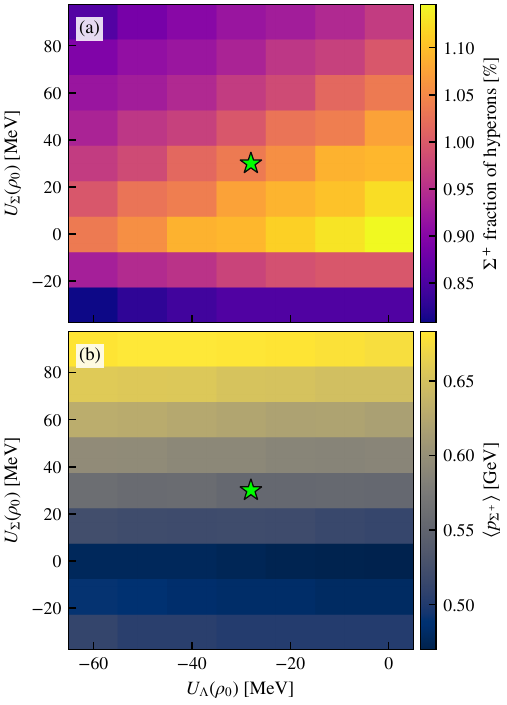}
\caption{The FSI $\Sigma^+$ tag in antineutrino quasi-elastic events over the
$(U_\Lambda,U_\Sigma)$ plane. (a)~$\Sigma^+$ as a fraction of all hyperons.
(b)~The mean escaping $\Sigma^+$ momentum. Both panels show FSI response to the
potential (DUNE RHC, $5\times10^4$ hyperons per node).}
\label{fig:sigmaplus}
\end{figure}

With the conversion and charge-exchange channels active, the simulation produces
$\Sigma^+$ in antineutrino quasi-elastic events through FSI. The $\Sigma^+$ yield is potential-dependent: at DUNE energy
it peaks near $U_\Sigma\!\approx\!0$ and is reduced both for attractive $U_\Sigma$
(exit trapping) and for strongly repulsive $U_\Sigma$ (the in-medium
$\Lambda\to\Sigma^+$ threshold rises with $U_\Sigma-U_\Lambda$); at SBND energy the
reduction on the repulsive side is within the Monte Carlo uncertainty. The escaping
$\langle p_{\Sigma^+}\rangle$ carries the $U_\Sigma$ exit shift
(Fig.~\ref{fig:sigmaplus}). After a kaon veto the $\Sigma^+$ rate is a probe of
in-medium hyperon dynamics and of the potential difference $U_\Sigma-U_\Lambda$ that
controls the $\Sigma^-$ onset in the star; its interpretation is limited by the
$\Sigma N\leftrightarrow\Lambda N$ cross sections [Sec.~\ref{sec:robust}, item~(a)].

Quantitatively ($10^5$ flux-folded
antineutrino quasi-elastic events per beam), a kaon-less post-FSI $\Sigma^+$ appears in
$f_{\Sigma^+}=0.80\pm0.03\%$ (SBND RHC) and $1.06\pm0.03\%$ (DUNE RHC) of
antineutrino quasi-elastic hyperon events, so
$\sigma(\Sigma^+)\simeq0.76\,(1.86)\times10^{-42}\,$cm$^2$/nucleon and, weighting by
the $\bar\nu$-CC fraction of the inclusive RHC sample ($0.72$ and $0.80$),
$\Sigma^+/{\rm CC}\simeq1.5\,(1.5)\times10^{-4}$ at SBND (DUNE) RHC. For the
near-detector samples of Table~\ref{tab:exposure} this is $\sim\!145$ produced
$\Sigma^+$ at SBND RHC and $\sim\!7.4\times10^3$ at DUNE RHC ($43$ and
$2.2\times10^3$ reconstructed at $30\%$). Its one competitor, the secondary
$\pi N\to\Sigma^+K$ feed-down (Sec.~\ref{sec:secondary}), exceeds the signal before
cuts ($S/B\simeq0.55\,(0.68)$ at SBND (DUNE) RHC) but always carries a companion
kaon. The signal is kaon-less, so a charged-kaon veto ($\sim\!69\%$ of feed-down
kaons are $K^+$) raises $S/B$ to $1.8\,(2.2)$; vetoing $K^0_S$ as well leaves the
events whose only kaon is a $K^0_L$ ($15.5\%$ of the feed-down) and gives
$S/B\simeq3.6\,(4.4)$. Events in
which a $\Lambda$ is captured (Sec.~\ref{sec:smc}) add a further source of vertex
activity: the hypernuclear weak decay at this mass number is mainly non-mesonic,
$\Lambda N\to NN$, with nucleons of about $80\MeV$ each and no $p\pi^-$
pair~\cite{alberico2002}. These nucleons enter kaon-less antineutrino events with no
reconstructed $\Lambda$, the sample in which the $\Sigma^+$ tag is selected; the
simulation drops captured hyperons and does not include them. On $\nu$ beams the
vertex makes $\Sigma^+$ directly and the feed-down is roughly $7\times$ larger (the
$\sim\!2\times$ higher $YK$ yield folded with the $\sim\!3.8\times$ higher $\Sigma^+$
fraction), so the tag is an antineutrino observable.

\section{Relativistic mean-field equations of state: GM1 and FSU2H}
\label{sec:rmf}
The toy EOS is schematic. We therefore also implement
nonlinear $\sigma$--$\omega$--$\rho$ relativistic mean-field (RMF) models with the
Boguta--Bodmer scalar self-couplings, for $n,p,\Lambda,\Sigma^-$ plus leptons. The
reference parametrisation is GM1~\cite{glendenning1991}, the set with which
Glendenning and Moszkowski related the hyperon couplings to the binding of the
$\Lambda$ in hypernuclei, the scheme of Eq.~\eqref{eq:anchor}; its hyperonic maximum
mass at the hypernuclear and $\Sigma$-atom depths is a standard reference point of
the hyperon-puzzle literature~\cite{Weissenborn:2011kb}. GM1 is stiff: its incompressibility is $K=301\MeV$,
its symmetry energy $J=32.5\MeV$ with slope $L=94\MeV$, and its $1.4\,\Msun$ star has
$R_{1.4}=14.0$\,km and $\Lambda_{1.4}=895$, above the GW170817 bound. We therefore add
FSU2H~\cite{Tolos:2016hhl,Tolos:2017lgv}, which was built for hyperonic stars with
the same anchoring scheme and adds an $\omega^4$ self-interaction ($\zeta$), an
$\omega$--$\rho$ coupling ($\Lambda_\omega$) and the hidden-strangeness $\phi$ meson:
$K=238\MeV$, $J=30.5\MeV$, $L=44.5\MeV$, $R_{1.4}=13.3$\,km and $\Lambda_{1.4}=763$.
Each baryon acquires a Dirac effective mass $m_B^{*}=m_B-g_{\sigma B}\sigma$ and a
vector self-energy $g_{\omega B}\omega_0+g_{\rho B}I_{3B}\rho_{03}+g_{\phi B}\phi_0$;
the meson mean fields satisfy
\begin{align}
  m_\sigma^2\sigma + g_{\sigma N}\big(b\,m_N S^2 + c\,S^3\big)
    &= \sum_B g_{\sigma B}\,n^{s}_B, \label{eq:rmfsigma}\\
  m_\omega^2\omega_0 + \tfrac{\zeta}{6}g_{\omega N}^4\omega_0^3
  + 2\Lambda_\omega g_{\rho N}^2 g_{\omega N}^2\rho_{03}^2\omega_0
    &= \sum_B g_{\omega B}\,n_B, \nonumber\\
  m_\rho^2\,\rho_{03} + 2\Lambda_\omega\, g_{\omega N}^2 g_{\rho N}^2\,\omega_0^2\,\rho_{03}
    &= \sum_B g_{\rho B}\,I_{3B}\,n_B, \nonumber\\
  m_\phi^2\,\phi_0 &= \sum_Y g_{\phi Y}\,n_Y, \label{eq:rmfvec}
\end{align}
with $S=g_{\sigma N}\sigma$, $b$ and $c$ the dimensionless coefficients of the scalar
self-couplings, $I_{3B}$ the third component of the isospin of baryon $B$, and
$\zeta=\Lambda_\omega=0$ and no $\phi$ field in GM1 and GM3. The $\phi$ couples
only to hyperons, with the SU(6) ratios
$g_{\phi\Lambda}=g_{\phi\Sigma}=-(\sqrt2/3)\,g_{\omega N}$; as in
Ref.~\cite{Tolos:2017lgv}, $g_{\phi\Lambda}$ is reduced by $20\%$ to reproduce the
$\Lambda\Lambda$ bond energy of $^{6}_{\Lambda\Lambda}$He. The fields are solved
self-consistently with $\beta$-equilibrium and charge neutrality at each density.
The nucleon couplings reproduce nuclear saturation (GM1: $E/A=-16.4\MeV$ at
$n_0=0.153\fm^{-3}$, $M^*/M=0.70$; FSU2H: $-16.3\MeV$ at $0.150\fm^{-3}$, $0.59$;
Table~\ref{tab:satprops}). All stars carry the SLy crust
(Appendix~\ref{sec:numerics}).

\begin{figure}[t]
\centering
\includegraphics[width=\columnwidth]{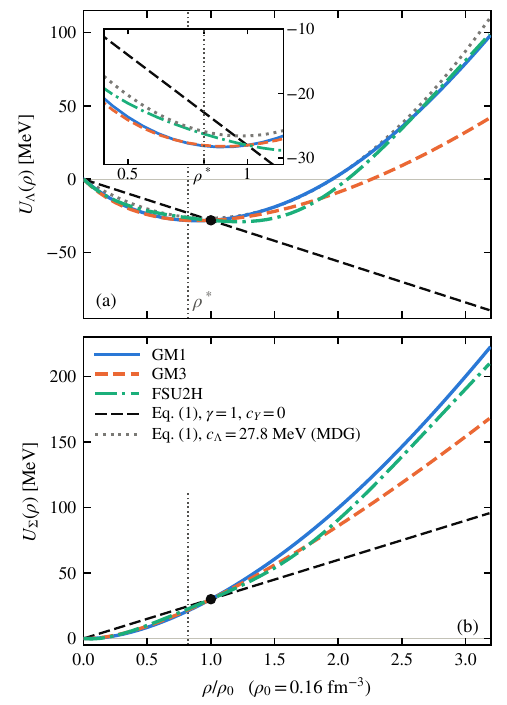}
\caption{Hyperon single-particle potentials at zero momentum in symmetric nuclear
matter for GM1, GM3 and FSU2H. The hyperon couplings are anchored at
$\rhoz=0.16\fm^{-3}$ to $U_\Lambda=-28\MeV$ and $U_\Sigma=+30\MeV$ (dots), with SU(6)
vector couplings and $c_Y=0$. The curves are compared with Eq.~\eqref{eq:potform} at
$\gamma=1$, $c_Y=0$ (black dashed) and, for the $\Lambda$, with the
Millener--Dover--Gal potential, Eq.~\eqref{eq:potform} with $\gamma=1$, $\beta=2$ and
$c_\Lambda=27.8\MeV$ (grey dotted). (a)~$U_\Lambda$; the inset enlarges
$0.4$--$1.15\rhoz$. (b)~$U_\Sigma$. The vertical line marks $\rho^*=0.82\rhoz$, the
density at which the neutrino forecast constrains $U_\Lambda$
(Sec.~\ref{sec:gamma}).}
\label{fig:rmfpot}
\end{figure}

\paragraph{Anchoring the hyperons to the established depths.} The hyperon vector
couplings are fixed by SU(6) quark counting
($x_\omega\equiv g_{\omega Y}/g_{\omega N}=2/3$; $x_\rho=0$ for $\Lambda$, $1$ for
$\Sigma$). The scalar coupling is then derived by requiring the
symmetric-matter potential at $\rhoz=0.16\fm^{-3}$ to equal the measured depth,
\begin{widetext}
\begin{equation}
  U_Y(\rhoz) = -\,g_{\sigma Y}\,\sigma(\rhoz) + g_{\omega Y}\,\omega_0(\rhoz)
  \;\Longrightarrow\;
  x_{\sigma Y} = \frac{x_\omega\,g_{\omega N}\omega_0(\rhoz) - U_Y(\rhoz)}
                      {g_{\sigma N}\,\sigma(\rhoz)} .
  \label{eq:anchor}
\end{equation}
\end{widetext}
Only the value of $U_Y$ at $0.16\fm^{-3}$ enters the RMF, through $x_{\sigma Y}$ in
Eq.~\eqref{eq:anchor}. The exponent $\gamma$ of Eq.~\eqref{eq:potform} does not
enter, and the $\phi$ field, which only hyperons source, vanishes in the symmetric
nucleon matter used for the anchoring. The field equations then fix the hyperon
potential at every other density,
\begin{equation}
  U_Y(\rho)=-g_{\sigma Y}\sigma + g_{\omega Y}\omega_0
  + g_{\rho Y} I_{3Y}\rho_{03} + g_{\phi Y}\phi_0 ,
  \label{eq:rmfUY}
\end{equation}
which coincides with Eq.~\eqref{eq:potform} only at the anchor
(Fig.~\ref{fig:rmfpot}). In symmetric matter the RMF $\Lambda$ potential is flatter
than Eq.~\eqref{eq:potform} with $\gamma=1$ below saturation, with a shallow minimum
near $0.9\rhoz$ in GM1 and GM3 and near $1.2\rhoz$ in FSU2H, and turns repulsive
above it. GM1 gives $U_\Lambda=-28.1$, $+2.6$ and $+79\MeV$ at $\rho^*=0.82\rhoz$,
$2\rhoz$ and $3\rhoz$, against $-23.0$, $-56$ and $-84\MeV$ from
Eq.~\eqref{eq:potform} with $c_\Lambda=0$. Read through Eq.~\eqref{eq:potform}, the
values at $2\rhoz$ and $3\rhoz$ correspond to an intrinsic turn-over
$c_\Lambda\simeq27$--$29\MeV$ in GM1 and $26$--$27\MeV$ in FSU2H, close to the
Millener--Dover--Gal value of Sec.~\ref{sec:potentials}. The RMF $\Sigma$ potential is
less repulsive than Eq.~\eqref{eq:potform} below saturation (by $3.5\MeV$ at $\rho^*$
in GM1) and steeper above it. In argon-like matter with $(N-Z)/A=0.1$, the $\rho$
meson splits $U_{\Sigma^-}$ and $U_{\Sigma^+}$ by $\pm7\MeV$ (GM1) and $\pm5\MeV$
(FSU2H) around $+30\MeV$ at $\rhoz$, and moves $U_\Lambda$ by less than $0.1\MeV$; the
generator uses one charge-independent $\Sigma$ depth (Sec.~\ref{sec:potentials}).

The turn-over enters the RMF as an additive term in the hyperon single-particle
energies,
\begin{equation}
  V_Y(n_B)=c_Y\big[(n_B/\rhoz)^\beta-n_B/\rhoz\big],
  \label{eq:vextra}
\end{equation}
with $\beta=2$, $\rhoz=0.16\fm^{-3}$ and $c_\Sigma=0$. It vanishes at $\rhoz$ and
leaves the anchoring unchanged. Since $V_Y$ depends on $n_B$, thermodynamic
consistency requires the rearrangement term $\sum_Y n_Y\,\partial V_Y/\partial n_B$ in
every baryon chemical potential. That term is common to all baryons and cancels in
$\beta$ equilibrium, so the composition is unchanged, and it adds
$n_B\sum_Y n_Y\,\partial V_Y/\partial n_B$ to the pressure. The $c_\Lambda$ of
Sec.~\ref{sec:bayes} is therefore additional repulsion on top of the intrinsic
turn-over of the RMF continuation.

Equation~\eqref{eq:potform}, the generator and the RMF anchoring all use $\rhoz=0.16\fm^{-3}$. The RMF parametrisations saturate at
$n_0=0.153\fm^{-3}$ (GM1, GM3) and $0.150\fm^{-3}$ (FSU2H); anchoring at $n_0$ instead
changes $U_\Lambda$ by about $0.2\MeV$ in GM1. The two-parameter Fermi density of
$^{40}$Ar in the generator reaches $0.176\fm^{-3}=1.1\rhoz$ at the centre, so the
production densities extend slightly above $\rhoz$.

The neutrino forecast constrains $U_\Lambda$ at $\rho^*$ (Sec.~\ref{sec:gamma}),
where the RMF and Eq.~\eqref{eq:potform} differ by $3$--$5\MeV$. Anchoring the RMF to
the Eq.~\eqref{eq:potform} values at $\rho^*$ instead of $\rhoz$ requires
$U_\Lambda(\rhoz)=-21.9\MeV$ and $U_\Sigma(\rhoz)=+34.1\MeV$ in GM1 ($-24.1$ and
$+32.4\MeV$ in FSU2H) and raises $\Mmax$ by $0.043\,\Msun$ ($0.018\,\Msun$). We take
this as the systematic of transferring the generator potential to the RMF.

\begin{figure}[t]
\centering
\includegraphics[width=\columnwidth]{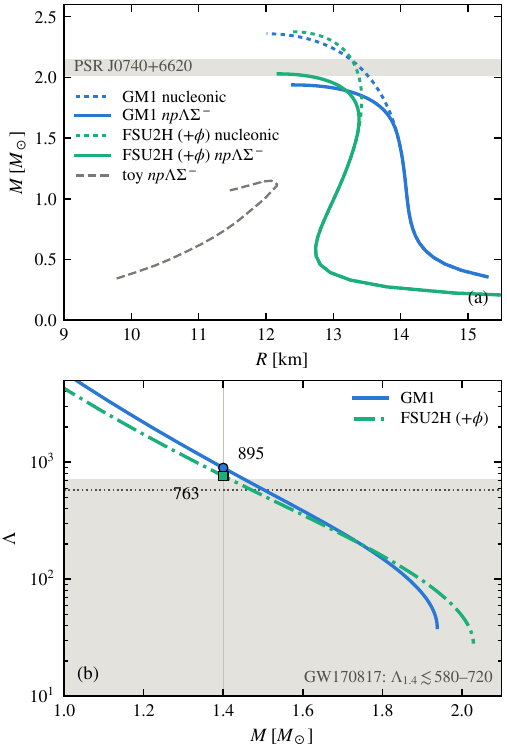}
\caption{RMF predictions at the anchored depths $U_\Lambda(\rhoz)=-28\MeV$ and
$U_\Sigma(\rhoz)=+30\MeV$ ($\rhoz=0.16\fm^{-3}$, $c_\Lambda=c_\Sigma=0$), with the SLy
crust. (a)~Mass--radius curves of GM1 and FSU2H($+\phi$) for nucleonic
matter (dotted) and the $np\Lambda\Sigma^-$ composition (solid), with the toy
hyperonic curve (dashed) for contrast; below the $\Lambda$ onset the two compositions
coincide. The band is the mass of PSR~J0740$+$6620~\cite{fonseca2021}.
(b)~The tidal deformability $\Lambda(M)$ along the hyperonic sequences;
the markers are $\Lambda_{1.4}=895$ (GM1) and $763$ (FSU2H), and the shaded region is
the GW170817 bound
$\Lambda_{1.4}\lesssim580$--$720$~\cite{ligo2017gw170817,abbott2018eos,abbott2019properties}.}
\label{fig:rmf}
\end{figure}

\paragraph{Mass--radius and tidal deformability.} We integrate the TOV equations
together with the relativistic tidal equation for the metric perturbation $y(r)$
\cite{hinderer2008}, yielding the $\ell=2$ Love number $k_2$ and the dimensionless
deformability $\Lambda=\tfrac{2}{3}k_2\,C^{-5}$ with compactness $C=GM/Rc^2$.
Table~\ref{tab:rmf} summarises the outcome at the anchored depths
$U_\Lambda(\rhoz)=-28\MeV$ and $U_\Sigma(\rhoz)=+30\MeV$ with $c_Y=0$. Hyperons lower
$\Mmax$ from $2.36$ to $1.94\,\Msun$ in GM1 and from $2.38$ to $2.03\,\Msun$ in FSU2H.
They do not appear in a $1.4\,\Msun$ star in either model, because its central
density, $0.33\fm^{-3}$, lies at or below the $\Lambda$ onset ($0.35$ and
$0.33\fm^{-3}$). So $R_{1.4}$ and $\Lambda_{1.4}$ are those of the nucleonic EOS and
do not depend on the hyperon potentials: $R_{1.4}=14.0$\,km and $\Lambda_{1.4}=895$
for GM1, and $13.3$\,km and $763$ for FSU2H. GM1 lies above the GW170817 bound
$\Lambda_{1.4}\lesssim580$--$720$~\cite{ligo2017gw170817,abbott2018eos,abbott2019properties},
like other GM1-class mean fields, and FSU2H lies just above its upper end. The softer
GM3 ($\Lambda_{1.4}=605$) lies below the upper end of the range, close to $580$, but
has a hyperonic $\Mmax$ of $1.69\,\Msun$ (Sec.~\ref{sec:eossys}). None of the three
sets meets $\Lambda_{1.4}<580$ together with a hyperonic $\Mmax\geq2\,\Msun$ at
$c_\Lambda=0$. A more repulsive $U_\Sigma$ raises $\Mmax$ mildly and pushes the
$\Sigma^-$ onset to higher density. These stellar quantities are properties of the
chosen parametrisation (Fig.~\ref{fig:rmf}).

\paragraph{The mean-field response.} We record the local slopes of the anchored maps at the physical point ($c_\Lambda=0$, $np\Lambda\Sigma^-$), in units of
$\Msun$ per $10\MeV$,
\begin{align}
  \text{GM1:}\quad
  \frac{d\Mmax}{dU_\Lambda} &= 0.068, &
  \frac{d\Mmax}{dU_\Sigma}  &= 0.005, \label{eq:gm1slope}\\
  \text{FSU2H:}\quad
  \frac{d\Mmax}{dU_\Lambda} &= 0.039, &
  \frac{d\Mmax}{dU_\Sigma}  &= 0.011. \label{eq:fsu2hslope}
\end{align}
In a mean field the anchored depth fixes $g_{\sigma Y}$ and hence both the onset
density and the hyperon fraction above it, so $\Mmax$ tracks $U_Y$ directly. The size
of the slope depends on the vector-meson content: the $\phi$ meson lowers
$d\Mmax/dU_\Lambda$ by $35$--$40\%$ in both models (GM1: $0.068\to0.044$). In FSU2H
the $\Sigma^-$ appears at $0.38\fm^{-3}$, against $0.63\fm^{-3}$ in GM1, and
$d\Mmax/dU_\Sigma$ is twice as large. For the corresponding dependence in a quarkyonic
equation of state see Ref.~\cite{companionPRC}.

\begin{table}[t]
\centering
\caption{RMF results at the anchored depths ($U_\Lambda(\rhoz)=-28$,
$U_\Sigma(\rhoz)=+30\MeV$ at $\rhoz=0.16\fm^{-3}$, $c_Y=0$): nucleonic and hyperonic
($np\Lambda\Sigma^-$) maximum mass in $\Msun$, and the radius and tidal
deformability of the $1.4\,\Msun$ star, with the SLy crust. The toy EOS of
Sec.~\ref{sec:eos} and the observational
anchors~\cite{fonseca2021,abbott2018eos,abbott2019properties} are shown for
comparison.}
\label{tab:rmf}
\begin{tabular}{lcccc}
\hline
 & $\Mmax^{\rm nuc}$ & $\Mmax^{\rm hyp}$ & $R_{1.4}$ [km] & $\Lambda_{1.4}$ \\
\hline
Toy EOS         & $2.45$ & $1.15$ & ---    & --- \\
GM1             & $2.36$ & $1.94$ & $14.0$ & $895$ \\
FSU2H ($+\phi$) & $2.38$ & $2.03$ & $13.3$ & $763$ \\
Observations    & $\ge 2.0$ & $\ge 2.0$ & --- & $\lesssim 580$--$720$ \\
\hline
\end{tabular}
\end{table}

\subsection{Equation-of-state systematic: GM1, GM3, FSU2H and the full octet}
\label{sec:eossys}
\begin{figure}[t]
\centering
\includegraphics[width=\columnwidth]{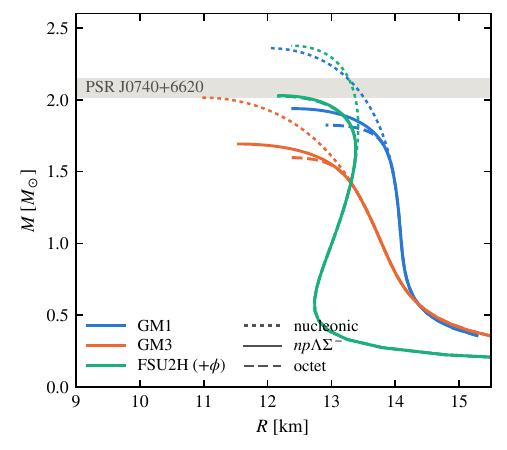}
\caption{The EOS-model systematic: mass--radius sequences for GM1, GM3 and
FSU2H($+\phi$), each nucleonic (dotted), with the $np\Lambda\Sigma^-$ benchmark
(solid) and with the full octet (dashed), at the anchored depths with $c_Y=0$ and the
SLy crust. The band is the mass of PSR~J0740$+$6620~\cite{fonseca2021}.}
\label{fig:eossys}
\end{figure}

\begin{figure}[t]
\centering
\includegraphics[width=\columnwidth]{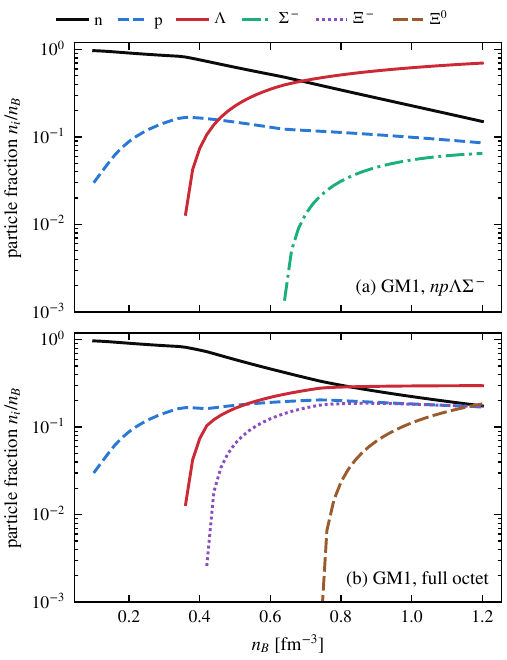}
\caption{$\beta$-equilibrium particle fractions $n_i/n_B$ versus baryon density
for GM1 at the physical depths. (a)~The $np\Lambda\Sigma^-$ benchmark.
(b)~The full octet, where the $\Xi^-$ appears in place of the
$\Sigma^-$.}
\label{fig:composition}
\end{figure}

The spread over RMF parametrisations and baryon content is an EOS-model systematic
that the same anchored $U_Y(\rhoz)$ must carry. Table~\ref{tab:rmfsys} compares GM1
($M^*/M=0.70$, $K=301\MeV$, $L=94\MeV$) with the softer GM3 ($0.78$, $240\MeV$,
$90\MeV$) and with FSU2H ($0.59$, $238\MeV$, $44.5\MeV$). Each has the benchmark
$np\Lambda\Sigma^-$ content and the full baryon octet ($\Sigma^0,\Sigma^+$ and the
$\Xi$ doublet, SU(6) couplings, $U_\Xi=-14\MeV$), and GM1 and FSU2H are given with
and without the $\phi$ meson. Figure~\ref{fig:composition} shows the GM1
compositions. In the GM1 and GM3 octets the $\Xi^-$ appears in place of the
$\Sigma^-$; in FSU2H both appear ($\Sigma^-$ at $0.38$, $\Xi^-$ at $0.41\fm^{-3}$).
Across them $\Mmax^{\rm hyp}$ spans
$1.60$--$2.05\,\Msun$, $R_{1.4}$ spans $13.3$--$14.0$\,km and $\Lambda_{1.4}$ spans
$605$--$895$ (Fig.~\ref{fig:eossys}).

This $\sim\!0.45\,\Msun$ band is the EOS-model systematic at fixed anchored
potential, and it is a lower bound on the theory uncertainty. The hyperon vector
couplings are held at their SU(6) values, and two classes of model are absent:
density-dependent RMF models with running meson--baryon couplings, and hyperonic
EOSs with chiral-EFT two- and three-body
forces~\cite{Haidenbauer:2016vfq,Logoteta:2019utx,Gerstung:2020ktv}. Both change the
supra-saturation stiffness that the neutrino data do not constrain.

\begin{table}[t]
\centering
\caption{Symmetric-nuclear-matter properties of the three parametrisations:
saturation density $n_0$ [fm$^{-3}$], $E/A$, incompressibility $K$, Dirac effective
mass $M^*/M$, symmetry energy $J$ and its slope $L$ (energies in MeV). The coupling
constants are those of Refs.~\cite{glendenning1991} (GM1, GM3) and
\cite{Tolos:2017lgv} (FSU2H).}
\label{tab:satprops}
\begin{tabular}{lcccccc}
\hline
set & $n_0$ & $E/A$ & $K$ & $M^*/M$ & $J$ & $L$ \\
\hline
GM1   & $0.153$ & $-16.4$ & $301$ & $0.70$ & $32.5$ & $94$ \\
GM3   & $0.153$ & $-16.3$ & $240$ & $0.78$ & $32.5$ & $90$ \\
FSU2H & $0.150$ & $-16.3$ & $238$ & $0.59$ & $30.5$ & $44.5$ \\
\hline
\end{tabular}
\end{table}

\begin{table}[t]
\centering
\caption{RMF systematic at the anchored depths ($c_Y=0$, SLy crust): nucleonic and
hyperonic $\Mmax$ [$\Msun$], $R_{1.4}$ [km] and $\Lambda_{1.4}$ for GM1, GM3 and
FSU2H, with and without the $\phi$ meson, for the $np\Lambda\Sigma^-$ benchmark and
the full octet ($U_\Xi=-14\MeV$). Moving the crust--core joint
(Appendix~\ref{sec:numerics}) lowers $R_{1.4}$ by up to $0.1$\,km and $\Lambda_{1.4}$
by up to $5$ (GM1, GM3) and $37$ (FSU2H).}
\label{tab:rmfsys}
\begin{tabular}{llcccc}
\hline
set & content & $\Mmax^{\rm nuc}$ & $\Mmax^{\rm hyp}$ & $R_{1.4}$ & $\Lambda_{1.4}$ \\
\hline
GM1               & $np\Lambda\Sigma^-$ & $2.36$ & $1.94$ & $14.0$ & $895$ \\
GM1               & full octet          & $2.36$ & $1.82$ & $14.0$ & $895$ \\
GM1$+\phi$        & $np\Lambda\Sigma^-$ & $2.36$ & $2.05$ & $14.0$ & $895$ \\
GM1$+\phi$        & full octet          & $2.36$ & $2.00$ & $14.0$ & $895$ \\
GM3               & $np\Lambda\Sigma^-$ & $2.01$ & $1.69$ & $13.3$ & $605$ \\
GM3               & full octet          & $2.01$ & $1.60$ & $13.3$ & $605$ \\
FSU2H ($+\phi$)   & $np\Lambda\Sigma^-$ & $2.38$ & $2.03$ & $13.3$ & $763$ \\
FSU2H ($+\phi$)   & full octet          & $2.38$ & $2.02$ & $13.3$ & $763$ \\
FSU2H (no $\phi$) & $np\Lambda\Sigma^-$ & $2.38$ & $1.89$ & $13.3$ & $763$ \\
FSU2H (no $\phi$) & full octet          & $2.38$ & $1.83$ & $13.3$ & $763$ \\
\hline
\end{tabular}
\end{table}

\section{Feed-down: hyperons from non-strange events}
\label{sec:secondary}
The overwhelming majority of (anti)neutrino interactions make no
strangeness at the vertex, yet their energetic pions can still produce a hyperon
while leaving the nucleus through the strong reaction $\pi N\to Y K$ (threshold
$\sqrt{s}\gtrsim 1.6\GeV$, $p_\pi\gtrsim 0.9\GeV/c$). Because non-strange events
outnumber strange ones by orders of magnitude, this secondary (feed-down)
production is simultaneously a background to the primary measurement and an
additional, independent source of in-medium hyperons that feel the same $U_Y$. We
quantify it with the secondary-production module: a non-strange pion source is
processed and every produced hyperon is transported through the same hyperon
cascade and the in-medium potential $U_Y$. We use realistic non-strange final states from
dedicated NuWro samples (v21.09.1, CC on argon, cascade off, the $\DS=1$ channel
excluded) on the NuMI $\nu_\mu$ and $\bar\nu_\mu$ fluxes with $10^6$ events each.

\begin{figure}[t]
\centering
\includegraphics[width=\columnwidth]{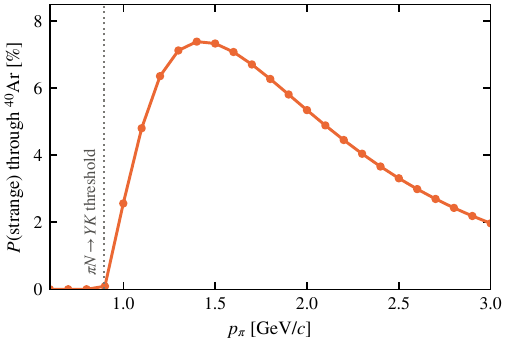}
\caption{Probability that a pion produces an associated $YK$ pair while leaving
$^{40}$Ar, versus pion momentum.}
\label{fig:sec}
\end{figure}

In the secondary-production module a pion is transported through the Woods--Saxon
nucleus and converts with probability
\begin{widetext}
\begin{equation}
  P=1-\exp\!\Big(-\!\int \rho(r)\,\sigma_{\pi N\to YK}(\sqrts)\,dx\Big),
  \qquad \sqrts=\sqrt{m_\pi^2+M_N^2+2E_\pi M_N},
\end{equation}
\end{widetext}
vanishing below the $\sim\!0.9\GeV/c$ threshold and reaching $\sim\!7\%$ at
$1.5\GeV/c$ (Fig.~\ref{fig:sec}); every produced pair conserves charge, baryon
number and net strangeness.

\begin{figure}[t]
\centering
\includegraphics[width=\columnwidth]{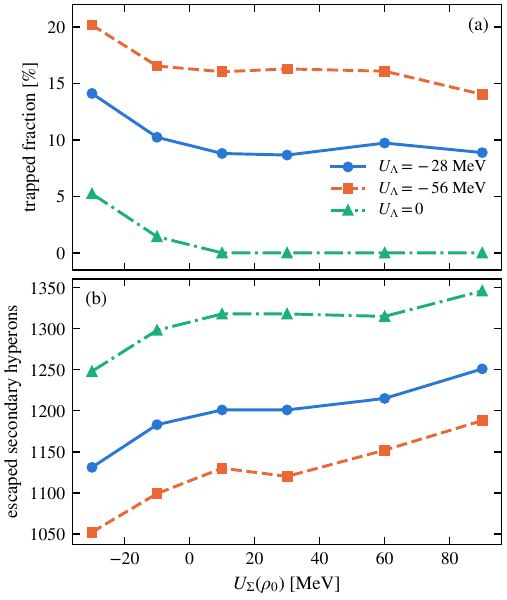}
\caption{Hyperons produced by FSI in non-strange events ($\pi N\to Y K$),
then transported through the in-medium potential. (a)~Their trapped fraction
versus $U_\Sigma$ for several $U_\Lambda$. (b)~The escaping secondary-hyperon
yield. The feed-down hyperons carry the same potential information as the primary
ones.}
\label{fig:secondaryfsi}
\end{figure}

Table~\ref{tab:feeddown} collects the absolute rates. Secondary feed-down deposits
a hyperon in $0.26\%$ ($\nu_\mu$) and $0.13\%$ ($\bar\nu_\mu$) of all CC events,
i.e.\ $8.3\times10^{-43}$ and $1.4\times10^{-43}\,{\rm cm^2/nucleon}$. The
$\nu_\mu$ rate is larger because its flux carries more, harder, $\pi^+$-rich pions
(also making the secondary $\Sigma^+$ fraction $\sim\!3$--$4\times$ higher in
$\nu_\mu$, relevant to the $\Sigma^+$-tag feasibility). Each feed-down pair carries
a kaon as well as the hyperon; $\sim\!69\%$ of these are charged $K^+$ (the
in-medium $K^0 p\to K^+ n$ charge exchange enriches the charged fraction). The
secondary process therefore also feeds a $K^+$ cross section of
$\sim\!5.6\times10^{-43}\,{\rm cm^2/nucleon}$ ($\nu_\mu$), about $4\%$ of the primary
exclusive $K^+$ anchored to Ref.~\cite{fatima2025} (Sec.~\ref{sec:methods}), and an
additional non-exclusive contribution to an inclusive CC-$K^+$ measurement. The
secondary hyperons respond to $U_Y$ exactly as the primaries
(Fig.~\ref{fig:secondaryfsi}) but at somewhat higher momentum, as the parent pion must
clear the $\sim\!1\GeV/c$ threshold.

\begin{table}[t]
\centering
\caption{Secondary ($\pi N\to YK$) hyperon feed-down from the NuWro non-strange
CC-on-argon samples ($10^6$ events each), normalised by the NuWro flux-averaged
total CC cross section. Bound and $\Sigma^+$ fractions are quoted at
$U_\Lambda=-28\MeV$ as $U_\Sigma$ runs from $-30$ to $+30\MeV$.}
\label{tab:feeddown}
\resizebox{\columnwidth}{!}{%
\begin{tabular}{lcc}
\hline
 & $\nu_\mu$ (FHC) & $\bar\nu_\mu$ (RHC) \\
\hline
pions / event ($\langle p_\pi\rangle$)        & $0.71$ ($0.51\GeV$) & $0.59$ ($0.43\GeV$) \\
$YK$ yield / CC event                          & $0.258\%$ & $0.132\%$ \\
$\sigma_{\rm tot,CC}$ [cm$^2$/Ar]             & $1.29\times10^{-38}$ & $4.21\times10^{-39}$ \\
$\sigma(YK)$ [cm$^2$/Ar]                       & $3.33\times10^{-41}$ & $5.54\times10^{-42}$ \\
$\sigma(YK)$ [cm$^2$/nucleon]                 & $8.31\times10^{-43}$ & $1.38\times10^{-43}$ \\
trapped frac. ($U_\Sigma:-30\to+30$)        & $0.142\to0.097$ & $0.141\to0.087$ \\
$\Sigma^+$ frac. ($U_\Sigma:-30\to+30$)     & $0.218\to0.268$ & $0.058\to0.071$ \\
\hline
\end{tabular}}
\end{table}

\section{Projected reach and a joint fit}
\label{sec:reach}

\subsection{Detector-level reach}
\label{sec:detreach}
\begin{figure}[t]
\centering
\includegraphics[width=\columnwidth]{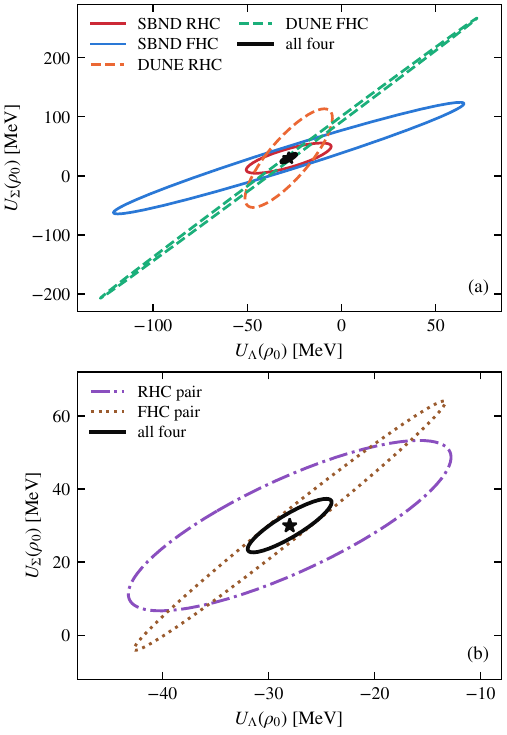}
\caption{Projected statistical $1\sigma$ reach on $U_\Lambda(\rhoz)$ and
$U_\Sigma(\rhoz)$, with $\rhoz=0.16\fm^{-3}$ as in Eq.~\eqref{eq:potform}, from the
reconstructed hyperon observables, per beam and combined, for the exposures of
Table~\ref{tab:exposure} with the normalisation model M2 and the two-body
normalisation profiled (Sec.~\ref{sec:detreach}), at fixed $\gamma=1$ and $c_Y=0$. The
ellipses are a forecast centred on the injected values $U_\Lambda(\rhoz)=-28\MeV$ and
$U_\Sigma(\rhoz)=+30\MeV$. The combined ellipse projects onto
$\delta U_\Lambda(\rhoz)=4.0\MeV$ and $\delta U_\Sigma(\rhoz)=7.3\MeV$
(Table~\ref{tab:stat}). The systematic uncertainties of Sec.~\ref{sec:robust} are not
included. (a)~Single beams and all four combined. (b)~The centre enlarged, with the
antineutrino (RHC) and neutrino (FHC) pairs.}
\label{fig:detector_reach}
\end{figure}

Each escaped hyperon is decayed weakly into its charged-track final state
(Table~\ref{tab:decays}) and passed through a schematic liquid-argon response
(Table~\ref{tab:detector}): per-track momentum smearing ($\sigma_p/p=5\%$),
proton/charged-pion thresholds, a flat $90\%$ track efficiency, and a
$\Lambda\to p\pi^-$ ``$V^0$'' tag (both daughters reconstructed, invariant mass
within $20\MeV$ of $m_\Lambda$). Running this over a $(U_\Lambda,U_\Sigma,\gamma)$ grid
for the four beams gives the observables---the reconstructed $\Lambda$-$V^0$
fraction $f_\Lambda$ per produced hyperon, its mean momentum
$\langle p_\Lambda\rangle$, and the FSI $\Sigma^+$ fraction $f_{\Sigma^+}$---whose
gradients, with statistical errors at the projected exposure, build a Fisher forecast
(Fig.~\ref{fig:detector_reach}). The gradients are taken from a least-squares
quadratic response surface over the full grid (finite differences between adjacent
nodes are limited by Monte Carlo noise); the response values and derivatives at the
truth point, together with the exposures and momentum spreads, are tabulated in
Appendix~\ref{sec:response}. The binomial trial count for the per-hyperon fractions
is the produced-hyperon sample, and the per-event $\Lambda$-momentum spread entering
the mean-momentum error is measured from the post-FSI spectra of each beam's own
production channel ($0.25$, $0.39$, $0.36$ and $0.58\GeV$ for SBND RHC, SBND FHC,
DUNE RHC and DUNE FHC). The $\Sigma^+$ fraction is counted at truth level---its
reconstruction is not modelled, and neither $\Sigma^+$ decay ($p\pi^0$, $n\pi^+$)
leaves the two-charged-prong $V^0$ that makes the $\Lambda$ tractable---so we
discount its binomial trial count by the same $30\%$ reconstruction fraction assumed
for the $\Sigma^+$ yields of Sec.~\ref{sec:sigmaplus}.

\paragraph{Normalisation model.} The number of produced hyperons, the denominator of
$f_\Lambda$ and $f_{\Sigma^+}$, is known only as well as the production cross
sections, the fluxes and the efficiencies. The fit therefore carries these as
nuisance parameters with Gaussian priors (model M2): $20\%$ on the quasi-elastic
cross section, common to the two antineutrino beams; $50\%$ on the associated cross
section, common to the two neutrino beams; $7.5\%$ on each beam's flux; and $3\%$ on
the $\Lambda$ and $30\%$ on the $\Sigma^+$ efficiency of each detector. The yields
respond only weakly to the potential, $d\ln f_\Lambda/dU_\Lambda=0.5$--$0.8\%$ per MeV
on the antineutrino beams, so with these priors the absolute yields carry little
$U_\Lambda$ information. The information comes from the mean $\Lambda$ momentum and
the $\Sigma^+/\Lambda$ ratio. For the quasi-elastic and associated templates alone,
without the two-body component introduced below, removing $\langle p_\Lambda\rangle$
raises the fixed-$\gamma$ $\delta U_\Lambda$ from $2.3$ to $23\MeV$, and removing
$f_\Lambda$ or $f_{\Sigma^+}$ raises it to $4.1$ and $4.0\MeV$. If the
produced-hyperon counts were known exactly, the same observables would give
$\delta U_\Lambda=0.3\MeV$ at fixed $\gamma$; that limit requires the
strange-production cross sections to better than $0.25\%$. Without the $\Sigma^+$
efficiency term (model M1) the width of $U_\Lambda(\rho^*)$ (Sec.~\ref{sec:gamma}) is
$0.92\MeV$ instead of $1.9\MeV$. Treating the $\Sigma^+$ fraction of each beam,
instead of the $\Sigma^+$ efficiency of each detector, as an independent $30\%$
nuisance parameter (model M3) gives $3.1\MeV$.
The forecast makes two further simplifications: each grid point is evaluated at a
representative beam energy rather than flux-folded, and the three observables are
treated as statistically independent. Flux folding lowers the quasi-elastic yields to
$0.54$ (SBND RHC) and $0.65$ (DUNE RHC) of the representative-energy values of
Table~\ref{tab:exposure}, which widens the statistical uncertainty by about $20\%$;
both simplifications are small against the systematic uncertainties of
Sec.~\ref{sec:robust}.

\paragraph{Two-body current.} In antineutrino interactions the strangeness-changing
two-body current, $\bar\nu_\mu NN\to\mu^+YN$~\cite{Nowak:2026twobody,Nowak:2026pheno},
adds hyperons to the quasi-elastic ones: $17\%$ of the quasi-elastic hyperon yield at
SBND RHC and $21\%$ at DUNE RHC. At the DUNE RHC energy its reconstructed $\Lambda$'s
are harder than the quasi-elastic ones, $\langle p_\Lambda\rangle=0.85$ against
$0.61\GeV$, and it gives $2.8$ times more $\Sigma^+$ per hyperon; at SBND RHC the
two components have similar spectra. We add it to both RHC templates, with the
response to $U_Y$ per hyperon of the quasi-elastic template and a $\pm30\%$
normalisation common to the two beams, profiled in every fit. In the fit with the
hyperon observables alone the data constrain this normalisation to a quarter of its
prior, and it widens the all-beam statistical precision to
$\delta U_\Lambda(\rhoz)=4.0\MeV$ at fixed $\gamma$ and $\delta U_\Lambda(\rho^*)=3.4\MeV$
(Table~\ref{tab:stat}). The two-body kinematics are model dependent
(Sec.~\ref{sec:robust}).

\paragraph{Backgrounds and reconstruction realism.} The forecast is signal-only. The
$\Lambda\to p\pi^-$ decay gives a
displaced ($c\tau\sim7.9\,$cm) $V^0$ vertex with a reconstructed invariant mass
within a narrow window of $m_\Lambda$, a topology that liquid-argon TPCs resolve and
that strongly suppresses beam-uncorrelated activity; the FSI $\Sigma^+$ tag
additionally vetoes events with a kaon, removing the associated-production
and single-kaon backgrounds. Cosmic rays are the leading concern at the surface SBND
detector, but beam timing, the in-time fiducial requirement and the displaced-$V^0$
plus kaon-veto topology reduce them. The sensitivity of the result to the proton
threshold and the momentum scale is given in Sec.~\ref{sec:robust}, item~(ii). A
quantitative cosmic and neutral-background rejection study requires a
detector-level analysis.

\begin{table}[t]
\centering
\caption{Implemented weak decay modes with their branching ratios (PDG values for
$\Lambda\to p\pi^-$ and $\Sigma^+\to p\pi^0$, the remainder assigned to the other
listed mode) and asymmetry parameters $\alpha$. The visible reconstruction channel is
$\Lambda\to p\pi^-$.}
\label{tab:decays}
\begin{tabular}{lcc}
\hline
mode & BR & $\alpha$ \\
\hline
$\Lambda\to p\pi^-$    & $0.639$ & $+0.732$ \\
$\Lambda\to n\pi^0$    & $0.361$ & --- \\
$\Sigma^+\to p\pi^0$   & $0.516$ & $-0.980$ \\
$\Sigma^+\to n\pi^+$   & $0.484$ & $+0.068$ \\
$\Sigma^-\to n\pi^-$   & $1.000$ & $-0.068$ \\
$\Sigma^0\to \Lambda\gamma$ & $1.000$ & (EM) \\
\hline
\end{tabular}
\end{table}

\begin{table}[t]
\centering
\caption{Parameters of the schematic liquid-argon response.}
\label{tab:detector}
\begin{tabular}{lc}
\hline
parameter & value \\
\hline
$\sigma_p/p$            & $5\%$ \\
proton KE thr. [MeV]   & $21$ \\
pion KE thr. [MeV]     & $10$ \\
track efficiency       & $0.90$ \\
mass window [MeV]      & $20$ \\
\hline
\end{tabular}
\end{table}

The two antineutrino beams together give $\delta U_\Lambda=15\MeV$ and
$\delta U_\Sigma=23\MeV$, the two neutrino beams $15$ and $34\MeV$
(Table~\ref{tab:stat}). Because the beams at one detector share its $\Lambda$ and
$\Sigma^+$ efficiencies, combining all four gives a statistical reach
$\delta U_\Lambda=4.0\MeV$, $\delta U_\Sigma=7.3\MeV$ at fixed $\gamma=1$
(Fig.~\ref{fig:detector_reach}); without the two-body component it is $2.3$ and
$7.0\MeV$. The Monte Carlo uncertainty of these widths, from a delete-one-seed
jackknife over the response surfaces, is $3$--$6\%$. These numbers are conditional on
$\gamma=1$.

\subsection{The low-density exponent and the decorrelation density}
\label{sec:gamma}
\begin{figure}[t]
\centering
\includegraphics[width=\columnwidth]{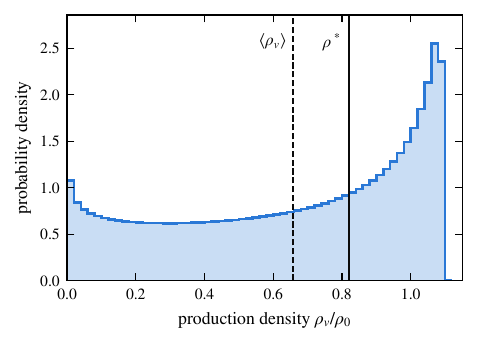}
\caption{Distribution of the hyperon production density $\rho_v/\rhoz$ in $^{40}$Ar
($\rhoz=0.16\fm^{-3}$) for production points drawn from $r^2\rho(r)$ of the
two-parameter Fermi density; it is the same for every channel and beam. The dashed
line marks the mean, $0.66\rhoz$, and the solid line the decorrelation density
$\rho^*=0.82\rhoz$ of the fit [Eq.~\eqref{eq:rhostar}].}
\label{fig:proddens}
\end{figure}

The hyperons are created throughout the nuclear volume. The production point is drawn
from $r^2\rho(r)$, so the production density $\rho_v$ follows the distribution of
Fig.~\ref{fig:proddens}, which is the same for every beam and channel: its mean is
$0.66\rhoz$ and its median $0.73\rhoz$, $34\%$ of the hyperons are born below
$0.5\rhoz$, and $21\%$ above $\rhoz$ (the central density of argon is $1.1\rhoz$). The
exit shift and the threshold shifts sample $U_Y(\rho_v)$, so to leading order the
observables respond to $\langle U_Y(\rho_v)\rangle=U_Y(\rhoz)\,
\langle(\rho_v/\rhoz)^\gamma\rangle$: the anchor and the exponent enter through one
combination, and holding $\gamma=1$ converts a two-parameter degeneracy into a
one-parameter measurement.

We quantify the effect by extending the reach grid to a third axis,
$\gamma\in\{0.6,1.0,1.4\}$, and building the Fisher matrix on
$(U_\Lambda,U_\Sigma,\gamma)$ and the normalisation parameters from a quadratic
response surface over all $27$ nodes per beam. The degeneracy direction depends on
the observable: at the truth point $dU_\Lambda/d\gamma=-6.3$ and $-6.1\MeV$ per unit
$\gamma$ for the SBND RHC and DUNE RHC $\Lambda$ yields, $-3.3$ and $-4.5\MeV$ for the
SBND FHC and DUNE FHC $\Lambda$ yields, $-5.1$ to $-6.9\MeV$ for the $\Lambda$ momenta
and $-11$ to $-15\MeV$ for the $\Sigma^+$ fraction. With the normalisations free within
their priors the yields contribute only through ratios. For the templates without the
two-body component the combination determines $\gamma$ only to $\delta\gamma\approx2$,
and
\begin{align}
  \delta U_\Lambda(\rhoz)&:\;
  2.3\MeV\;\xrightarrow[\ \text{over }\gamma\ ]{\text{marginalise}}\;\gtrsim10\MeV ,
  \nonumber\\
  \delta U_\Lambda(\rho^*)&=1.9\MeV ,\qquad \rho^*=(0.82\pm0.02)\,\rhoz ,
  \label{eq:gammadeg}
\end{align}
where the $\gamma$-free width at $\rhoz$ has not converged in the Monte Carlo
statistics of the response surfaces. $\delta U_\Sigma$ grows from $7.0$ to
$\sim\!10\MeV$. A $\pm0.2$ prior on $\gamma$ gives $\delta U_\Lambda(\rhoz)=2.6\MeV$.
With the two-body normalisation profiled, $\delta U_\Lambda(\rho^*)=3.4\MeV$ at
$x^*=0.86$ (Table~\ref{tab:stat}), and the $\gamma$-free $\delta U_\Sigma$ is $18\MeV$.

\paragraph{The decorrelation density.} The quantity the data determine without an
assumption on $\gamma$ is the potential at the density where the fitted
$U_\Lambda(x\rhoz)=U_\Lambda(\rhoz)\,x^\gamma$ is uncorrelated with $\gamma$.
Linearising about $\gamma=1$, $\delta U_\Lambda(x\rhoz)=x\,\delta U_\Lambda(\rhoz)+
U_0\,x\ln x\,\delta\gamma$ with $U_0=-28\MeV$, and the covariance with $\gamma$
vanishes at
\begin{equation}
  x^* = \exp\!\Big(-\frac{C_{U\gamma}}{U_0\,C_{\gamma\gamma}}\Big) ,
  \label{eq:rhostar}
\end{equation}
where $C$ is the covariance matrix of the $\gamma$-free fit. At $\rho^*=x^*\rhoz$ the
width is the fixed-$\gamma$ width scaled by $x^*$,
$\delta U_\Lambda(\rho^*)=x^*\,\delta U_\Lambda|_{\gamma=1}$, and with the hyperon
observables alone it lies within $2\times10^{-3}$ in $x$ of the minimum of
$\delta U_\Lambda(x\rhoz)$; in the configurations H$+$PC and HM0$+$PC, which profile the
production model, the two differ by up to $0.005$, and in HM$+$PC by $0.015$. For M2 without
the two-body component $x^*=0.82$ and $\delta U_\Lambda(\rho^*)=1.9\MeV$, determined
to $5\%$ by the Monte Carlo statistics of the grid. The value of $x^*$ follows from
the distribution of Fig.~\ref{fig:proddens}: for the family $U_0x^\gamma$ at fixed
$\langle U_Y(\rho_v)\rangle$, $U(x\rhoz)$ is stationary in $\gamma$ at
$\ln x=\langle x_v\ln x_v\rangle/\langle x_v\rangle$ with $x_v=\rho_v/\rhoz$, which gives $x=0.78$, and the
weighting by escape and reconstruction moves this to $0.82$. The observables and
nuisance parameters of a fit move it further: $x^*=0.86$ with the two-body normalisation profiled
(Table~\ref{tab:stat}), and $0.81$, $0.67$ and $0.82$ in the configurations H$+$PC,
HM$+$PC and HM0$+$PC of Sec.~\ref{sec:robust}, which profile the production model. We
take $U_\Lambda(\rho^*)$ as the primary result, with $\rho^*=0.82\rhoz$, the
decorrelation density of HM0$+$PC. $U_\Lambda(\rhoz)$ is an extrapolation within
Eq.~\eqref{eq:potform} and is quoted at fixed $\gamma=1$ or with a prior on $\gamma$.

\begin{table}[t]
\centering
\caption{Statistical precision of the forecast with the hyperon observables, under
the normalisation model M2 (Sec.~\ref{sec:detreach}) and with the normalisation of the
two-body component profiled (configuration H of Sec.~\ref{sec:robust}), in MeV, for
all four beams and for the antineutrino (RHC) and neutrino (FHC) pairs alone:
$\delta U_\Lambda(\rhoz)$ and $\delta U_\Sigma(\rhoz)$ at fixed $\gamma=1$, and the
width of $U_\Lambda$ at the decorrelation density $\rho^*=x^*\rhoz$ with $\gamma$ free.
The two-body component enters the RHC templates only. Without it the all-beam entries are
$2.33$, $7.00$, $0.82$ and $1.91$, a prior $\gamma=1.0\pm0.2$ gives
$\delta U_\Lambda(\rhoz)=2.57\MeV$, and with $\gamma$ free
$\delta U_\Lambda(\rhoz)\gtrsim10\MeV$ (not converged); the RHC-pair entries are
$6.05$, $20.4$, $0.80$ and $4.82$. The Monte Carlo uncertainty of the all-beam widths
is $3$--$6\%$.}
\label{tab:stat}
\begin{tabular}{lcccc}
\hline
beams & $\delta U_\Lambda(\rhoz)$ & $\delta U_\Sigma(\rhoz)$ & $x^*$ &
$\delta U_\Lambda(\rho^*)$ \\
\hline
all four        & $3.96$ & $7.32$ & $0.86$ & $3.42$ \\
RHC pair        & $15.2$ & $23.3$ & $0.80$ & $12.2$ \\
FHC pair        & $14.7$ & $34.1$ & $0.78$ & $11.4$ \\
\hline
\end{tabular}
\end{table}

To test the dependence on the functional form we generate the truth node with six
other density dependences, all with $U_\Lambda(\rhoz)=-28\MeV$, and fit them with the
power-law templates (Table~\ref{tab:shapes}): $\gamma=0.6$ and $1.4$; the turn-over of
Eq.~\eqref{eq:potform} with $\beta=2$ and $c_\Lambda=15$ and $30\MeV$; the
Millener--Dover--Gal form of Sec.~\ref{sec:potentials} scaled to $-28\MeV$ at
$\rhoz$~\cite{Millener:1988hp}; and a saturating form $U_0\,2x/(1+x)$. At the decorrelation
density the fitted potential reproduces the true one to $1.4\MeV$ or better. At $\rhoz$ with
$\gamma$ fixed the
offsets reach $4.5\MeV$, and they change sign between forms. The extracted
$U_\Lambda(\rhoz)$ is therefore conditional on the assumed functional form, while
$U_\Lambda(\rho^*)$ is stable to the precision of this test. The Millener--Dover--Gal
form, carried as the density-shape item of Table~\ref{tab:systematics}, gives
$+0.9$, $+1.9$ and $+1.5\MeV$ in the configurations H$+$PC, HM$+$PC and HM0$+$PC.
The reconstructed $\Lambda$'s that carry the information have mean momenta of
$0.4$--$1.0\GeV/c$ (Table~\ref{tab:response}), while the hypernuclear $U_\Lambda$
refers to $k\approx0$. A momentum-dependent $U_\Lambda$ with effective mass
$m^*/m=0.8$ ($0.9$) at $\rhoz$ moves $U_\Lambda(\rho^*)$ by $-60$ ($-27$)$\MeV$ in
configuration H, and by $-11$ ($-5$)$\MeV$ without the two-body component. These shifts
are not included in the totals of Table~\ref{tab:systematics}. The fitted
$U_\Lambda$ is an effective potential at the momenta of the escaping hyperons, and
its comparison with the hypernuclear value at $k\approx0$ requires a model of the
momentum dependence.

\begin{table}[t]
\centering
\caption{Fitted minus true $U_\Lambda$, in MeV, when the truth node is generated with
another density dependence (all with $U_\Lambda(\rhoz)=-28\MeV$) and fitted with the
power-law templates ($c_\Lambda=0$) in configuration H (Table~\ref{tab:stat}): at the
decorrelation density of configuration H, $0.86\rhoz$, with $\gamma$ free, and at $\rhoz$ with
$\gamma=1$. The errors are Monte Carlo uncertainties.}
\label{tab:shapes}
\begin{tabular}{lcc}
\hline
density dependence & $U_\Lambda(0.86\rhoz)$ & $U_\Lambda(\rhoz)$, $\gamma=1$ \\
\hline
$\gamma=0.6$                         & $-1.4\pm0.3$ & $-3.3\pm0.4$ \\
$\gamma=1.4$                         & $+0.4\pm0.2$ & $+2.1\pm0.2$ \\
turn-over, $c_\Lambda=15\MeV$        & $-0.5\pm0.3$ & $-2.6\pm0.4$ \\
turn-over, $c_\Lambda=30\MeV$        & $-0.4\pm0.4$ & $-4.5\pm0.4$ \\
Millener--Dover--Gal                 & $-0.4\pm0.4$ & $-4.4\pm0.4$ \\
saturating, $2x/(1+x)$               & $-1.1\pm0.3$ & $-3.2\pm0.4$ \\
\hline
\end{tabular}
\end{table}

With $\gamma$ free the neutrino data do not improve on the $\pm4\MeV$ hypernuclear
prior on $U_\Lambda(\rhoz)$. In the joint fit of Sec.~\ref{sec:bayes} the posterior on the
potentials spreads $\Mmax$ at fixed $c_\Lambda$ by $0.02$--$0.03\,\Msun$ (68\% half-width)
with statistical errors and by $0.02$--$0.05\,\Msun$ with the systematic uncertainties. This
is comparable to the
$0.043\,\Msun$ by which anchoring the RMF at $\rho^*$ instead of $\rhoz$ moves $\Mmax$
in GM1 and the $0.018\,\Msun$ in FSU2H (Sec.~\ref{sec:rmf}), and an order of magnitude
below the EOS-model systematic of Sec.~\ref{sec:eossys}.

\paragraph{On an SBND antineutrino run.} The soft BNB antineutrino beam (SBND RHC)
supplies the cleanest quasi-elastic hyperon sample---$\sim\!77\%$ of its strange
final states are quasi-elastic hyperons (Table~\ref{tab:comp})---together with the
$\Sigma^+$ tag. Whether SBND will run in reverse-horn (antineutrino) mode is not yet
decided. On its own SBND RHC does not constrain $U_\Lambda$ once the quasi-elastic
normalisation is free within its prior ($\delta U_\Lambda=23\MeV$ at fixed $\gamma$).
It shares that normalisation and the two-body normalisation with DUNE RHC, and the
pair gives $\delta U_\Lambda=15\MeV$ and $\delta U_\Sigma=23\MeV$, against $23$ and
$83\MeV$ for DUNE RHC alone. In configuration HM0$+$PC of Sec.~\ref{sec:robust} the
total uncertainty on $U_\Lambda(\rho^*)$ is $31\MeV$ for DUNE RHC alone, for the RHC
pair and for all four beams. The pair narrows the profiled width
of $U_\Lambda(\rho^*)$ from $20$ to $13\MeV$, and the potential-convention item of
Sec.~\ref{sec:robust}, item~(b), grows from $+6.0$ to $+19.3\MeV$.

\subsection{A joint fit and the maximum-mass band}
\label{sec:bayes}
A dense-grid Bayesian fit over $(U_\Lambda(\rhoz),U_\Sigma(\rhoz),c_\Lambda)$
combines the projected neutrino constraint with Gaussian priors from hypernuclear
$\Lambda$ binding ($U_\Lambda=-30\pm4\MeV$ \cite{gal2016}) and from $\Sigma$ atoms and
$(\pi^-,K^+)$ data ($U_\Sigma=+30\pm20\MeV$ \cite{saha2004}), and with a prior on the
additional repulsion $c_\Lambda$ of Eq.~\eqref{eq:vextra} taken from chiral effective
field theory.

No measurement fixes $c_\Lambda$. $\Lambda$ directed flow in Au+Au collisions at
$\sqrt{s_{NN}}=3$--$30\GeV$ is reproduced by chiral-EFT $\Lambda$ potentials with and
without $\Lambda NN$ forces, and is insensitive to their density
dependence~\cite{nara2022,ohnishi2022}. The prior therefore comes from Brueckner
calculations of $U_\Lambda(p=0,\rho)$ with the SU(3) chiral NLO13 hyperon--nucleon
interaction~\cite{Haidenbauer:2016vfq,Gerstung:2020ktv}. In Ref.~\cite{Gerstung:2020ktv}
the $\Lambda NN$ force is taken in the decuplet-saturation approximation, its two
contact constants are fixed by $U_\Lambda(\rhoz)=-30\MeV$ in symmetric matter, and of
the allowed pairs those giving $U_\Lambda(3\rhoz)>80\MeV$ in neutron matter are kept;
with the cutoff variation the band reaches down to $70\MeV$ there. The calculation
converges up to $3.5\rhoz$. Table~\ref{tab:clambda} lists these bands,
the earlier results of Ref.~\cite{Haidenbauer:2016vfq}, which end at $2.2\rhoz$, and
the RMF potentials at $c_\Lambda=0$.

\begin{table}[t]
\centering
\caption{$\Lambda$ single-particle potential at zero momentum, in MeV, at $2\rhoz$ and
$3\rhoz$ ($\rhoz=0.16\fm^{-3}$) in symmetric nuclear matter (SNM) and neutron matter
(PNM). Chiral EFT: Brueckner calculations with the NLO13 hyperon--nucleon interaction
with two-body forces (2BF) and with $\Lambda NN$ forces added (2+3BF). The ranges come
from the cutoff and, in Ref.~\cite{Gerstung:2020ktv}, from the choice of $\Lambda NN$
contact constants; they are read from Fig.~7 of Ref.~\cite{Gerstung:2020ktv} and
Fig.~1 of Ref.~\cite{Haidenbauer:2016vfq} to within $1\MeV$. RMF: GM1 and FSU2H($+\phi$)
anchored to $U_\Lambda(\rhoz)=-28\MeV$ in SNM, $c_\Lambda=0$. Last two rows: range of
$c_\Lambda$ (MeV) for which the RMF potential plus Eq.~\eqref{eq:vextra} lies inside
the 2+3BF band of Ref.~\cite{Gerstung:2020ktv}.}
\label{tab:clambda}
\setlength{\tabcolsep}{2pt}\small
\begin{tabular}{lcccc}
\hline
 & \multicolumn{2}{c}{SNM} & \multicolumn{2}{c}{PNM} \\
 & $2\rhoz$ & $3\rhoz$ & $2\rhoz$ & $3\rhoz$ \\
\hline
2BF \cite{Gerstung:2020ktv}   & $[-21,-17]$ & $[15,23]$ & $[-10,-6]$ & $[41,52]$ \\
2+3BF \cite{Gerstung:2020ktv} & $[-6,5]$    & $[59,92]$ & $[-1,5]$   & $[70,92]$ \\
2BF \cite{Haidenbauer:2016vfq}   & $[-11,-6]$ & --- & $[10,12]$ & --- \\
2+3BF \cite{Haidenbauer:2016vfq} & $[4,8]$    & --- & $[18,21]$ & --- \\
GM1   & $2.6$  & $79.0$ & $14.7$ & $97.6$ \\
FSU2H & $-4.9$ & $81.0$ & $8.7$  & $96.1$ \\
\hline
$c_\Lambda$, GM1   & $[-4.1,1.1]$ & $[-3.4,2.2]$ & $[-7.9,-4.7]$ & $[-4.6,-1.0]$ \\
$c_\Lambda$, FSU2H & $[-0.4,4.9]$ & $[-3.7,1.9]$ & $[-4.9,-1.7]$ & $[-4.4,-0.8]$ \\
\hline
\end{tabular}
\end{table}

The term of Eq.~\eqref{eq:vextra} depends only on $n_B$ and adds $2c_\Lambda$ at
$2\rhoz$ and $6c_\Lambda$ at $3\rhoz$ in any medium. At $c_\Lambda=0$ both RMF
continuations lie inside the $\Lambda NN$ band in symmetric matter at $2\rhoz$ and
$3\rhoz$; in neutron matter they lie above it, by $3$ to $9\MeV$ at $2\rhoz$ and by
$4.5$ to $6\MeV$ at $3\rhoz$. Matching at $3\rhoz$, inside the converged range of the
chiral calculation and where the term already adds $6c_\Lambda$, gives
$-4.6\le c_\Lambda\le2.2\MeV$ for GM1 and $-4.4\le c_\Lambda\le1.9\MeV$ for FSU2H over
the two media; at $2.5\rhoz$ and $3.5\rhoz$ these limits move by at most $1\MeV$.
$\beta$-stable matter, with proton fractions of $0.10$--$0.17$ at $2$--$3.5\rhoz$ in
these models, lies between the two media and close to neutron matter. With two-body
forces alone the chiral band would require $c_\Lambda=-11$ to $-7\MeV$ at $3\rhoz$.
At $2\rhoz$ the central values of the two chiral calculations differ by $6$ to
$19\MeV$, and the $\Lambda NN$ results of
Ref.~\cite{Haidenbauer:2016vfq} correspond to $c_\Lambda=0.6$--$3.0\MeV$ for GM1 and
$4.3$--$6.6\MeV$ for FSU2H. We take for both models a uniform prior
$-5\le c_\Lambda\le2.5\MeV$. Because the $\Lambda NN$ constants of
Ref.~\cite{Gerstung:2020ktv} were chosen for maximal repulsion at $3\rhoz$, the upper
end of this range is the most repulsive value supported there; weaker $\Lambda NN$
forces lower $c_\Lambda$. Anchoring the RMF at $U_\Lambda(\rhoz)=-30\MeV$, the depth
imposed in Ref.~\cite{Gerstung:2020ktv}, instead of $-28\MeV$ moves the limits up by
$0.6$--$0.7\MeV$.

\begin{figure}[t]
\centering
\includegraphics[width=\columnwidth]{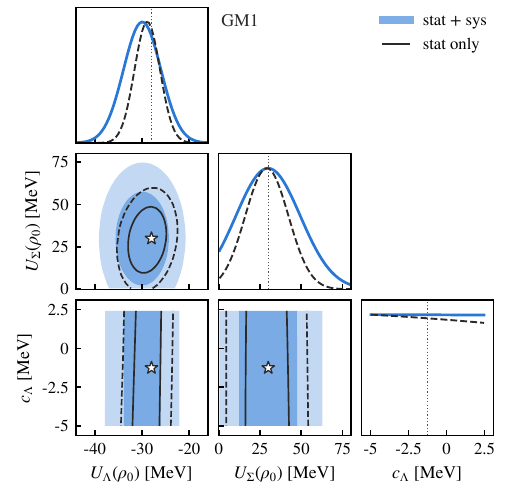}
\caption{Joint posterior on $(U_\Lambda(\rhoz),U_\Sigma(\rhoz),c_\Lambda)$ for GM1.
Filled contours ($68\%$ and $95\%$): neutrino forecast in configuration HM0$+$PC with
its systematic uncertainties (Sec.~\ref{sec:robust}); lines: configuration H with
statistical errors only (solid $68\%$, dashed $95\%$). Both are combined with the priors
$U_\Lambda(\rhoz)=-30\pm4\MeV$, $U_\Sigma(\rhoz)=+30\pm20\MeV$ and $c_\Lambda$ uniform
in $[-5,+2.5]\MeV$ (Table~\ref{tab:clambda}). Stars mark the injected values
($-28\MeV$, $+30\MeV$, $c_\Lambda=-1.25\MeV$). The posterior on the potentials does not
depend on the EOS model, and the $c_\Lambda$ marginal follows its prior.}
\label{fig:bayes}
\end{figure}

\begin{figure}[t]
\centering
\includegraphics[width=\columnwidth]{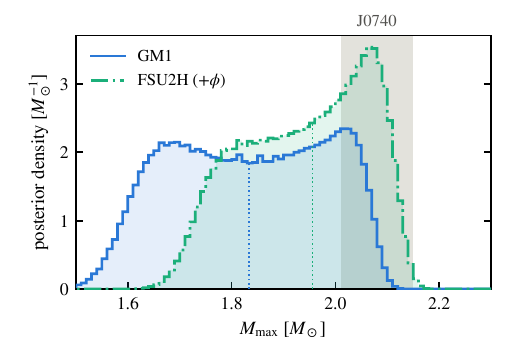}
\caption{Posterior on the neutron-star maximum mass from the joint fit of
Fig.~\ref{fig:bayes} (HM0$+$PC with systematic uncertainties), for GM1 and
FSU2H($+\phi$), with the medians dotted and the mass of
PSR~J0740$+$6620~\cite{fonseca2021} shaded. The prior on $c_\Lambda$ is uniform in
$[-5,+2.5]\MeV$, the range in which the RMF $\Lambda$ potential stays within
chiral-EFT results with $\Lambda NN$ forces at $3\rhoz$ (Table~\ref{tab:clambda}).
The neutrino data do not constrain $c_\Lambda$, so the posterior is this prior mapped
through Fig.~\ref{fig:mmaxcl}.}
\label{fig:bayesmmax}
\end{figure}

Each grid point carries an RMF $\Mmax$ from the three-dimensional tables
$\Mmax(U_\Lambda,U_\Sigma,c_\Lambda)$ of GM1 and FSU2H, with crust and rearrangement
term and nodes every $2.5\MeV$ in $c_\Lambda$ from $-10$ to $30\MeV$, so the posterior
on the potential (Fig.~\ref{fig:bayes}) propagates to a posterior on the maximum mass
(Fig.~\ref{fig:bayesmmax}). Between the nodes $\Mmax$ is interpolated with monotone
cubic splines, which reproduce direct RMF solutions to $0.001\,\Msun$.

\begin{figure}[t]
\centering
\includegraphics[width=\columnwidth]{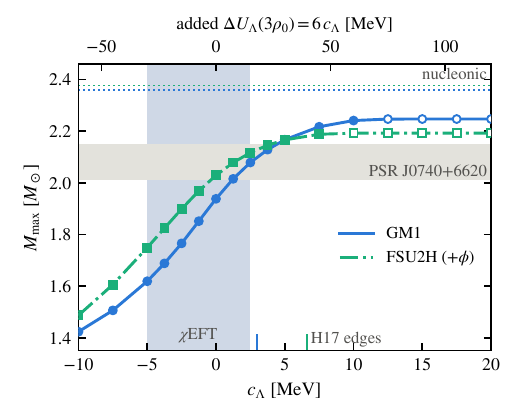}
\caption{Maximum mass against the coefficient $c_\Lambda$ of the additional
turn-over, Eq.~\eqref{eq:vextra}, at the anchored depths $U_\Lambda(\rhoz)=-28\MeV$
and $U_\Sigma(\rhoz)=+30\MeV$, for GM1 and FSU2H($+\phi$) ($np\Lambda\Sigma^-$, SLy
crust, rearrangement term included, $c_\Sigma=0$). The shaded interval is the
chiral-EFT range $-5\le c_\Lambda\le2.5\MeV$ (Table~\ref{tab:clambda}); the ticks (H17 edges) mark
the upper ends of the ranges corresponding to the $\Lambda NN$ results of
Ref.~\cite{Haidenbauer:2016vfq} at $2\rhoz$, $3.0\MeV$ (GM1) and $6.6\MeV$ (FSU2H).
Filled markers: $\Lambda$ present at the centre of the maximum-mass star. Open
markers: the maximum-mass core contains nucleons and $\Sigma^-$ only. Dotted lines:
nucleonic maximum masses. The horizontal band is the mass of
PSR~J0740$+$6620~\cite{fonseca2021}. The top axis gives the repulsion the term adds at
$3\rhoz$, $6c_\Lambda$.}
\label{fig:mmaxcl}
\end{figure}

The neutrino likelihood is the $\gamma$-marginalised constraint on $U_\Lambda$ at the
decorrelation density of Sec.~\ref{sec:gamma}, together with the correlated
$\gamma$-free constraint on $U_\Sigma(\rhoz)$. It is used once with statistical errors
only, from configuration H (Table~\ref{tab:stat}), and once with the
one-standard-deviation systematic uncertainties of Table~\ref{tab:systematics} added in
quadrature, from the configurations HM0$+$PC and HM$+$PC of Sec.~\ref{sec:robust}. It
depends on $c_\Lambda$ through the sub-saturation part of the turn-over,
$c_\Lambda(x^2-x)=-0.15\,c_\Lambda$ at $x=0.82$ (Sec.~\ref{sec:potentials}), and we
evaluate it at $U_\Lambda(x\rhoz)+c_\Lambda(x^2-x)$, equivalently at
$U_\Lambda(\rhoz)+k\,c_\Lambda$ with $k=x-1$: $-0.18$ for HM0$+$PC, $-0.13$ for H and
$-0.33$ for HM$+$PC. Over the prior range this coupling is below $1.7\MeV$. The fitted
$U_\Lambda(\rhoz)$ responds to an anchored turn-over, on the $c_\Lambda=15$ and
$30\MeV$ truth nodes of Table~\ref{tab:shapes}, with $-0.18$ and $-0.15$ per unit
$c_\Lambda$, close to $k$. With statistical errors the neutrino term constrains
$U_\Sigma(\rhoz)$ to $\pm18\MeV$ with $\gamma$ free, against the $\pm20\MeV$ prior.
The fit gives $U_\Lambda(\rhoz)=-28.9\pm2.8\MeV$ and $U_\Lambda(\rho^*)=-23.6\pm2.3\MeV$
with statistical errors, and $-30.0\pm4.0\MeV$ and $-24.4\pm3.3\MeV$ with the
systematic uncertainties, the values the priors alone give; $U_\Sigma(\rhoz)$ is
$29.6\pm12.6$ and $32.4\pm17.2\MeV$.

Figure~\ref{fig:mmaxcl} shows how $\Mmax$ depends on $c_\Lambda$. At the anchored
depths GM1 gives $1.62$, $1.77$, $1.94$ and $2.08\,\Msun$ at $c_\Lambda=-5$, $-2.5$, $0$
and $+2.5\MeV$, and FSU2H gives $1.75$, $1.90$, $2.03$ and $2.12\,\Msun$. Throughout
this range the $\Lambda$ is present at the centre of the maximum-mass star. A negative
$c_\Lambda$ lowers the $\Lambda$ onset, from $0.347$ to $0.330\fm^{-3}$ in GM1 at
$c_\Lambda=-5\MeV$. The mass of PSR~J0740$+$6620, $2.08\,\Msun$, is reached at
$c_\Lambda\approx2.5\MeV$ in GM1, the upper end of the chiral range, and at $1.3\MeV$
in FSU2H. Above that range $\Mmax$ rises to a plateau, $2.25\,\Msun$ (GM1) for
$c_\Lambda\geq12.5\MeV$ and $2.19\,\Msun$ (FSU2H) for $c_\Lambda\geq10\MeV$, where the
$\Lambda$ is absent from the core of the maximum-mass star; the plateau is the maximum
mass of nucleons plus $\Sigma^-$, below the nucleonic $2.36$ and $2.38\,\Msun$.

With the systematic uncertainties the joint fit gives $\Mmax=1.83$ ($68\%$ interval
$1.66$--$2.00$)~$\Msun$ for GM1 and $1.96$ ($1.80$--$2.07$)~$\Msun$ for FSU2H, and
probabilities of $0.17$ and $0.39$ for $\Mmax>2\,\Msun$ ($0.011$ and $0.13$ for
$\Mmax>2.08\,\Msun$). With statistical errors only, and with the priors alone, the
medians agree with these to $0.004\,\Msun$. These intervals are the $c_\Lambda$ prior
mapped through Fig.~\ref{fig:mmaxcl}; at fixed $c_\Lambda$ the potentials spread $\Mmax$
by at most $0.05\,\Msun$ (Sec.~\ref{sec:gamma}). With the upper end of the prior at the
$\Lambda NN$ results of Ref.~\cite{Haidenbauer:2016vfq}, $3.0\MeV$ for GM1 and
$6.6\MeV$ for FSU2H, the medians are $1.85$ ($1.67$--$2.03$) and $2.05$
($1.84$--$2.16$)~$\Msun$, with probabilities of $0.22$ and $0.60$ for
$\Mmax>2\,\Msun$. The maximum-mass stars have central densities of
$0.61$--$0.92\fm^{-3}$, $3.8$--$5.8\rhoz$, beyond the range of the chiral calculations,
so the results also depend on the $\rho^2$ form of Eq.~\eqref{eq:vextra} above
$3.5\rhoz$. Within GM1 and FSU2H the maximum mass of hyperonic matter is set by the
$\Lambda NN$-type repulsion $c_\Lambda$, which the neutrino data do not constrain;
narrowing the band requires information on the high-density sector (heavy-ion,
multi-messenger).

\subsection{Systematic uncertainties}
\label{sec:robust}
The baseline forecast uses the exposure of Table~\ref{tab:exposure}, M2 and the
profiled two-body normalisation. Shifts are quoted on $U_\Lambda(\rho^*)$ with $\gamma$
free, which equals $x^*$ times the fixed-$\gamma$ shift of $U_\Lambda(\rhoz)$,
and on $U_\Sigma(\rhoz)$ at fixed $\gamma$. Each systematic is a variant of the truth
node fitted with the nominal templates and scaled to one standard deviation. Four
configurations differ in the observables and in the treatment of the production model.
In H the fit uses the hyperon observables of Sec.~\ref{sec:detreach}, and the
production-model uncertainties of item~(d) below enter as biases. In H$+$PC they are
profiled as nuisance parameters, together with the composition of the DUNE samples
(PC: production and composition). HM$+$PC adds, for each beam, the mean momentum, mean
angle and mean $Q^2$ of the muon (M) in the events with a reconstructed $\Lambda$, correlated with the hyperon observables
of the same events, and profiles a $2\%$ muon momentum scale per detector. The muon
means respond to the production model, and also to
$U_\Lambda$ through the $\Lambda$ selection: a $\Lambda$ near the trapping or proton
threshold comes from a low-$Q^2$ event with an energetic muon. HM0$+$PC sets this
response to zero, so that the muons calibrate the production model only. The beams are
taken at their representative energies, with $Q^2$ from the true neutrino energy. The items that are not profiled enter in quadrature,
together with the quadratic remainder of the profiled production pairs (curvature).
Table~\ref{tab:systematics} lists them for the three configurations that profile the
production model.
(i)~\emph{Decay asymmetry.} Replacing isotropic decays with the parity-violating
distribution $(1+\alpha\,\mathbf{P}\!\cdot\!\hat{\mathbf{n}})$
(Table~\ref{tab:decays}, $|\mathbf{P}|=0.5$) changes the statistical
$\delta U_\Lambda$ by a factor $0.90\pm0.09$ and $\delta U_\Sigma$ by $0.81\pm0.10$ in
the likelihood with known normalisations: the $V^0$ yield and momentum barely respond
to the decay angle.
(ii)~\emph{Detector model.} A $2\%$ scale on the reconstructed proton and pion
momenta shifts $U_\Lambda(\rho^*)$ by $+11.0\MeV$ in HM0$+$PC, $+4.8\MeV$ in H$+$PC and
$+1.1\MeV$ in HM$+$PC, and $U_\Sigma$ by $32$--$45\MeV$; the muon momentum scale is
profiled separately. A $\pm1\MeV$ change of the $21\MeV$ proton kinetic-energy
threshold shifts $U_\Lambda(\rho^*)$ by at most $2.6\MeV$.
(iii)~\emph{Exposure basis.} The exposure of Table~\ref{tab:exposure}---the
hyperon cross section at the representative energy relative to the inclusive CC rate,
times the stated near-detector samples---predicts $6.8\times10^3$ reconstructed
$\Lambda$ at SBND RHC, and $2.4\times10^5$ and $1.8\times10^5$ at DUNE RHC and FHC.

\begin{figure}[t]
\centering
\includegraphics[width=\columnwidth]{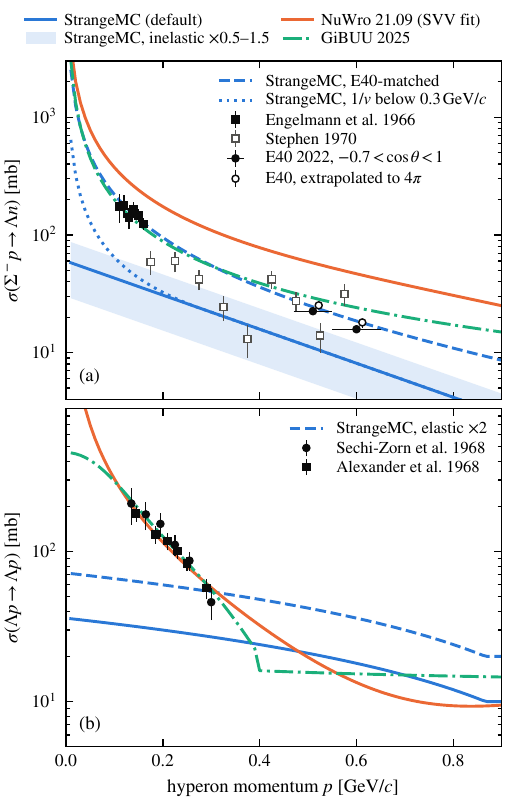}
\caption{Hyperon--nucleon cross sections of the StrangeMC cascade against data and
against the parametrisations used in NuWro (the fit of Ref.~\cite{singh2006}) and
GiBUU~\cite{Buss:2011mx}. (a)~$\Sigma^-p\to\Lambda n$: default (solid), the
inelastic $\pm50\%$ band of Sec.~\ref{sec:robust}, item~(a), and the E40-matched and
$1/v$ variants; data from Refs.~\cite{Engelmann:1966npz,Stephen:1970phd} and J-PARC
E40~\cite{miwa2022} (filled: measured range $-0.7<\cos\theta<1$; open: extrapolated
to the full solid angle with the mean of the three most backward bins).
(b)~$\Lambda p$ elastic: default and doubled; data from
Refs.~\cite{Sechi-Zorn:1968mao,Alexander:1968acu}.}
\label{fig:yn}
\end{figure}

\paragraph*{Physics systematics.}
(a)~\emph{Hyperon--nucleon cross sections (dominant on $U_\Sigma$).} The $YN$ cross
sections of the cascade are effective forms, compared with data in
Fig.~\ref{fig:yn}. The default $\Sigma^-p\to\Lambda n$ cross section is $11.0$ and
$8.1\,$mb at $0.51$ and $0.60\GeV/c$, about half of the J-PARC E40 values,
$22.5\pm0.7\pm0.7$ and $15.8\pm0.8\pm0.5\,$mb over
$-0.7<\cos\theta<1$~\cite{miwa2022}; the fit of Ref.~\cite{singh2006} used in
NuWro~\cite{thorpe2021} lies $2.6$--$3.0$ times above them. Between $0.13$ and
$0.25\GeV/c$ the default $\Lambda N$ elastic cross section is $3$--$7$ times below
the $\Lambda p$ data~\cite{Sechi-Zorn:1968mao,Alexander:1968acu}. Scaling all
inelastic channels by $\pm50\%$ moves $f_{\Sigma^+}$ by $+18/{-}31\%$ at SBND RHC and
$+23/{-}38\%$ at DUNE RHC. It moves the reconstructed-$\Lambda$ fraction by at most
$2.6\%$, while the truth-level trapped fraction moves by $8$--$31\%$: in $f_\Lambda$
the change in $\Sigma\!\to\!\Lambda$ conversion compensates the change in trapping.
We carry five items. (1)~Once matched to the E40 data the $\Sigma N\to\Lambda N$
conversion is known to $15\%$; the $\pm50\%$ regenerations of all inelastic channels,
scaled to $\pm15\%$, give this item, which therefore also moves charge exchange and
$\Lambda N\to\Sigma N$ by $15\%$. (2)~The inelastic channels that E40 does not
constrain, charge exchange and $\Lambda N\to\Sigma N$, are given $\pm50\%$, taken as
half the difference between a regeneration with all inelastic channels doubled and
one with the conversion refitted to the E40 and low-momentum
data~\cite{miwa2022,Engelmann:1966npz},
$\sigma(\Sigma^-p\to\Lambda n)=21.5\,{\rm mb}\times(c/v)\,e^{-p/0.632\GeV}$. (3)~The
templates use the default cascade, so the full shift of a cascade with the refitted
conversion and with the channels closed by the charge of the struck nucleon treated as
null collisions is carried as a template offset. (4)~The $\Lambda N$ elastic cross
section doubled and (5)~a $1/v$ rise of the conversion below $0.3\GeV/c$ enter at
their full shifts. In HM0$+$PC the five items move $U_\Lambda(\rho^*)$ by $-0.5$,
$-1.6$, $+14.0$, $+10.1$ and $+1.9\MeV$, $17.5\MeV$ in quadrature, and in H$+$PC by
$+1.7$, $-9.7$, $+13.9$, $-13.5$ and $+1.1\MeV$, $21.7\MeV$ in quadrature
(Table~\ref{tab:systematics}). The refitted conversion alone lowers $f_{\Sigma^+}$ by
$36\%$ and $28\%$ at SBND RHC and DUNE RHC, while doubling all inelastic channels
raises it by $20\%$ and $30\%$, and the two act on $U_\Lambda(\rho^*)$ in opposite
directions through $f_{\Sigma^+}$. Treating the charge-blocked channels as null
collisions without the refitted conversion is part of item~(3). On $U_\Sigma$ the
$YN$ items give $66$--$86\MeV$ in quadrature, an order of magnitude above the
statistical width, and a competitive $U_\Sigma$ needs external $YN$ input, such as the
E40 data.

\emph{The fitted $U_\Sigma$.} In nuclear matter the $\Sigma$ couples strongly to
the open $\Lambda N$ channel, and a microscopic $\Sigma$ self-energy is complex and
energy dependent~\cite{Friedman:2007zza,Petschauer:2015nea,Haidenbauer:2019boi}.
StrangeMC divides it into a real, momentum-independent potential $U_\Sigma(\rho)$,
which acts through Eq.~\eqref{eq:exitshift} and through the threshold shifts of the
conversion channels, and an absorptive part carried by the free-space
$\Sigma N\to\Lambda N$ cross sections of the cascade, with Pauli blocking of the
recoil nucleon. The fitted $U_\Sigma$ is therefore a parameter of this transport
prescription, tied to the conversion cross sections with which it is extracted, and
is not the real part of a microscopic optical potential. The $\Lambda$ momentum
spectrum, which carries most of the $U_\Sigma$ information, is shaped by
$\Sigma\!\to\!\Lambda$ conversion and hence by the same cross sections.
Table~\ref{tab:wequiv} expresses the absorption in the cascade as an equivalent
imaginary potential $W=\tfrac12\hbar v\rho\,\sigma_{\Sigma N\to\Lambda N}$. The default
cross sections give $W=2.5$--$3.8\MeV$ for $\Sigma^-$ at $\rhoz$, below the
$10$--$15\MeV$ imaginary depth of the $t\rho$ fit to $\Sigma^-$-atom data reviewed in
Ref.~\cite{Friedman:2007zza}; the E40-matched conversion gives $7$--$13\MeV$. The KEK
$(\pi^-,K^+)$ spectra were fitted with an imaginary depth of $40\MeV$, to which they
are only weakly sensitive~\cite{Friedman:2007zza}. A $U_\Sigma$ that rises with
momentum by $12\MeV$ at $0.5\GeV/c$ and $\rhoz$ moves the fitted $U_\Sigma$ by
$+36\MeV$ without the two-body component; in configuration H the shift appears in
$U_\Lambda(\rho^*)$, $-27\MeV$, and $U_\Sigma$ moves by $-6\MeV$. This shift is not included in
Table~\ref{tab:systematics}.

\begin{table}[t]
\centering
\caption{Equivalent absorptive potential
$W=\tfrac12\hbar v\rho\,\sigma_{\Sigma N\to\Lambda N}$ at $\rho=\rhoz$ in argon,
before Pauli blocking, in MeV: $\Sigma^-$ converts on protons ($Z/A=0.45$), $\Sigma^0$
on all nucleons. Default and E40-matched cross sections of the cascade, and the fit
of Ref.~\cite{singh2006} used in NuWro. The $t\rho$ fit to $\Sigma^-$-atom data
reviewed in Ref.~\cite{Friedman:2007zza} has an imaginary depth of $10$--$15\MeV$.}
\label{tab:wequiv}
\begin{tabular}{lcccc}
\hline
$p$ [GeV/$c$]             & $0.1$  & $0.2$  & $0.3$  & $0.5$  \\
\hline
default, $\Sigma^-$       & $2.5$  & $3.6$  & $3.8$  & $3.1$  \\
default, $\Sigma^0$       & $5.7$  & $8.0$  & $8.5$  & $6.9$  \\
E40-matched, $\Sigma^-$   & $13.0$ & $11.1$ & $9.5$  & $6.9$  \\
E40-matched, $\Sigma^0$   & $14.5$ & $12.4$ & $10.6$ & $7.7$  \\
NuWro, $\Sigma^-$         & $22.1$ & $20.4$ & $18.9$ & $16.2$ \\
NuWro, $\Sigma^0$         & $24.1$ & $22.2$ & $20.7$ & $17.8$ \\
\hline
\end{tabular}
\end{table}

(b)~\emph{Transport prescription and capture criterion.} The baseline applies the
potential as a single exit-energy shift evaluated at the production density
[Eq.~\eqref{eq:exitshift}]; an alternative, force-integrated gradient transport,
follows $-\nabla U_Y$ along the trajectory. The two coincide for hyperons that do not
rescatter and differ for those that do. Gradient transport raises the truth-level
trapped fraction at the truth point, on the same seeds, from $0.158$ to $0.191$ (SBND RHC) and from
$0.111$ to $0.131$ (DUNE RHC). With the full difference as one standard deviation,
fitting exit-shift templates to gradient-transport pseudo-data moves
$U_\Lambda(\rho^*)$ by $-7.1\MeV$ in H$+$PC and by $-0.5$ and $-0.6\MeV$ in HM$+$PC and HM0$+$PC,
and $U_\Sigma$ by $+5$ to $+13\MeV$. Evaluating the exit shift at the density of the
last scattering instead of the production density moves $U_\Lambda(\rho^*)$ by
$+1.0\MeV$ in HM0$+$PC and $+1.7\MeV$ in H$+$PC.
In the convention of the NuWro cascade, in which the local potential is subtracted from
the hyperon kinetic energy at production and added back on exit (Sec.~\ref{sec:methods}), a
hyperon that does not rescatter leaves with its production momentum and cannot be
captured. An event whose $\Sigma$ has less kinetic energy than the local $U_\Sigma$ cannot
be started and is replaced, as in NuWro: $1.4\%$ of the events at SBND RHC, $0.8\%$ at
DUNE RHC and below $0.01\%$ in the FHC beams. Regenerated in this convention on the same
seeds, the truth node has a trapped fraction of $0.106$ instead of $0.158$ at SBND RHC and
$0.076$ instead of $0.111$ at DUNE RHC, and $f_\Lambda$ rises by $0.011$--$0.040$ in the
four beams. The two-body component of the RHC pseudo-data is regenerated in the same
convention, with $5.1\%$ of its events replaced at SBND RHC and $0.8\%$ at DUNE RHC.
Fitted with the exit-shift templates, the pseudo-data move $U_\Lambda(\rho^*)$ by
$+18.0\pm2.6\MeV$ in HM0$+$PC, $+14.4\pm1.0\MeV$ in H$+$PC and $+25.4\pm1.1\MeV$ in
HM$+$PC, and $U_\Sigma$ by $+53$, $+37$ and $+56\MeV$. Like the other items of
Table~\ref{tab:systematics} these are linear estimates at the truth point, and the implied
$U_\Sigma$ lies outside the range of the template grid. The minimum $\chi^2$ of the shifted
fit is $38$ in HM0$+$PC and $24$ in H, against at most $18$ and $19$ for the other items,
and $4$ in H$+$PC, so in HM0$+$PC and H the data would distinguish the templates of the two conventions.
With templates also made in the NuWro convention, including regenerations at
$\gamma=0.6$ and $1.4$, HM0$+$PC has $x^*=0.85\pm0.10$ and a profiled width of
$U_\Lambda(\rho^*)$ of $9.7\pm1.6\MeV$ (Monte Carlo errors), against $0.82$ and $10.0\MeV$
with the exit shift, and a profiled width of $U_\Lambda(\rhoz)$ at fixed $\gamma$ of
$11.1\pm1.3\MeV$, against $12.2\MeV$.
The shift enters Table~\ref{tab:systematics} as a transport item, with the full difference
as one standard deviation, and it is the largest item in HM0$+$PC and HM$+$PC.
Eq.~\eqref{eq:exitshift} is the baseline. In the energy-conserving treatments of
Refs.~\cite{Sobczyk:2019uej,Hernandez:2025snr} a hyperon produced with momentum
$\mathbf{k}=\mathbf{p}+\mathbf{q}$ that leaves along a real potential with its energy
conserved reaches the detector with $E_Y(\mathbf{k})+V$, as with
Eq.~\eqref{eq:exitshift}, and is bound if this is below $m_Y$, while the NuWro
convention gives it its vertex energy $E_Y(\mathbf{k})$ and, inside the nucleus, a
momentum different from $\mathbf{k}$. At fixed lepton kinematics the same treatments
give the hyperon the asymptotic energy $E_h+\omega$, as the NuWro convention does, and
Eq.~\eqref{eq:exitshift}, which does not take $V$ from the lepton, gives
$E_h+\omega+V$ (Sec.~\ref{sec:smc}). The baseline assumes that $V$ in the vertex leaves
the distribution of $\mathbf{k}$ nearly unchanged, which Ref.~\cite{Sobczyk:2019uej}
supports at the level of the single-differential and integrated cross sections.
The energy threshold for capture has no effect at reconstruction level. Three variants
were tested: capture of every $\Lambda$ that leaves with less than $2\MeV$, or less
than $5\MeV$, of kinetic energy, and capture only of $\Lambda$'s bound by more than
$5\MeV$. They move the truth-level trapped fraction at SBND RHC from $0.159$ to
$0.173$, $0.194$ and $0.122$, and $U_\Lambda(\rho^*)$ by $0$, $+0.1\pm0.1$ and $0\MeV$
in configuration H, because the $\Lambda$'s they reclassify give decay protons below
the reconstruction threshold (Sec.~\ref{sec:smc}). With the exit shift, the $U_\Lambda$
dependence of the reconstructed observables comes from the shifted spectrum crossing the
detector thresholds, so the detector response of item~(ii) enters at the same level. A
hypernuclear-structure treatment of capture, as in cascade models with explicit
hypernucleus formation and de-excitation~\cite{Hirtz:2019rqe}, would change the
truth-level trapped fraction and the vertex activity of captured events
(Sec.~\ref{sec:sigmaplus}) but not the reconstructed $\Lambda$ observables; the
$U_\Lambda$ result depends on how the potential acts on the hyperons that escape.
(c)~\emph{EOS model.} Across GM1, GM3 and FSU2H, with and without the $\phi$ meson,
and for the $np\Lambda\Sigma^-$ and octet contents, $\Mmax^{\rm hyp}$ spans
$1.60$--$2.05\,\Msun$ at $c_\Lambda=0$ (Sec.~\ref{sec:eossys}). Only FSU2H($+\phi$)
($2.03$ and $2.02\,\Msun$) and GM1$+\phi$ ($2.05$ and $2.00\,\Msun$) reach
$2\,\Msun$ without additional repulsion. This systematic acts on the stellar
inference only; the extracted $U_Y$ does not depend on the EOS.
(d)~\emph{Production model.} The quasi-elastic and associated normalisations are
nuisance parameters of M2; the shape of the production model enters through the
$\Lambda$ momentum and the $\Sigma^+/\Lambda$ ratio. Table~\ref{tab:prod} lists the
priors. For the quasi-elastic channel they are $20\%$ on $f_2(0)$, the spread between
SU(3) and lattice or data values; $0.15\GeV$ on the axial mass; the switch from dipole
to BBBA05 vector form factors; $|g_2/g_1|=0.3$; and the SU(3) breaking of $f_1(0)$ and
$g_1(0)$. For the associated channel they are a factor $1.4$ on the $\Lambda K$:$\Sigma K$
ratio, $20\%$ on the axial mass, $100\%$ of the $N^*$ strength, the range between
Born-level and resonance-dominated models~\cite{fatima2025}, and $50\%$ of the $Y^*$
share. The deep-inelastic and associated yields of the DUNE samples carry $25\%$ and
$50\%$ priors. Entered as biases (configuration H) the priors move
$U_\Lambda(\rho^*)$ by $69\MeV$ in quadrature, led by the $N^*$ strength ($+46\MeV$), the
associated axial mass ($-32\MeV$), the $\Lambda K$:$\Sigma K$ ratio ($-26\MeV$) and the
$Y^*$ share ($-23\MeV$). Under M2 the $\Sigma^+$ fractions of the neutrino beams
calibrate the $\Sigma^+$ efficiency of each detector, so the $\Sigma K$:$\Lambda K$
ratio enters $U_\Lambda(\rho^*)$ through the antineutrino $\Sigma^+/\Lambda$ ratio.
Profiled with the hyperon observables (H$+$PC), the priors widen
$\delta U_\Lambda(\rho^*)$ from $3.4$ to $14.5\MeV$. The muon means constrain the
associated axial mass to $5\%$ of its prior, the $N^*$ strength to $29\%$, the
quasi-elastic form factors to $28\%$ and the DUNE shares to $5$--$30\%$, and in
HM0$+$PC the width is $10.0\MeV$. The quadratic remainder of the profiled pairs adds
$1.3$--$6.9\MeV$ (Table~\ref{tab:systematics}). Of the priors, the $20\%$
quasi-elastic normalisation matters most: at $10\%$ or $40\%$ the HM0$+$PC total is
$27.9$ or $34.5\MeV$. Halving all production priors together gives $30.8\MeV$
(Table~\ref{tab:requirements}), because tighter shape priors absorb less of the residual
shifts; the potential-convention item of item~(b) grows from $18.0$ to $21.3\MeV$.
(e)~\emph{Flux shape.} The beam spectra are gamma-shape approximations of the real
LBNF/BNB fluxes; a $\pm15\%$ shift in $\langle E\rangle$ moves the flux-averaged
quasi-elastic cross section by $\pm14\%$ at SBND RHC and $\pm5\%$ at DUNE RHC, which
the flux and cross-section nuisance parameters absorb. On the SBND beam-simulation
flux (BNB) the associated rate per CC event is a factor $2.6$ lower, and the
quasi-elastic hyperon rate a factor $1.5$ lower, than on the gamma-shape
approximation, and the flux-averaged exclusive CC $K^+$ cross section is a factor $5$
lower, $2.8\times10^{-42}$ against $1.4\times10^{-41}\cm^2$ (Sec.~\ref{sec:methods}).
The forecast has no flux-shape nuisance. Such a nuisance would act on the muon means
much as the muon momentum scale does, so the constraint of the HM configurations on
the production model is optimistic.
(f)~\emph{Initial state.} The default global Fermi gas with the optical-potential
removal energy (Sec.~\ref{sec:methods}) is replaced, in full regenerations of the
truth node on all four beams, by the argon spectral function of NuWro, by a local Fermi gas,
by a constant separation energy of $30\MeV$, and by $e_{\rm rem}\pm5\MeV$. As
Eq.~\eqref{eq:omega} implies, extra binding goes mainly into the lepton: at fixed
centre-of-mass angle the hyperon laboratory energy rises by $0.2$--$0.4\MeV$ per MeV
of extra removal energy, where a one-to-one degeneracy with $U_Y$ would lower it by
$1\MeV$. The larger of the spectral-function and local-Fermi-gas shifts, the constant
separation energy and half the difference of $e_{\rm rem}\pm5\MeV$ enter as
one-standard-deviation items. In HM0$+$PC they move $U_\Lambda(\rho^*)$ by $+7.2$,
$+3.6$ and $-0.1\MeV$, $8.0\MeV$ in quadrature, and in H$+$PC and HM$+$PC the group is
$3.2$ and $3.7\MeV$; on $U_\Sigma$ it is $11$--$17\MeV$. In the configurations that
profile the production model $e_{\rm rem}\pm5\MeV$ moves $U_\Lambda(\rho^*)$ by less
than $1\MeV$, against $5\MeV$ for a one-to-one degeneracy.
In a local Fermi gas both the nucleon binding and $U_Y$ follow $\rho(r)$. To test
whether the two self-energies can trade off locally, a further variant places the
hyperon production point at the radius at which the struck nucleon was drawn, so
that both are evaluated at the same density; relative to the local Fermi gas with an
independent production point it moves $U_\Lambda(\rho^*)$ by $-0.4\MeV$ in HM0$+$PC
and by $+4.8$ and $+4.1\MeV$ in H$+$PC and HM$+$PC. In HM0$+$PC the initial state is
the largest item after the potential convention of item~(b) when the $YN$ and production
uncertainties are reduced (Table~\ref{tab:requirements}).

\begin{table}[t]
\centering
\caption{Production-model priors (one standard deviation), the shift of
$U_\Lambda(\rho^*)$, in MeV, when each enters as a bias with the hyperon observables
(configuration H), and the posterior width over the prior width when the priors are
profiled with the hyperon observables (H$+$PC) and with the muon means (HM0$+$PC). The
form-factor prior is the switch from dipole to BBBA05 vector form factors. The
composition priors (DIS: deep-inelastic share) are absolute shares of the strange
sample before selection, from $25\%$ on the deep-inelastic and $50\%$ on the associated
yield. The two cross-section
normalisations are profiled in every configuration.}
\label{tab:prod}
\setlength{\tabcolsep}{3pt}
\begin{tabular}{@{}llrcc@{}}
\hline
 & prior & bias & \multicolumn{2}{c}{post/prior} \\
 & & H & H$+$PC & HM0$+$PC \\
\hline
QE norm. & $20\%$ & --- & $0.69$ & $0.54$ \\
assoc.\ norm. & $50\%$ & --- & $0.19$ & $0.16$ \\
QE $f_2(0)$ & $20\%$ & $+2.6$ & $0.96$ & $0.62$ \\
QE $M_A$ & $0.15\GeV$ & $+2.6$ & $0.98$ & $0.52$ \\
QE form factors & BBBA05 & $+3.8$ & $0.74$ & $0.28$ \\
QE $g_2$ & $0.3\,g_1$ & $-0.7$ & $1.00$ & $0.99$ \\
QE $f_1(0)$ & $3\%$ & $+0.7$ & $1.00$ & $0.99$ \\
QE $g_1(0)$ & $4$--$5\%$ & $+0.3$ & $1.00$ & $1.00$ \\
$\Lambda K$:$\Sigma K$ ratio & $\times1.4^{\pm1}$ & $-25.6$ & $0.52$ & $0.50$ \\
assoc.\ $M_A$ & $20\%$ & $-31.6$ & $0.79$ & $0.05$ \\
$N^*$ strength & $100\%$ & $+46.3$ & $0.59$ & $0.29$ \\
$Y^*$ share & $50\%$ & $-22.7$ & $0.87$ & $0.49$ \\
DIS, DUNE RHC & $6.2\%$ & $+12.4$ & $0.86$ & $0.26$ \\
assoc., DUNE RHC & $8.3\%$ & $+17.9$ & $0.73$ & $0.30$ \\
DIS, DUNE FHC & $4.7\%$ & $+0.8$ & $1.00$ & $0.05$ \\
\hline
\end{tabular}
\end{table}

\paragraph*{Two-body component.} Left out of the templates, the two-body hyperons
would bias $U_\Lambda(\rho^*)$ by $-29\MeV$ in HM0$+$PC, $-16\MeV$ in HM$+$PC,
$-10\MeV$ in H$+$PC and $-90\MeV$ in H, and raise the minimum $\chi^2$ of the fit to
$34$, $21$, $4$ and $283$; in H$+$PC the bias would go unnoticed. Profiling the
normalisation costs little: the profiled width changes from $9.6$ to $10.0\MeV$ in
HM0$+$PC, from $3.3$ to $4.5\MeV$ in HM$+$PC and from $15.0$ to $14.5\MeV$ in H$+$PC,
and a $\pm50\%$ or per-beam normalisation changes it by at most $0.4\MeV$. With the
normalisation entered as a $\pm30\%$ bias instead, the totals are $41.7$, $32.2$ and
$32.2\MeV$; with the two-body hyperons treated as a background that does not respond
to $U_Y$, $35.5$, $31.4$ and $34.4\MeV$. The two-body kinematics come from the
generator's effective V$-$A weight on three-body phase space and are model dependent.
Folding with the flux, instead of the representative energy, changes the difference
between the two-body and quasi-elastic $\langle p_\Lambda\rangle$ by $+106\MeV$ at SBND
RHC and $-3\MeV$ at DUNE RHC. Applied to the two-body component as a
one-standard-deviation item, this hyperon-side shift moves $U_\Lambda(\rho^*)$ by
$+0.7\MeV$ in HM0$+$PC, $+9.5\MeV$ in HM$+$PC and $-4.1\MeV$ in H$+$PC. The same
folding moves the mean muon momentum of the tagged SBND RHC events by about $21\MeV$,
many times its statistical error. In HM$+$PC this muon-side shift would move
$U_\Lambda(\rho^*)$ by $-49\MeV$, at a minimum $\chi^2$ of $59$. An analysis that uses
the muon response to $U_\Lambda$ therefore needs a flux-shape or two-body-kinematics
nuisance; with the full two-body kinematics profiled the HM$+$PC total is $38\MeV$, and
the HM0$+$PC total stays at $31\MeV$.

\begin{table*}[t]
\centering
\caption{One-standard-deviation systematic items, in MeV, on $U_\Lambda(\rho^*)$, with
$\gamma$ free at the decorrelation density $\rho^*$ of each configuration (Monte Carlo
standard error in parentheses), and on $U_\Sigma(\rhoz)$ at fixed $\gamma$, for the
configurations of Sec.~\ref{sec:robust} that profile the production model. For
HM$+$PC the statistics and profiled rows are the minimum widths, at $0.66\rhoz$. The
statistics row includes the M2 nuisances and the muon momentum scale, with the two-body
normalisation fixed (profiled, configuration H gives $3.4\MeV$, Table~\ref{tab:stat}); the
profiled row adds the production, composition and two-body normalisation nuisances. The
group rows are quadrature sums of the items above them. The totals add
the profiled width, the items and the quadratic remainder of the profiled production
pairs (curvature) in quadrature; their errors are Monte Carlo standard errors from the
per-seed scatter of the items. Items without parentheses have a Monte Carlo standard error
of $0.1\MeV$ or less. LFG: local Fermi gas; MDG: Millener--Dover--Gal. With the production
model entered as biases
(configuration H) the totals are $93.3\pm2.8$ and $217\pm5\MeV$.}
\label{tab:systematics}
\setlength{\tabcolsep}{4pt}
\begin{tabular}{@{}llrrrrrr@{}}
\hline
 & & \multicolumn{2}{c}{HM0$+$PC} & \multicolumn{2}{c}{H$+$PC} & \multicolumn{2}{c}{HM$+$PC} \\
item & $1\sigma$ & $U_\Lambda(\rho^*)$ & $U_\Sigma$ & $U_\Lambda(\rho^*)$ & $U_\Sigma$ & $U_\Lambda(\rho^*)$ & $U_\Sigma$ \\
\hline
$x^*$ & & \multicolumn{2}{c}{$0.82$} & \multicolumn{2}{c}{$0.81$} & \multicolumn{2}{c}{$0.67$} \\
statistics & & $1.5$ & $5.8$ & $1.7$ & $6.0$ & $0.4$ & $1.8$ \\
profiled & & $10.0$ & $15.9$ & $14.5$ & $24.0$ & $4.5$ & $10.3$ \\[2pt]
$\Sigma N\to\Lambda N$ conversion & $\pm15\%$ & $-0.5$\,($0.4$) & $+25$ & $+1.7$\,($0.3$) & $+25$ & $-3.6$\,($0.2$) & $+21$ \\
other inelastic channels & $\pm50\%$ & $-1.6$\,($1.0$) & $-57$ & $-9.7$\,($0.7$) & $-60$ & $+9.2$\,($0.5$) & $-44$ \\
E40 conversion with null collisions & offset & $+14.0$\,($2.8$) & $+58$ & $+13.9$\,($0.8$) & $+45$ & $-4.1$\,($1.5$) & $+35$ \\
$\Lambda N$ elastic $\times2$ & full & $+10.1$\,($2.9$) & $-14$ & $-13.5$\,($1.3$) & $-31$ & $+1.3$\,($1.5$) & $-26$ \\
$1/v$ conversion below $0.3\GeV/c$ & full & $+1.9$\,($2.1$) & $+5$ & $+1.1$\,($0.7$) & $+0$ & $+0.1$\,($1.0$) & $+3$ \\
\emph{$YN$ group} & & \emph{17.5} & \emph{86} & \emph{21.7} & \emph{85} & \emph{10.7} & \emph{66} \\[2pt]
gradient force & full & $-0.6$\,($3.2$) & $+13$ & $-7.1$\,($1.2$) & $+5$ & $-0.5$\,($1.6$) & $+13$ \\
exit at last-scatter density & full & $+1.0$\,($0.4$) & $-4$ & $+1.7$\,($0.1$) & $-5$ & $+0.4$\,($0.2$) & $-5$ \\
potential at start (NuWro) & full & $+18.0$\,($2.6$) & $+53$ & $+14.4$\,($1.0$) & $+37$ & $+25.4$\,($1.1$) & $+56$ \\
\emph{transport group} & & \emph{18.0} & \emph{55} & \emph{16.1} & \emph{38} & \emph{25.4} & \emph{57} \\[2pt]
spectral function or LFG & larger & $+7.2$\,($2.7$) & $+5$ & $-2.4$\,($1.3$) & $-15$ & $-3.6$\,($1.2$) & $-7$ \\
$e_{\rm rem}\pm5\MeV$ & $\frac12$ diff. & $-0.1$\,($1.2$) & $+5$ & $-0.3$\,($0.4$) & $+2$ & $+0.7$\,($0.6$) & $+6$ \\
constant separation energy & full & $+3.6$\,($1.5$) & $-8$ & $+2.1$\,($0.9$) & $-6$ & $+0.9$\,($1.0$) & $-11$ \\
\emph{initial-state group} & & \emph{8.0} & \emph{11} & \emph{3.2} & \emph{17} & \emph{3.7} & \emph{15} \\[2pt]
$p$, $\pi$ momentum scale & $2\%$ & $+11.0$ & $+32$ & $+4.8$ & $+38$ & $+1.1$ & $-45$ \\
proton threshold & $\pm1\MeV$ & $-1.7$ & $+7$ & $-2.6$ & $+7$ & $+0.0$ & $+9$ \\
\emph{detector group} & & \emph{11.1} & \emph{33} & \emph{5.5} & \emph{38} & \emph{1.1} & \emph{46} \\[2pt]
LFG vertex at nucleon radius & full & $-0.4$\,($2.7$) & $+2$ & $+4.8$\,($1.0$) & $+15$ & $+4.1$\,($1.1$) & $+5$ \\
density shape (MDG) & full & $+1.5$\,($0.3$) & $-7$ & $+0.9$\,($0.1$) & $-6$ & $+1.9$\,($0.2$) & $-8$ \\
two-body kinematics & full & $+0.7$ & $+35$ & $-4.1$ & $+25$ & $+9.5$ & $+44$ \\
curvature & & $6.9$ & $20$ & $1.3$ & $3$ & $4.1$ & $9$ \\
\hline
total & & $31.1\pm2.3$ & $117\pm3$ & $32.0\pm0.9$ & $108\pm2$ & $30.4\pm1.0$ & $110\pm2$ \\
\hline
\end{tabular}
\end{table*}

\paragraph*{Totals.} With the production model entered as biases (H) the total on
$U_\Lambda(\rho^*)$ is $93\MeV$, of which $69\MeV$ is production: with the two-body
normalisation free, the hyperon-only fit leans on the neutrino beams and takes on
their associated-production uncertainties, and the RHC pair alone gives $38\MeV$.
Profiling the production model (H$+$PC) gives $32.0\MeV$ at $0.81\rhoz$. With the muon
means calibrating it (HM0$+$PC) the total is $31.1\pm2.3\MeV$ at $\rho^*=0.82\rhoz$:
$1.5\MeV$ statistical, $10.0\MeV$ with the profiled nuisances, and $29\MeV$ from the
residual items and the curvature. The largest items are the convention of the NuWro
cascade ($+18.0\MeV$), the template offset of the E40-matched cascade ($+14.0\MeV$), the
hadron momentum scale ($+11.0\MeV$), the doubled $\Lambda N$ elastic cross section
($+10.1\MeV$) and the initial state ($+7.2\MeV$). Doubling the one-standard-deviation
scale of the quasi-elastic normalisation prior, of the template offset, of the momentum
scale or of the elastic item raises the HM0$+$PC total to $34$--$40\MeV$, and doubling
the convention item raises it to $44\MeV$; templates made with the E40-matched cascade
would remove the offset and give $27.8\MeV$. The Monte Carlo noise of the residual
items, the quadrature of their per-seed standard errors, is $7.6\MeV$ in HM0$+$PC and
raises its total by about $1\MeV$. Using the muon response to $U_\Lambda$ (HM$+$PC)
moves the decorrelation density to $0.67\rhoz$, where the total is $30.4\pm1.0\MeV$; the
width is taken at its minimum, $0.66\rhoz$. At a common $0.82\rhoz$ the totals are
$31.1\MeV$ (HM0$+$PC), $33.4\MeV$ (H$+$PC) and $46.2\MeV$ (HM$+$PC). The muon response
to $U_\Lambda$ comes from the $\Lambda$ selection, through the slow $\Lambda$'s lost
to capture or to the proton threshold, and in the convention of the NuWro cascade it falls to $0.24$--$0.42$ of its
nominal value at the RHC beams. In HM$+$PC the convention item is therefore the largest,
$+25.4\MeV$ at $0.67\rhoz$ and $+32.4\MeV$ at $0.82\rhoz$, and the muon response hardly
reduces the total. HM$+$PC also rests on that response, taken from the variants with
common random numbers, and on the absence of a flux-shape nuisance. For
$U_\Sigma(\rhoz)$ the totals are $108$, $110$ and $117\MeV$ in H$+$PC, HM$+$PC and
HM0$+$PC, and $217\MeV$ in H, against the $\pm20\MeV$ of $\Sigma$-atom analyses, with the
$YN$ items the largest term; the neutrino data do not constrain $U_\Sigma$.

\begin{table}[t]
\centering
\caption{Total uncertainty on $U_\Lambda$, in MeV, at the decorrelation density of
each configuration when one-standard-deviation scales are reduced. $YN$: the five items
of Sec.~\ref{sec:robust}, item~(a), including the template offset; production: the
priors of Table~\ref{tab:prod}, with the curvature remainder scaled by the square of
the factor; momentum scale: the hadron item of item~(ii). The potential-convention item
of item~(b) is re-evaluated in each scenario but not scaled; the last row leaves it out.}
\label{tab:requirements}
\setlength{\tabcolsep}{4pt}
\begin{tabular}{@{}lccc@{}}
\hline
 & HM0$+$PC & H$+$PC & HM$+$PC \\
 & $0.82\rhoz$ & $0.81\rhoz$ & $0.67\rhoz$ \\
\hline
Table~\ref{tab:systematics} & $31.1$ & $32.0$ & $30.4$ \\
E40 cascade in templates & $27.8$ & $28.9$ & $30.1$ \\
$YN\times\frac12$ & $27.2$ & $25.9$ & $28.9$ \\
production $\times\frac12$ & $30.8$ & $27.0$ & $28.3$ \\
$YN$, production $\times\frac12$ & $27.6$ & $21.5$ & $26.7$ \\
$YN$ exact & $25.8$ & $23.5$ & $28.4$ \\
$YN$ exact, production $\times\frac14$ & $22.9$ & $18.4$ & $23.2$ \\
same, momentum scale $1\%$ & $21.2$ & $17.6$ & $23.1$ \\
same, convention settled & $11.6$ & $9.5$ & $6.6$ \\
\hline
\end{tabular}
\end{table}

\paragraph*{Requirements.} Table~\ref{tab:requirements} gives the total when the
uncertainties are reduced. Halving the $YN$ items ($\Sigma^-p\to\Lambda n$ known to
$7.5\%$, the other inelastic channels to $25\%$, the elastic and offset items halved)
and the production priors (quasi-elastic normalisation $10\%$, associated $25\%$) gives
$27.6\MeV$ at $0.82\rhoz$ and $26.7\MeV$ at $0.67\rhoz$. With the $YN$ cross sections known
exactly, the production priors at a quarter and a $1\%$ momentum scale, the HM0$+$PC
total is $21.2\MeV$, set mainly by the potential convention ($17.7\MeV$) and the initial
state ($8.0\MeV$); with the convention settled, by an energy-conserving
treatment of vertex and propagation or by the data, it is $11.6\MeV$, set by the initial state,
the momentum scale and the profiled width. The hypernuclear constraint at $\rho^*$ is
$\pm3.3\MeV$. No scenario
of Table~\ref{tab:requirements} approaches it, and with present $YN$ and production
input the neutrino measurement is a check of $U_\Lambda$ below saturation with
different systematics. $U_\Sigma$ stays above $45\MeV$ with the convention item and above $37\MeV$
without it. Part of the
$YN$ uncertainty could be calibrated by the data themselves: a one-standard-deviation
shift of the unconstrained inelastic channels would appear as a minimum $\chi^2$ of
$15$--$19$ in H and the HM configurations, and the doubled elastic cross section as
$6$--$11$. This requires $YN$ nuisance parameters, which the forecast does not include.
The potential convention of item~(b) would appear as $38$ in HM0$+$PC, and a fit with
both template sets would test it.

\begin{table}[t]
\centering
\caption{Real-exposure yield estimate: the calibrated hyperon cross section per
nucleon, its ratio to the inclusive CC cross section (taken as the standard
$0.68\,(0.33)\times10^{-38}\,E\,\mathrm{cm^2}$ per nucleon for $\nu$ ($\bar\nu$)),
the stated near-detector CC sample, and the resulting produced-hyperon count. The
reconstructed-$\Lambda$ sample is this times the $24$--$29\%$ reconstruction
fraction. Here $E$ is the representative energy at which the cross sections are
evaluated, not the flux mean $\langle E_\nu\rangle$ of Table~\ref{tab:comp}. The FHC
rows carry the associated model spread.}
\label{tab:exposure}
\begin{tabular}{lccccc}
\hline
beam & $E$ & $\sigma_{\rm hyp}$ & $\sigma_{\rm hyp}/\sigma_{\rm CC}$ &
$N_{\rm CC}$ & $N_{Y}$ \\
\hline
SBND RHC  & $0.8$ & $7.6\times10^{-41}$ & $2.8\times10^{-2}$ & $1\times10^{6}$ & $2.8\times10^{4}$ \\
SBND FHC  & $1.6$ & $3.2\times10^{-41}$ & $2.9\times10^{-3}$ & $5\times10^{6}$ & $1.5\times10^{4}$ \\
DUNE RHC  & $3.0$ & $1.9\times10^{-40}$ & $1.9\times10^{-2}$ & $5\times10^{7}$ & $9.5\times10^{5}$ \\
DUNE FHC  & $3.0$ & $1.3\times10^{-40}$ & $6.6\times10^{-3}$ & $1\times10^{8}$ & $6.6\times10^{5}$ \\
\hline
\end{tabular}
\end{table}

\section{Related work}
\label{sec:related}
On the neutrino side, the elementary and
nuclear hyperon-production cross sections have been developed extensively
\cite{singh2006,akbar2014,fatima2016,fatima2021,alam2013}, and nuclear effects and
hyperon FSI implemented in generators \cite{thorpe2021}; the closest match is the
recent calculation of antineutrino-induced hyperon production off nuclei
\cite{benitez2024}, which includes a $\Lambda$-nucleus potential and the
$\Sigma\!\to\!\Lambda$ conversion but stops at cross sections, with no connection
to the EOS. On the astrophysical side, the
``measure-the-potential$\,\to\,$EOS$\,\to\,\Mmax$'' logic is developed using
other probes: heavy-ion $\Lambda$ directed flow~\cite{ohnishi2022,nara2022},
and chiral-EFT hyperon potentials tested against hypernuclei and heavy-ion
flow~\cite{Jinno:2025lfu}.

An earlier treatment of hyperon transport after weak production is that of
Sobczyk \emph{et al.}~\cite{Sobczyk:2019uej}, who computed antineutrino quasi-elastic
$\Lambda$, $\Sigma^0$ and $\Sigma^-$ production on $^{16}$O, with total cross
sections also for $^{12}$C and $^{40}$Ca, using a local Fermi gas and two realistic
hole spectral functions, and propagated the hyperons through a Monte Carlo cascade
with the $YN$ cross-section parametrisation of Ref.~\cite{singh2006}.
Table~\ref{tab:sobczyk} sets the two treatments side by side. In their calculation a
single potential $V=-30\,(\rho/\rhoz)\MeV$ for all hyperons enters only the vertex
$\delta$ function; the hyperons propagate on straight lines without a potential, and
propagation stops below $30\MeV$ of kinetic energy with the hyperon counted as
escaping, so there is no capture, and spectra below $50\MeV$ are not interpreted. They
find that the cascade produces $\Sigma^+$, enhances $\Lambda$ production and reduces
$\Sigma^0$ and $\Sigma^-$ by more than half at $1\GeV$, and that cascade effects
exceed the differences between the spectral functions and the local Fermi gas. The
present work extends this to $^{40}$Ar; to species-resolved potentials
$U_Y(\rho)$ that act on the escaping hyperon [Eq.~\eqref{eq:exitshift}] and on the
conversion thresholds; to capture, through a transport-level criterion in the
low-energy region that Ref.~\cite{Sobczyk:2019uej} leaves out; to the associated and
single-kaon channels and the flux-folded SBND and DUNE beams; and to a detector-level
fit for $U_Y$.

\begin{table}[t]
\centering
\caption{Weak hyperon production with intranuclear transport in
Ref.~\cite{Sobczyk:2019uej} and in this work. LFG and GFG: local and global Fermi gas; SF:
spectral function.}
\label{tab:sobczyk}
\begin{tabular}{@{}lll@{}}
\hline
 & Ref.~\cite{Sobczyk:2019uej} & this work \\
\hline
\pcell{0.20}{channels} &
  \pcell{0.37}{$\bar\nu$ QE $\Lambda,\Sigma^0,\Sigma^-$; $\nu$ QE $\Lambda_c$} &
  \pcell{0.37}{$\bar\nu$ QE; associated $YK$ and single kaon ($\nu,\bar\nu$)} \\
\pcell{0.20}{target} &
  \pcell{0.37}{$^{16}$O ($^{12}$C, $^{40}$Ca), $N=Z$} &
  \pcell{0.37}{$^{40}$Ar} \\
\pcell{0.20}{beam} &
  \pcell{0.37}{$E_{\bar\nu}=1$--$3\GeV$} &
  \pcell{0.37}{SBND and DUNE fluxes} \\
\pcell{0.20}{initial state} &
  \pcell{0.37}{LFG; two hole spectral functions} &
  \pcell{0.37}{GFG with $V(k_F,p)$; LFG, NuWro Ar SF, $e_{\rm rem}$ variants} \\
\pcell{0.20}{hyperon potential} &
  \pcell{0.37}{$-30\,\rho/\rhoz\MeV$ for all $Y$, in the vertex $\delta$ only} &
  \pcell{0.37}{$U_Y(\rho)$ per species, at exit, as $-\nabla U_Y$ or from the start of the
    cascade; conversion thresholds shifted} \\
\pcell{0.20}{$YN$ cross sections} &
  \pcell{0.37}{parametrisation~\cite{singh2006}} &
  \pcell{0.37}{effective forms (Fig.~\ref{fig:yn})} \\
\pcell{0.20}{low energy} &
  \pcell{0.37}{propagation stopped at $T<30\MeV$, hyperon kept} &
  \pcell{0.37}{capture if $E_{\rm out}\le m_Y$} \\
\pcell{0.20}{output} &
  \pcell{0.37}{cross sections} &
  \pcell{0.37}{$f_\Lambda$, $f_{\Sigma^+}$, $\langle p_\Lambda\rangle$; fit for $U_Y$} \\
\hline
\end{tabular}
\end{table}

The accelerator-neutrino probe accesses a different production-density and momentum
regime and can provide a largely independent pull on $U_\Lambda$ below saturation, a
low-density input that enters the neutron-star EOS only through a specified
high-density model. The independence is not complete: the $YN$ cross sections, one of
its leading systematics, also enter the hypernuclear, $\Sigma$-atom and heavy-ion
extractions.

\section{Methods: the StrangeMC simulation}
\label{sec:methods}
Each event is
built in layers: a bound nucleon is drawn from the nuclear ground state; a
strangeness-production channel and current are selected in proportion to the
absolute cross section folded with the beam flux; the produced hyperon feels the
in-medium potential $U_Y(\rho)$ of Sec.~\ref{sec:potentials}; and every hadron is
transported through an intranuclear cascade. The default nuclear model, used for
all results unless stated otherwise, is a global Fermi gas for $^{40}$Ar
($k_F=217\MeV$ for protons, $230\MeV$ for neutrons) with the removal energy
$e_{\rm rem}=M_N-\sqrt{p^2+M_N^2}-V(k_F,p)$ of the momentum-dependent optical
potential $V(k_F,p)$ of Ref.~\cite{Juszczak:2005wk} ($V=-59\MeV$ at $p=0$ for
protons); $e_{\rm rem}$ averages $37\,(40)\MeV$ for protons (neutrons). The
production point is drawn from $r^2\rho(r)$ of the two-parameter Fermi density
($c=3.53\fm$, $a=0.541\fm$, central density $0.176\fm^{-3}=1.1\rhoz$), independently
of the struck nucleon. The alternatives used in Sec.~\ref{sec:robust}, item~(f), are a
local Fermi gas, the spectral function tabulated for argon in
NuWro~21.09.1~\cite{juszczak2005}, a constant separation energy and a $\pm5\MeV$ shift
of $e_{\rm rem}$. The nuclear and cascade machinery is adapted from the LUNAR proton-decay
package~\cite{Nowak:2026czc}.

\paragraph{Production channels.} Six channels span the Cabibbo-suppressed ($\DS=1$)
and Cabibbo-favoured ($\DS=0$) processes (Table~\ref{tab:channels}). The $\DS=1$
quasi-elastic hyperon and single-kaon channels use explicit Dirac matrix elements
with $SU(3)$ form factors; the associated, $\phi$ and deep-inelastic channels use
effective currents, the last hadronised with \textsc{Pythia}\,8.
\begin{table*}[t]
\centering
\begin{tabular}{@{}llcll@{}}
\toprule
channel & process & $\DS$ & model & ref. \\ \midrule
\texttt{qe\_hyperon}  & $\nubar N\!\to\!\ell^+ Y$ & 1 & $SU(3)$ V$-$A form factors & \cite{singh2006,thorpe2021}\\
\texttt{single\_kaon} & $\nu/\nubar N\!\to\!\ell N' K$ & 1 & chiral contact, \emph{calibrated} & \cite{alam2010}\\
\texttt{associated}   & $\nu/\nubar N\!\to\!\ell Y K$, $\nu N\!\to\!\nu Y K$ & 0 & V$-$A + Breit--Wigner, \emph{calibrated} & \cite{alam2013,fatima2025}\\
\texttt{phi}          & $\nu N\!\to\!\nu N\,\phi$ & 0 & NC diffractive, $\phi\!\to\!K\bar K$ & ---\\
\texttt{dis\_strange} & $\nu N\!\to\!\nu N + K\bar K\ldots$ & 0 & \textsc{Pythia}\,8 fragmentation & \cite{pythia83}\\
\texttt{sigma\_star}  & $\nu/\nubar N\!\to\!\ell\,\Sigma^*(1385)K$ & 0 & legacy (subset of \texttt{associated}) & ---\\
\bottomrule
\end{tabular}
\caption{Strangeness-production channels. In \texttt{associated}, $Y$ includes
$\Sigma(1385)$, $\Lambda(1405)$ and $\Lambda(1520)$, and the $N^*(1650,1710,1720)\to YK$
resonances contribute below $W=2.5\GeV$, above which strange final states are
generated as deep-inelastic.}
\label{tab:channels}
\end{table*}

\begin{figure}[t]
\centering
\includegraphics[width=\columnwidth]{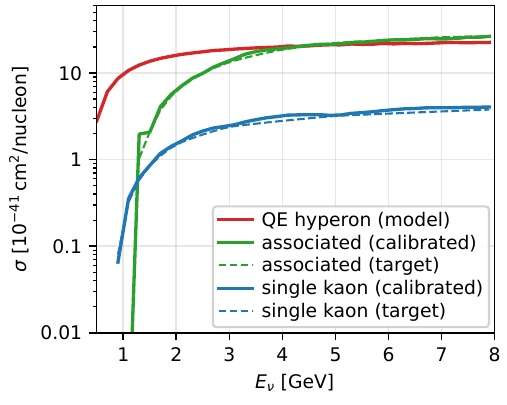}
\caption{Per-nucleon cross sections on argon versus neutrino energy (logarithmic
scale): the quasi-elastic hyperon prediction, and the calibrated single-kaon and associated
channels (solid) tracking their published targets (dashed; above $3\GeV$ the
associated target extrapolates the published curve).}
\label{fig:xsec}
\end{figure}

\begin{figure}[t]
\centering
\includegraphics[width=\columnwidth]{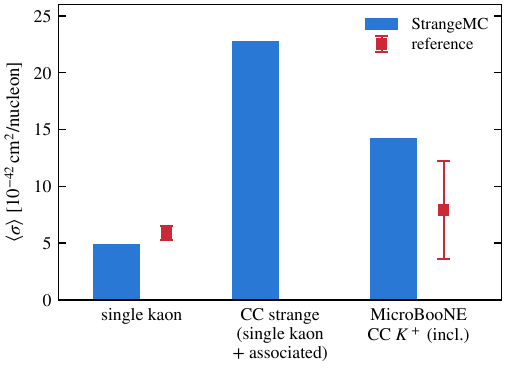}
\caption{BNB flux-averaged cross sections: StrangeMC (bars) versus the
single-kaon model of Ref.~\cite{alam2010} (Table~II) and the inclusive MicroBooNE
CC $K^+$-on-Ar measurement~\cite{microboone2025} (points with error bars). The bars
are averages over the gamma-shape BNB flux of the generator. On the SBND
beam-simulation flux the single-kaon and $K^+$ averages are $1.0$--$1.2\times10^{-42}$
and $2.8$--$3.4\times10^{-42}\cm^2$, and the single-kaon value of
Ref.~\cite{alam2010} refers to the MiniBooNE flux.}
\label{fig:fluxavg}
\end{figure}

\paragraph{Cross sections and calibration.} Absolute per-nucleon cross sections are
Monte Carlo integrals of $|M|^2$ over RAMBO phase space (Fig.~\ref{fig:xsec}). The
single-kaon channel is calibrated to Ref.~\cite{alam2010}; the associated channel
is anchored to the full-$YK$ model of Ref.~\cite{fatima2025} (the digitised
$\nu_\mu n\to\mu^-\Lambda K^+$ curve, converted to an isoscalar per-nucleon cross
section and scaled by the $\Sigma K$ channel ratios of Ref.~\cite{alam2013},
$\times1.8$), giving $1.34\times10^{-40}\cm^2$ at $3\GeV$; the quasi-elastic
hyperon channel is an absolute prediction. It uses the $SU(3)$ form factors of
Ref.~\cite{singh2006} with $f_1,f_2>0$ and $g_1/f_1=-0.735$ for $p\to\Lambda$ and
$+0.337$ for the $\Sigma$ channels; on free nucleons its cross section is
$0.93$--$0.98$ times that of \textsc{NuWro}~21.09.1~\cite{juszczak2005,thorpe2021}
over $0.5$--$5\GeV$, and on argon it is $0.75$, $0.98$, $1.60$ and
$1.87\times10^{-40}\cm^2$ per nucleon at $0.8$, $1$, $2$ and $3\GeV$. The associated
model spread is a factor $\sim\!3$--$5$ (the dynamical coupled-channel $\Lambda K^+$
curve is $\sim\!2.8\times10^{-41}\cm^2$ at $3\GeV$), and our anchor sits at the upper,
full-channel edge of that spread; at $3\GeV$,
$\sigma_{\rm QE}/\sigma_{\rm assoc}\simeq1.4$.
As an independent check, we compare the flux-averaged exclusive CC $K^+$ cross
section with the inclusive MicroBooNE measurement,
$(7.93\pm4.29)\times10^{-42}\cm^2$~\cite{microboone2025} (Fig.~\ref{fig:fluxavg}).
On the gamma-shape BNB flux of the generator it is $1.4\times10^{-41}\cm^2$,
$1.4\sigma$ above the central value; on the SBND beam-simulation flux of
Sec.~\ref{sec:robust}, item~(e), it is $2.8\times10^{-42}\cm^2$, $1.2\sigma$ below.
The inclusive measurement also collects DIS and secondary $K^+$, which the exclusive
set omits, and neither flux is the one at the MicroBooNE detector, so the comparison
does not discriminate within the associated model spread.

\paragraph{Flux folding.} Events are drawn from gamma-shape approximations of the
DUNE (LBNF) and SBND (BNB) $\nu_\mu/\bar\nu_\mu$ spectra in both horn polarities,
each with its wrong-sign component (a $\sim\!10$--$20\%$ contamination of the
reverse-horn beams); a tabulated flux may also be loaded. Each channel is selected
$\propto\Phi(E)\,\sigma(E)$, so the sample is distributed as $\Phi(E)\,\sigma(E)$.
The resulting flux-folded strangeness composition (fractions \emph{among} strange
final states) is given in Table~\ref{tab:comp}.
\begin{table}[t]
\centering
\begin{tabular}{@{}lcccc@{}}
\toprule
beam & $\langle E_\nu\rangle$ [GeV] & single-kaon & assoc. & QE hyp. \\ \midrule
DUNE FHC ($\nu$)    & 3.8 & 10\% & 85\% & \phantom{0}5\% \\
DUNE RHC ($\nubar$) & 3.6 & \phantom{0}6\% & 45\% & 49\% \\
SBND FHC ($\nu$)    & 1.9 & 13\% & 75\% & 12\% \\
SBND RHC ($\nubar$) & 1.6 & \phantom{0}4\% & 19\% & 77\% \\
\bottomrule
\end{tabular}
\caption{Flux-folded strangeness composition (fractions \emph{among strange final
states}, which are themselves at the per cent level or below of the total CC
rate), with FSI, both potentials, and the associated normalisation anchored to
Ref.~\cite{fatima2025} (see text; the associated shares carry that channel's
factor $\sim\!3$--$5$ model spread). The associated column includes the
neutral-current mode. The neutral-current $\phi$ and DIS channels contribute less
than one per cent. The shares are rounded by the largest-remainder method, so that
each row adds to $100\%$; their Monte Carlo uncertainty is below one point.}
\label{tab:comp}
\end{table}

The forecast uses the channels of Table~\ref{tab:channels} and, in the RHC
templates, the strangeness-changing two-body current (Sec.~\ref{sec:detreach}).
StrangeMC also contains charged-current deep-inelastic strange production, which is
not in the templates; its share of the DUNE samples enters the forecast through the
composition priors of Table~\ref{tab:prod}. With the two-body current and deep-inelastic
production included, the quasi-elastic share of the
strange final states is $66\%$ at SBND RHC and $18\%$ at DUNE RHC, where deep-inelastic
production supplies $60\%$, so a DUNE near-detector analysis has to model or reject
the deep-inelastic component. The two-body current makes $\Sigma^+$ at the weak vertex
($\bar\nu_\mu pp\to\mu^+\Sigma^+n$). With the generator's reference weight for this
channel, which includes the pair census of Ref.~\cite{Nowak:2026twobody}, primary
$\Sigma^+$ that leave the nucleus are $12\%$ ($10\%$) of the kaon-less $\Sigma^+$ at SBND
(DUNE) RHC, and two-body events supply $31\%$ ($36\%$) of them; in the
flux-folded generator sample of Ref.~\cite{Nowak:2026pheno} the SBND figures are
$12\pm2\%$ and $28\%$.

\paragraph{Intranuclear transport and validation.} Produced hadrons are transported
from a shared vertex: mesons and nucleons through the cascade of LUNAR~\cite{Nowak:2026czc}, and
hyperons through a dedicated cascade with $YN$ elastic, $\Sigma\!\to\!\Lambda$
conversion and charge-exchange cross sections, inside which $U_Y(\rho)$ shifts the
exit energy (or, optionally, acts as a continuous $-\nabla U_Y$ gradient force, or is
subtracted from the kinetic energy at the start of the cascade and added back on exit,
as in NuWro).

\begin{figure}[t]
\centering
\includegraphics[width=\columnwidth]{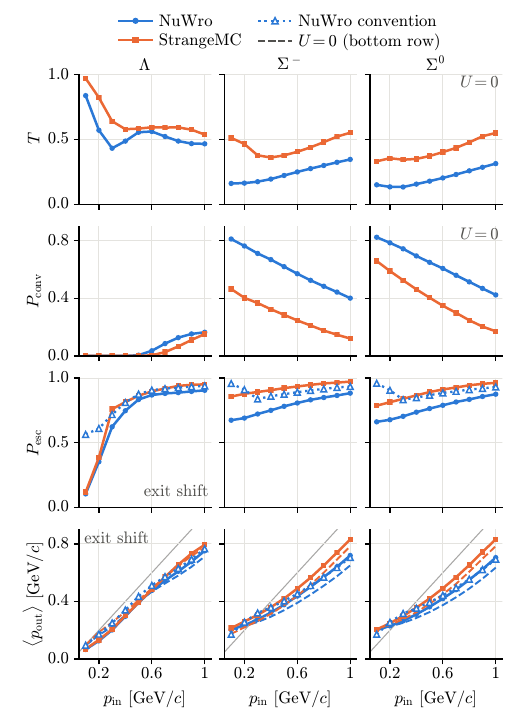}
\caption{Particle gun in argon: NuWro~21.09.1 (circles) and StrangeMC (squares), for
$\Lambda$, $\Sigma^-$ and $\Sigma^0$ injected with momentum $p_{\rm in}$. From the top:
probability $T$ of leaving without a collision and probability $P_{\rm conv}$ that the
outgoing hyperon is a $\Lambda$ ($\Sigma$ for $\Lambda$ input), with the potentials off;
escape probability $P_{\rm esc}$ and mean outgoing momentum $\langle p_{\rm out}\rangle$ at
$U_\Lambda=-28$ and $U_\Sigma=+30\MeV$, with the exit shift of Eq.~\eqref{eq:exitshift}
(solid; applied to the NuWro events offline) and in NuWro's own convention (dotted). The
dashed lines in the bottom row are $\langle p_{\rm out}\rangle$ with the potentials off,
and the thin line is $\langle p_{\rm out}\rangle=p_{\rm in}$.}
\label{fig:gun}
\end{figure}

\paragraph{Comparison with the NuWro cascade.} We compare the StrangeMC hyperon
cascade with the NuWro cascade~\cite{golan2012,thorpe2021} (version 21.09.1) in a
particle-gun configuration. $\Lambda$, $\Sigma^-$ and $\Sigma^0$ start at points drawn
from the argon density of each code, with fixed momenta $p_{\rm in}$ between $0.1$ and
$1.0\GeV/c$, $10^5$ per point (Table~\ref{tab:gun}, Fig.~\ref{fig:gun}). With the
potentials off, NuWro transmits $16$--$35\%$ of the $\Sigma^-$ without a collision,
against $36$--$55\%$ in StrangeMC, and converts $81\%$ of them to a $\Lambda$ at
$0.1\GeV/c$ and $40\%$ at $1\GeV/c$, against $47\%$ and $12\%$. For $\Sigma^0$ the
conversion probabilities fall from $82$ to $42\%$ and from $66$ to $17\%$. For $\Lambda$
the NuWro transmission is $0.68$--$0.86$ times that of StrangeMC below $0.4\GeV/c$ and
$0.81$--$0.95$ times above. These differences come from the cross sections: a
straight-line estimate that folds each code's $YN$ cross sections with its argon
density, the Fermi motion of the target and Pauli blocking reproduces the transmission
of both codes at all thirty points to within $0.016$. NuWro uses the fit of
Ref.~\cite{singh2006}, whose $\Sigma^-p\to\Lambda n$ cross section is $2.6$--$3.0$ times
the E40 data and whose $\Lambda N$ and $\Sigma N$ elastic cross sections rise as $1/p$ at
low momentum (Fig.~\ref{fig:yn}).

The two codes apply the potential differently. NuWro subtracts the local potential
from the hyperon kinetic energy at the start of the cascade and adds it back on exit, so
a hyperon that does not scatter leaves with its initial momentum, and a $\Lambda$ whose
kinetic energy falls below the depth of the local potential is captured. StrangeMC applies Eq.~\eqref{eq:exitshift}. With
Eq.~\eqref{eq:exitshift} applied to the NuWro events offline, at the start radius of each
hyperon, the $\Lambda$ capture probabilities of the two codes agree to $0.01$ at
$0.1\GeV/c$ and $0.03$ at $0.2\GeV/c$, where capture is fixed by the density at the start
point, and the derivative $\partial P_{\rm esc}/\partial U_\Lambda$ agrees to $10\%$ except
at $0.3\GeV/c$ and above $0.6\GeV/c$, where NuWro's is $1.15$--$1.6$ times larger. Above
$0.3\GeV/c$ only scattered $\Lambda$'s are captured. The capture probability per scattered
$\Lambda$ agrees to $0.05$ up to $0.6\GeV/c$, and NuWro captures $1.2$--$2.0$ times more
$\Lambda$'s because more of them scatter. A $\Sigma$ is captured only after conversion to a
$\Lambda$, and NuWro captures $1.6$--$4.1$ times more of them. The mean momentum of the
escaping $\Lambda$'s responds less to $U_\Lambda$ in NuWro, by a factor $0.5$--$0.7$ at
$0.3$--$0.5\GeV/c$, because more of them have been slowed to near the capture threshold,
where a change of $U_\Lambda$ adds or removes slow $\Lambda$'s as well as shifting the
others. StrangeMC has the NuWro convention as an option: the hyperon starts the cascade with the
local potential subtracted from its kinetic energy and carries that potential, reset to the local
potential of the new species at a $\Sigma\leftrightarrow\Lambda$ conversion, until it
leaves, and it is captured when a scattering leaves it unable to escape. In this
convention, in both codes, every hyperon that does not scatter leaves with its initial
momentum. The $\Lambda$ escape probability at $0.1$ and $0.3\GeV/c$ is $0.56$ and $0.72$
in NuWro and $0.77$ and $0.80$ in StrangeMC, against $0.12$ and $0.76$ with the exit
shift (Table~\ref{tab:gunstart}), and between $0.1$ and $0.5\GeV/c$ the potential
raises the mean outgoing $\Lambda$ momentum, by up to $40\MeV/c$ in NuWro and $20\MeV/c$
in StrangeMC, where the exit shift lowers it by $16$--$53\MeV/c$. The capture probability
per scattered $\Lambda$ differs between the codes by at most $0.05$, and the NuWro escape
probability follows to within $0.025$ from its own probability of leaving without a
collision and the StrangeMC capture probability, so the remaining differences come from
the cross sections. A truth-node regeneration in this convention moves
$U_\Lambda(\rho^*)$ by $+18.0\pm2.6\MeV$ and $U_\Sigma$ by $+53\MeV$ in configuration
HM0$+$PC [Sec.~\ref{sec:robust}, item~(b)]; it is an item of
Table~\ref{tab:systematics}.

The cross-section differences fall along the $YN$ variants of Sec.~\ref{sec:robust},
item~(a). In the gun, the E40-matched conversion raises the StrangeMC $\Sigma^-$
conversion probability to the NuWro value at $0.1\GeV/c$ and covers $50$--$70\%$ of the
difference at $0.3$--$0.8\GeV/c$, and the doubled $\Lambda N$ elastic cross section gives
the NuWro $\Lambda$ transmission at $0.3\GeV/c$. In the hyperon-only fit without the
two-body component these two variants move $U_\Lambda(\rho^*)$ by $+20$ and $-8\MeV$.
NuWro also produces more $\Sigma^+$, through $\Sigma^0p\to\Sigma^+n$ and
$\Lambda p\to\Sigma^+n$, than any variant, including the doubling of all inelastic
channels, which moves $U_\Lambda(\rho^*)$ by $-15\MeV$. The conversion and the $\Sigma^+$
production act on $U_\Lambda(\rho^*)$ in opposite directions through $f_{\Sigma^+}$, so
the NuWro cross sections lie outside the range of the variants without fixing the sign
of the shift they would cause.

\begin{table}[t]
\centering
\caption{Particle gun in argon: NuWro~21.09.1 (N) and StrangeMC (S). $T$, probability
of leaving without a collision, and $P_{\rm conv}$, probability that the outgoing
hyperon is a $\Lambda$ ($\Sigma$ for $\Lambda$ input), with the potentials off;
$P_{\rm esc}$, escape probability, and $\langle p_{\rm out}\rangle$ [GeV/$c$], mean
outgoing momentum, with the exit shift of Eq.~\eqref{eq:exitshift} at
$U_\Lambda=-28$ and $U_\Sigma=+30\MeV$, applied to the NuWro events offline.
$10^5$ hyperons per point; statistical errors are below $0.002$.}
\label{tab:gun}
\setlength{\tabcolsep}{2.6pt}
\begin{tabular}{@{}llcccccccc@{}}
\hline
 & $p_{\rm in}$ & \multicolumn{2}{c}{$T$} & \multicolumn{2}{c}{$P_{\rm conv}$} & \multicolumn{2}{c}{$P_{\rm esc}$} & \multicolumn{2}{c}{$\langle p_{\rm out}\rangle$} \\
 & [GeV/$c$] & N & S & N & S & N & S & N & S \\
\hline
$\Lambda$ & $0.2$ & $0.57$ & $0.82$ & $0.00$ & $0.00$ & $0.35$ & $0.39$ & $0.127$ & $0.133$ \\
 & $0.5$ & $0.56$ & $0.58$ & $0.01$ & $0.00$ & $0.83$ & $0.86$ & $0.393$ & $0.392$ \\
 & $0.8$ & $0.49$ & $0.59$ & $0.13$ & $0.07$ & $0.89$ & $0.94$ & $0.609$ & $0.656$ \\[3pt]
$\Sigma^-$ & $0.2$ & $0.16$ & $0.47$ & $0.76$ & $0.40$ & $0.69$ & $0.88$ & $0.235$ & $0.261$ \\
 & $0.5$ & $0.22$ & $0.38$ & $0.62$ & $0.29$ & $0.78$ & $0.92$ & $0.378$ & $0.421$ \\
 & $0.8$ & $0.30$ & $0.48$ & $0.48$ & $0.18$ & $0.85$ & $0.96$ & $0.569$ & $0.650$ \\[3pt]
$\Sigma^0$ & $0.2$ & $0.13$ & $0.36$ & $0.79$ & $0.59$ & $0.68$ & $0.81$ & $0.231$ & $0.249$ \\
 & $0.5$ & $0.18$ & $0.37$ & $0.65$ & $0.40$ & $0.76$ & $0.89$ & $0.362$ & $0.421$ \\
 & $0.8$ & $0.26$ & $0.48$ & $0.51$ & $0.25$ & $0.83$ & $0.94$ & $0.551$ & $0.653$ \\
\hline
\end{tabular}
\end{table}

\begin{table}[t]
\centering
\caption{Particle gun in argon at $U_\Lambda=-28$ and $U_\Sigma=+30\MeV$ with the potential
in the convention of the NuWro cascade: NuWro~21.09.1 (N) and StrangeMC in the same
convention (S$_{\rm st}$), and StrangeMC with the exit shift of Eq.~\eqref{eq:exitshift}
(S$_{\rm sh}$). $P_{\rm esc}$, escape probability; $T$, probability of leaving without a
collision; $\Delta\langle p\rangle$, change of the mean outgoing momentum from $U_Y=0$
[MeV/$c$]. $\Sigma$'s below $0.3\GeV/c$, whose start points are redrawn where their
kinetic energy is below the local $U_\Sigma$, are left out. $10^5$ hyperons per point;
statistical errors are below $0.003$ and $1.5\MeV/c$.}
\label{tab:gunstart}
\setlength{\tabcolsep}{2.4pt}
\begin{tabular}{@{}llcccccccc@{}}
\hline
 & $p_{\rm in}$ & \multicolumn{3}{c}{$P_{\rm esc}$} & \multicolumn{2}{c}{$T$} & \multicolumn{3}{c}{$\Delta\langle p\rangle$} \\
 & [GeV/$c$] & N & S$_{\rm st}$ & S$_{\rm sh}$ & N & S$_{\rm st}$ & N & S$_{\rm st}$ & S$_{\rm sh}$ \\
\hline
$\Lambda$ & $0.1$ & $0.56$ & $0.77$ & $0.12$ & $0.51$ & $0.75$ & $+1$ & $+0.1$ & $-29$ \\
 & $0.2$ & $0.61$ & $0.75$ & $0.39$ & $0.42$ & $0.65$ & $+7$ & $+0.9$ & $-53$ \\
 & $0.3$ & $0.72$ & $0.80$ & $0.76$ & $0.45$ & $0.59$ & $+25$ & $+8$ & $-49$ \\
 & $0.5$ & $0.87$ & $0.88$ & $0.86$ & $0.61$ & $0.59$ & $+40$ & $+20$ & $-16$ \\
 & $0.8$ & $0.93$ & $0.95$ & $0.94$ & $0.50$ & $0.60$ & $+49$ & $+25$ & $+10$ \\[2pt]
$\Sigma^-$ & $0.3$ & $0.84$ & $0.92$ & $0.89$ & $0.17$ & $0.46$ & $+63$ & $+44$ & $+42$ \\
 & $0.5$ & $0.87$ & $0.95$ & $0.92$ & $0.21$ & $0.37$ & $+64$ & $+43$ & $+47$ \\
 & $0.8$ & $0.92$ & $0.97$ & $0.96$ & $0.29$ & $0.46$ & $+60$ & $+29$ & $+49$ \\[2pt]
$\Sigma^0$ & $0.3$ & $0.83$ & $0.89$ & $0.84$ & $0.14$ & $0.34$ & $+65$ & $+41$ & $+22$ \\
 & $0.5$ & $0.87$ & $0.93$ & $0.89$ & $0.17$ & $0.36$ & $+68$ & $+40$ & $+37$ \\
 & $0.8$ & $0.91$ & $0.96$ & $0.94$ & $0.25$ & $0.46$ & $+65$ & $+31$ & $+46$ \\
\hline
\end{tabular}
\end{table}

\paragraph{Consistency with independent calculations.}
(i)~\emph{Yield.} The predicted antineutrino-induced
hyperon yield is $\mathcal{O}(10^4)$ at SBND (Table~\ref{tab:exposure}), as in the
independent nuclear-model estimate of Ref.~\cite{benitez2024}. (ii)~\emph{Channel
balance.} The Cabibbo-suppressed quasi-elastic reaction dominates the antineutrino
strange final state at SBND ($\sim\!77\%$ in SBND RHC, Table~\ref{tab:comp}), as in
the NuWro implementation~\cite{thorpe2021} and the nuclear-model calculations
\cite{akbar2014,benitez2024}.
(iii)~\emph{FSI direction.} The in-medium
$\Sigma N\!\to\!\Lambda N$ conversion enhances the $\Lambda$ yield at the expense of
$\Sigma$, as in every hyperon-FSI treatment cited
here~\cite{Sobczyk:2019uej,thorpe2021,akbar2014,benitez2024}; it is the same mechanism
that feeds the FSI $\Sigma^+$ tag. The absolute $YN$ conversion and charge-exchange
rates differ between codes by factors of a few (Fig.~\ref{fig:yn}) and are the leading
$U_\Sigma$ systematic (Table~\ref{tab:transportcompare}).

\begin{table}[t]
\centering
\caption{Comparison of the StrangeMC hyperon-FSI phenomenology with independent
calculations and data. The first three rows are consistency checks; the fourth
compares the $\Sigma^-p\to\Lambda n$ cross section with the E40
data~\cite{miwa2022}; the fifth is the $\Sigma^-\!\to\!\Lambda$ conversion
probability at $0.5\GeV/c$ with the potentials off in the particle-gun comparison
with NuWro (Table~\ref{tab:gun}).}
\label{tab:transportcompare}
\begin{tabular}{@{}lll@{}}
\hline
feature & StrangeMC & independent \\
\hline
SBND $\bar\nu$ hyperon yield & $\mathcal{O}(10^4)$ & $\mathcal{O}(10^4)$~\cite{benitez2024} \\
QE share, SBND RHC & $\sim\!77\%$ & QE-dominated~\cite{thorpe2021,benitez2024} \\
$\Sigma\!\to\!\Lambda$ enh.\ $\Lambda$ & yes & yes~\cite{Sobczyk:2019uej,akbar2014,benitez2024} \\
$\sigma(\Sigma^-p\to\Lambda n)$/E40 & $0.5$ & $2.6$--$3.0$ (NuWro) \\
$P(\Sigma^-\!\to\!\Lambda)$, gun & $0.29$ & $0.62$ (NuWro) \\
\hline
\end{tabular}
\end{table}

\paragraph{Data availability.} StrangeMC and the analysis scripts and data files
underlying this work---the production channels and cross-section calibration, the
intranuclear cascade, the RMF/TOV equation-of-state solver, and the detector-reach,
Bayesian-fit and systematics pipelines---are available from the author on reasonable
request. The response derivatives sufficient to reproduce the Fisher
sensitivities are given in Appendix~\ref{sec:response}.

\section{Conclusions}
\label{sec:conclusions}
Charged-current (anti)neutrino interactions on argon produce $\Lambda$ and $\Sigma$
hyperons inside the nucleus, and the hyperons reach the detector after crossing
nuclear matter.

\begin{table}[t]
\centering
\caption{What the projected SBND and DUNE data constrain. Uncertainties are one
standard deviation; the configurations are defined in Sec.~\ref{sec:robust} and the
systematic items listed in Table~\ref{tab:systematics}.}
\label{tab:conclusions}
\begin{tabular}{@{}ll@{}}
\hline
\multicolumn{2}{@{}l}{\emph{Directly constrained}}\\
\pcell{0.30}{$U_\Lambda(\rho^*)$, $\rho^*=0.82\rhoz$} &
  \pcell{0.64}{$\pm1.5\MeV$ (stat.), $\pm10.0\MeV$ with the production model profiled,
    $\pm31\MeV$ in total (HM0$+$PC); hypernuclear prior $\pm3.3\MeV$} \\ \addlinespace[2pt]
\multicolumn{2}{@{}l}{\emph{Constrained with an assumption}}\\
\pcell{0.30}{$U_\Lambda(\rhoz)$} &
  \pcell{0.64}{$\pm38\MeV$ in total at $\gamma=1$ (HM0$+$PC); with $\gamma$ free no
    improvement on the hypernuclear $\pm4\MeV$; depends on the functional form} \\ \addlinespace[2pt]
\multicolumn{2}{@{}l}{\emph{Not determined}}\\
\pcell{0.30}{$U_\Sigma(\rhoz)$} &
  \pcell{0.64}{$\pm6\MeV$ (stat., HM0$+$PC) against totals of $108$--$117\MeV$, mainly
    from the $YN$ cross sections; a transport parameter} \\
\pcell{0.30}{$U_Y$ above $\rhoz$, $c_\Lambda$} &
  \pcell{0.64}{not constrained; $c_\Lambda$ shifts $U_\Lambda(\rho^*)$ by
    $-0.15\,c_\Lambda$ in Eq.~\eqref{eq:potform}} \\ \addlinespace[2pt]
\multicolumn{2}{@{}l}{\emph{Inferred through an EOS model}}\\
\pcell{0.30}{$\Mmax$} &
  \pcell{0.64}{$1.94$ (GM1) and $2.03\,\Msun$ (FSU2H$+\phi$) for $np\Lambda\Sigma^-$ at
    $c_\Lambda=0$; $1.62$--$2.08$ and $1.75$--$2.12\,\Msun$ for $-5\le
    c_\Lambda\le2.5\MeV$; joint fit $1.83$ ($1.66$--$2.00$) and $1.96$
    ($1.80$--$2.07$)~$\Msun$, set by the $c_\Lambda$ prior} \\
\pcell{0.30}{hyperon onset, core composition} &
  \pcell{0.64}{set by the RMF parametrisation and $c_\Lambda$ at the anchored
    $U_Y(\rhoz)$} \\
\pcell{0.30}{$R_{1.4}$, $\Lambda_{1.4}$} &
  \pcell{0.64}{$14.0$\,km and $895$ (GM1), $13.3$\,km and $763$ (FSU2H); set by the
    nucleonic EOS} \\
\hline
\end{tabular}
\end{table}

The quantity the neutrino observables determine is $U_\Lambda$ at the decorrelation
density, $\rho^*=0.82\rhoz$ when the muon kinematics of the $\Lambda$-tagged events
calibrate the production model (Table~\ref{tab:conclusions}). The hyperons are produced throughout the nuclear
volume, with a mean production density of $0.66\rhoz$, and at $\rho^*$ the fitted
potential does not depend on the low-density exponent $\gamma$ and changes by at most
$1.4\MeV$ between the six density dependences tested (configuration H; $0.9$--$1.9\MeV$ for
the Millener--Dover--Gal form in the configurations that profile the production model).
The information comes mainly
from the mean momentum of the reconstructed $\Lambda$'s, whose spectrum the exit
potential shifts relative to the detector thresholds; the truth-level trapped fraction
is not a fitted observable, and moving the capture threshold by a few MeV does not affect the result.
The convention by which the cascade applies the potential to hyperons that do not rescatter
moves $U_\Lambda(\rho^*)$ by $+18\MeV$, the largest item of the total
(Sec.~\ref{sec:robust}, item~(b)).
$U_\Lambda(\rhoz)$ follows only with an assumption on $\gamma$.

The statistical precision at $\rho^*$ is $1.5\MeV$. Profiling the production model,
with priors on the quasi-elastic form factors, the associated channel mix and the
composition of the DUNE samples, widens it to $10.0\MeV$, and the hyperon--nucleon cross
sections, the transport prescription, the nucleon initial state, the detector response
and the two-body current add $29\MeV$, for a total of $\pm31\MeV$. With the muon means
responding to $U_\Lambda$ through the $\Lambda$ selection the total is hardly smaller,
$30\MeV$ at $0.67\rhoz$. Halving the hyperon--nucleon and production uncertainties gives
about $28\MeV$ at $0.82\rhoz$ and $27\MeV$ at $0.67\rhoz$, and the potential convention,
the initial state, the momentum scale and the profiled width leave a floor near $21\MeV$
at $0.82\rhoz$, or $12\MeV$ with the convention settled
(Table~\ref{tab:requirements}). The hypernuclear constraint at the same density is
$\pm3.3\MeV$, so the neutrino measurement is a check of $U_\Lambda$ below saturation
with different systematics. The largest hyperon--nucleon items come from a
$\Sigma^-p\to\Lambda n$ cross section that is half of the J-PARC E40 data and a
$\Lambda N$ elastic cross section $3$--$7$ times below the data at low momentum;
templates made with the E40-matched cascade would lower the total to $28\MeV$.

The fitted $U_\Sigma$ is uncertain by about $110\MeV$ in every configuration that
profiles the production model, against a statistical $6\MeV$ in HM0$+$PC and the $\pm20\MeV$ of
$\Sigma$-atom analyses. $U_\Sigma$ enters the transport as a real, momentum-independent
potential, with the $\Sigma$ absorption carried by the cascade cross sections, and with
present hyperon--nucleon input the neutrino data do not determine it.

The stellar numbers are conditional on the EOS model and on the prior on $c_\Lambda$.
With the hyperon couplings anchored to $U_\Lambda(\rhoz)=-28\MeV$ and
$U_\Sigma(\rhoz)=+30\MeV$, GM1 gives $\Mmax=1.94\,\Msun$ for $np\Lambda\Sigma^-$ matter
and FSU2H with the $\phi$ meson $2.03\,\Msun$. A turn-over $c_\Lambda$ within the
chiral-EFT results with $\Lambda NN$ forces, $-5\le c_\Lambda\le2.5\MeV$, gives
$1.62$--$2.08\,\Msun$ in GM1 and $1.75$--$2.12\,\Msun$ in FSU2H, and the mass of
PSR~J0740$+$6620 is reached only at the repulsive end of that range. A joint fit with a
uniform prior over it gives $\Mmax=1.83$ ($1.66$--$2.00$)~$\Msun$ and $1.96$
($1.80$--$2.07$)~$\Msun$, with probabilities of $0.17$ and $0.39$ for $\Mmax>2\,\Msun$;
the neutrino data do not move these values, which are the $c_\Lambda$ prior mapped
through $\Mmax(c_\Lambda)$. The neutrino measurement tests the $\Lambda N$ anchor of the
$\Lambda$ potential below saturation, and whether hyperonic matter in these models
supports $2\,\Msun$ depends on the strength of the $\Lambda NN$ repulsion, which it does
not constrain. At fixed $c_\Lambda$ the potentials spread $\Mmax$ by $0.02$--$0.05\,\Msun$,
an order of magnitude below the $0.45\,\Msun$ spread between EOS models. The radius and
tidal deformability of a $1.4\,\Msun$ star are properties of the nucleonic EOS.

The extraction is a fit of transport-model templates, and the fitted value depends on
the production model, the hyperon--nucleon cross sections and the transport prescription
with which the templates are made. It samples $U_\Lambda$ below saturation with
reconstructed $\Lambda$'s of mean momentum $0.4$--$1.0\GeV/c$, through a reaction in
which the hyperon is produced inside the nucleus. SBND antineutrino data enter this
constraint through the quasi-elastic and two-body normalisations they share with DUNE
antineutrino data: in HM0$+$PC the total is $31\MeV$ with DUNE RHC alone and with the RHC
pair, whose smaller profiled width is offset by a larger potential-convention item.

\appendix
\section{Numerical methods}
\label{sec:numerics}
\begin{table*}[t]
\centering
\caption{Observable values and $(U_\Lambda,U_\Sigma)$ response derivatives at the
truth point, per beam: the reconstructed-$\Lambda$ fraction per produced hyperon
$f_\Lambda$, the $\Sigma^+$ fraction $f_{\Sigma^+}$ (truth level), and the mean
reconstructed $\Lambda$ momentum $\langle p_\Lambda\rangle$. Derivatives
$\partial_\Lambda\equiv\partial/\partial U_\Lambda$ and
$\partial_\Sigma\equiv\partial/\partial U_\Sigma$ are from the quadratic response
surfaces; $N_{\rm reco}$ and $\sigma_p$ complete the Fisher inputs (see text).
The values are at the representative energy $E$ of Table~\ref{tab:exposure}; the
flux-folded $\Sigma^+$ fractions of Sec.~\ref{sec:sigmaplus} are per quasi-elastic
event over the full flux. The Monte Carlo uncertainty is one or two units of the last
digit for $f_\Lambda$,
$\langle p_\Lambda\rangle$ and their derivatives, and $0.7$--$2.6\times10^{-6}\,$MeV$^{-1}$
on the $f_{\Sigma^+}$ derivatives.}
\label{tab:response}
\begin{tabular}{lcc ccc ccc ccc}
\hline
 & & & \multicolumn{3}{c}{$f_\Lambda$\ \ [$\partial$: $10^{-3}\,$MeV$^{-1}$]}
 & \multicolumn{3}{c}{$f_{\Sigma^+}$ [$10^{-3}$]\ \ [$\partial$: $10^{-6}\,$MeV$^{-1}$]}
 & \multicolumn{3}{c}{$\langle p_\Lambda\rangle$ [GeV]\ \ [$\partial$: $10^{-3}\,$GeV\,MeV$^{-1}$]}\\
beam & $N_{\rm reco}$ & $\sigma_p$ [GeV]
 & $f_\Lambda$ & $\partial_\Lambda$ & $\partial_\Sigma$
 & $f_{\Sigma^+}$ & $\partial_\Lambda$ & $\partial_\Sigma$
 & $\langle p_\Lambda\rangle$ & $\partial_\Lambda$ & $\partial_\Sigma$ \\
\hline
SBND RHC & $6.8\times10^{3}$ & $0.25$ & $0.242$ & $+1.89$ & $+0.07$ & $5.35$ & $+7.3$  & $-7.1$  & $0.437$ & $-0.19$ & $+0.17$ \\
SBND FHC & $4.2\times10^{3}$ & $0.39$ & $0.286$ & $+0.71$ & $+0.07$ & $279$  & $+34.2$ & $-34.4$ & $0.707$ & $-0.31$ & $+0.31$ \\
DUNE RHC & $2.4\times10^{5}$ & $0.36$ & $0.256$ & $+1.32$ & $+0.07$ & $10.8$ & $+17.9$ & $-17.7$ & $0.606$ & $-0.76$ & $+0.22$ \\
DUNE FHC & $1.8\times10^{5}$ & $0.58$ & $0.277$ & $+0.48$ & $+0.09$ & $324$  & $+23.6$ & $-24.2$ & $0.985$ & $-0.67$ & $+0.29$ \\
\hline
\end{tabular}
\end{table*}
\paragraph{RMF mean-field solution.} At each $n_B$ the meson mean fields
$(\sigma,\omega_0,\rho_{03})$, and $\phi_0$ in FSU2H, and the composition are found by a damped fixed-point
iteration. For a trial scalar field $S=g_{\sigma N}\sigma$ the effective masses are
$m_B^*=m_B-x_{\sigma B}S$; the composition is solved at fixed $S$ by the nested
bisection of Sec.~\ref{sec:eos}, with each Fermi momentum from
$k_{F,B}^2=(\mu_n-q_B\mu_e-x_{\omega B}W-x_{\rho B}I_{3B}R-V_B)^2-m_B^{*2}$
($W\!=\!g_{\omega N}\omega_0$, $R\!=\!g_{\rho N}\rho_{03}$, $V_B$ the optional turn-over
term, Eq.~\eqref{eq:vextra}, and $-x_{\phi B}F$ with $F=g_{\omega N}\phi_0$ in
FSU2H). The vector fields follow algebraically in GM1 and GM3; in FSU2H, $W$ and $R$
are solved jointly by bisection in $W$ with $R(W)$ explicit, and $F$ follows
algebraically. $S$ is updated by a Newton step (GM1, GM3) or found by bisection on
the physical branch $0\le S<\min_B m_B/x_{\sigma B}$ (FSU2H) of
$S/c_\sigma^2+b\,m_N S^2+c\,S^3=\sum_B x_{\sigma B}\,n^s_B$
[Eq.~\eqref{eq:rmfsigma} divided by $g_{\sigma N}$], with $c_\sigma=g_{\sigma N}/m_\sigma$ and scalar density
$n^s_B=\tfrac{m_B^*}{2\pi^2}[k_F E_F^*-m_B^{*2}\ln\frac{k_F+E_F^*}{m_B^*}]$. The
fields are mixed at $0.5$ and converge in $\lesssim\!20$ passes. With $c_Y\neq0$ the pressure includes the
rearrangement term $n_B\sum_Y n_Y\,\partial V_Y/\partial n_B$ (Sec.~\ref{sec:rmf}).

\paragraph{Crust.} Below the core every RMF star carries the SLy
crust~\cite{Douchin:2001sv} in the four-piece polytrope fit of
Ref.~\cite{Read:2008iy}, from $10^3\,$g\,cm$^{-3}$. The crust is joined to the
uniform RMF matter where the two $\varepsilon(P)$ relations cross ($n\geq0.05\fm^{-3}$),
so that $P$ and $\varepsilon$ are continuous. The crossing lies at $0.065$ (GM1),
$0.060$ (GM3) and $0.110\fm^{-3}$ (FSU2H). The joint is not unified with the
core~\cite{Fortin:2016hny}: moving it to $0.08$--$0.10\fm^{-3}$ lowers $R_{1.4}$ by up
to $0.08$, $0.10$ and $0.04$\,km and $\Lambda_{1.4}$ by up to $5$, $5$ and $37$, and
changes $\Mmax$ by less than $10^{-3}\,\Msun$.

\paragraph{TOV and tidal integration.} The sequence integrates
Eqs.~\eqref{eq:tov1}--\eqref{eq:tov2} by adaptive fourth-order Runge--Kutta in
geometric units, with step $dr=\min(10\,{\rm m},\,0.02\,P/|dP/dr|)$, alongside the
relativistic tidal equation for $y(r)$ \cite{hinderer2008},
\begin{equation}
  r\,y'(r) + y(r)^2 + y(r)\,F(r) + r^2 Q(r) = 0,
\end{equation}
with $F,Q$ functions of $(m,P,\varepsilon,dP/d\varepsilon)$. The relation
$\varepsilon(P)$ and $dP/d\varepsilon$ are taken from one monotone cubic
interpolation in $\ln P$--$\ln\varepsilon$. The surface value $y_R$ and the
compactness $C$ give $k_2$ and $\Lambda=\tfrac23 k_2 C^{-5}$; $\Lambda_{1.4}$ is
converged to $0.02\%$ in the step size.

\paragraph{Hard process and decays.} Hard-scattering final states are distributed
by the corrected RAMBO $n$-body phase space~\cite{rambo}; hyperon and meson decays
use the isotropic (or parity-violating) two-body kinematics. The intranuclear
hyperon transport steps through the Woods--Saxon medium in $0.05\fm$ increments,
applying the $YN$ cross sections and, in force-integrated mode, the gradient force
$-\nabla U_Y$ (conserving $\sqrt{p^2+m^2}+U_Y$ to $<\!10^{-4}\GeV$).

\section{Response derivatives}
\label{sec:response}
For the forecast of Sec.~\ref{sec:detreach}, Table~\ref{tab:response} lists, for each
beam, the three observables evaluated at the truth point $(U_\Lambda,U_\Sigma)=(-28,+30)\MeV$ and their derivatives from
the least-squares quadratic response surfaces fitted to grids pooled over $20$, $12$,
$32$ and $32$ Monte Carlo seeds for SBND RHC, SBND FHC, DUNE RHC and DUNE FHC, with
$2\times10^4$ hyperons per node and seed, together with the reconstructed-$\Lambda$
exposure $N_{\rm reco}=N_Y f_\Lambda$ from Table~\ref{tab:exposure} and the per-event momentum spread
$\sigma_p$ measured on each beam's production channel (quasi-elastic on the RHC rows,
associated on the FHC rows). These suffice to rebuild the fixed-$\gamma$ Fisher
matrix independently of StrangeMC: each fraction enters with binomial variance
$f(1-f)/N_{\rm prod}$, where $N_{\rm prod}=N_{\rm reco}/f_\Lambda$ is the
produced-hyperon sample and the $\Sigma^+$ trials carry the $30\%$ reconstruction
fraction of Sec.~\ref{sec:sigmaplus}, and the mean momentum enters with variance
$\sigma_p^2/N_{\rm reco}$. The nuisance parameters of M2 (Sec.~\ref{sec:detreach})
add one column each to the design matrix, $+f$ on the
fractions they scale, with Gaussian priors: the quasi-elastic ($20\%$, both RHC rows)
and associated ($50\%$, both FHC rows) cross sections, the flux of each beam
($7.5\%$) and the $\Lambda$ efficiency of each detector ($3\%$) scale $f_\Lambda$ and
$f_{\Sigma^+}$; the $\Sigma^+$ efficiency of each detector ($30\%$) scales
$f_{\Sigma^+}$ alone. With these columns the table reproduces, to $2\%$, the widths
without the two-body component given in the caption of Table~\ref{tab:stat}.

\begin{acknowledgments}
Although the author is a member of the DUNE and SBND collaborations, all views
presented here are his and not those of the collaborations as a whole. This work was
supported by the Science and Technology Facilities Council (STFC) Lancaster EPP
Consolidated Grant 2025--2029: UKRI2846.
\end{acknowledgments}

\bibliographystyle{apsrev4-2}
\bibliography{refs}

@phdthesis{nowakphd,
  author  = {Nowak, J. A.},
  title   = {Konstruowanie generatora oddzia{\l}ywa{\'n} neutrin},
  school  = {University of Wroc{\l}aw},
  year    = {2006},
  note = {in Polish}
}

@article{singh2006,
  author  = {Singh, S. K. and Vicente Vacas, M. J.},
  title   = {Weak quasielastic production of hyperons},
  journal = {Phys. Rev. D},
  volume  = {74},
  pages   = {053009},
  year    = {2006},
  doi     = {10.1103/PhysRevD.74.053009},
  eprint  = {hep-ph/0606235},
  archivePrefix = {arXiv}
}

@article{alam2010,
  author  = {Rafi Alam, M. and Ruiz Simo, I. and Sajjad Athar, M. and Vicente Vacas, M. J.},
  title   = {Weak kaon production off the nucleon},
  journal = {Phys. Rev. D},
  volume  = {82},
  pages   = {033001},
  year    = {2010},
  doi     = {10.1103/PhysRevD.82.033001},
  eprint  = {1004.5484},
  archivePrefix = {arXiv}
}

@article{alam2013,
  author  = {Alam, M. Rafi and Ruiz Simo, I. and Athar, M. Sajjad and Alvarez-Ruso, L. and Vicente Vacas, M. J.},
  title   = {Weak production of strange particles off the nucleon},
  journal = {AIP Conf. Proc.},
  volume  = {1663},
  pages   = {070004},
  year    = {2015},
  doi     = {10.1063/1.4919485},
  eprint  = {1303.5924},
  archivePrefix = {arXiv},
  primaryClass = {hep-ph}
}

@article{fatima2025,
  author  = {Fatima, A. and Sajjad Athar, M. and Singh, S. K.},
  title   = {Charged current neutrino and antineutrino induced associated particle production from nucleons},
  journal = {Phys. Rev. D},
  year    = {2025},
  eprint  = {2507.20754},
  archivePrefix = {arXiv},
  primaryClass = {hep-ph},
  volume = {112},
  pages = {073004},
  doi = {10.1103/6c5c-sl4t}
}

@article{microboone2025,
  author  = {{MicroBooNE Collaboration}},
  title   = {First measurement of charged-current muon-neutrino-induced $K^+$ production on argon using the MicroBooNE detector},
  journal = {Phys. Rev. Lett.},
  volume  = {135},
  pages   = {251804},
  year    = {2025},
  eprint  = {2503.00291},
  archivePrefix = {arXiv},
  primaryClass = {hep-ex},
  doi = {10.1103/96vx-gq2g}
}

@article{Nowak:2026czc,
    author = "Nowak, Jaroslaw",
    title = "{LUNAR: a Monte Carlo generator for bound-nucleon decay in liquid argon}",
    eprint = "2606.30872",
    archivePrefix = "arXiv",
    primaryClass = "hep-ph",
    month = "6",
    year = "2026",
  journal = "Phys. Rev. D",
  doi = "10.1103/mw7s-jl1c"
}

@misc{Nowak:2026hyperonPRL,
  author  = {Nowak, J. A.},
  title   = {Accelerator neutrinos as a probe of in-medium hyperon potentials},
  year    = {2026},
  eprint  = {2607.09255},
  archivePrefix = {arXiv}
}

@article{demorest2010,
  author  = {Demorest, P. B. and Pennucci, T. and Ransom, S. M. and Roberts, M. S. E. and Hessels, J. W. T.},
  title   = {A two-solar-mass neutron star measured using {Shapiro} delay},
  journal = {Nature},
  volume  = {467},
  pages   = {1081},
  year    = {2010},
  doi     = {10.1038/nature09466},
  eprint  = {1010.5788},
  archivePrefix = {arXiv}
}

@article{antoniadis2013,
  author  = {Antoniadis, J. and others},
  title   = {A massive pulsar in a compact relativistic binary},
  journal = {Science},
  volume  = {340},
  pages   = {1233232},
  year    = {2013},
  doi     = {10.1126/science.1233232},
  eprint  = {1304.6875},
  archivePrefix = {arXiv}
}

@article{fonseca2021,
  author  = {Fonseca, E. and others},
  title   = {Refined Mass and Geometric Measurements of the High-mass {PSR J0740+6620}},
  journal = {Astrophys. J. Lett.},
  volume  = {915},
  pages   = {L12},
  year    = {2021},
  doi     = {10.3847/2041-8213/ac03b8},
  eprint  = {2104.00880},
  archivePrefix = {arXiv}
}

@article{chatterjee2016,
  author  = {Chatterjee, D. and Vida\~{n}a, I.},
  title   = {Do hyperons exist in the interior of neutron stars?},
  journal = {Eur. Phys. J. A},
  volume  = {52},
  pages   = {29},
  year    = {2016},
  doi     = {10.1140/epja/i2016-16029-x},
  eprint  = {1510.06306},
  archivePrefix = {arXiv}
}

@article{tolos2020,
  author  = {Tolos, L. and Fabbietti, L.},
  title   = {Strangeness in nuclei and neutron stars},
  journal = {Prog. Part. Nucl. Phys.},
  volume  = {112},
  pages   = {103770},
  year    = {2020},
  doi     = {10.1016/j.ppnp.2020.103770},
  eprint  = {2002.09223},
  archivePrefix = {arXiv}
}

@article{burgio2021,
  author  = {Burgio, G. F. and Schulze, H.-J. and Vida\~{n}a, I. and Wei, J.-B.},
  title   = {Neutron stars and the nuclear equation of state},
  journal = {Prog. Part. Nucl. Phys.},
  volume  = {120},
  pages   = {103879},
  year    = {2021},
  eprint  = {2105.03747},
  archivePrefix = {arXiv},
  doi = {10.1016/j.ppnp.2021.103879}
}

@article{gal2016,
  author  = {Gal, A. and Hungerford, E. V. and Millener, D. J.},
  title   = {Strangeness in nuclear physics},
  journal = {Rev. Mod. Phys.},
  volume  = {88},
  pages   = {035004},
  year    = {2016},
  doi     = {10.1103/RevModPhys.88.035004},
  eprint  = {1605.00557},
  archivePrefix = {arXiv}
}

@article{saha2004,
  author  = {Saha, P. K. and others},
  title   = {{$\Sigma$}-nucleus potential studied with the $(\pi^-,K^+)$ reaction on medium-to-heavy nuclear targets},
  journal = {Phys. Rev. C},
  volume  = {70},
  pages   = {044613},
  year    = {2004},
  doi     = {10.1103/PhysRevC.70.044613},
  eprint  = {nucl-ex/0405031},
  archivePrefix = {arXiv}
}

@article{bednarek2012,
  author  = {Bednarek, I. and Haensel, P. and Zdunik, J. L. and Bejger, M. and Ma\'{n}ka, R.},
  title   = {Hyperons in neutron-star cores and a 2\,{$M_\odot$} pulsar},
  journal = {Astron. Astrophys.},
  volume  = {543},
  pages   = {A157},
  year    = {2012},
  doi     = {10.1051/0004-6361/201118560},
  eprint  = {1111.6942},
  archivePrefix = {arXiv}
}

@article{ohnishi2022,
  author  = {Ohnishi, A. and Jinno, A. and Murase, K. and Nara, Y.},
  title   = {Directed flow of {$\Lambda$} from heavy-ion collisions and hyperon puzzle of neutron stars},
  journal = {EPJ Web Conf.},
  volume  = {271},
  pages   = {08006},
  year    = {2022},
  doi     = {10.1051/epjconf/202227108006},
  eprint  = {2210.17202},
  archivePrefix = {arXiv}
}

@article{nara2022,
  author  = {Nara, Y. and Jinno, A. and Murase, K. and Ohnishi, A.},
  title   = {Directed flow of {$\Lambda$} in high-energy heavy-ion collisions and {$\Lambda$} potential in dense nuclear matter},
  journal = {Phys. Rev. C},
  volume  = {106},
  pages   = {044902},
  year    = {2022},
  doi     = {10.1103/PhysRevC.106.044902},
  eprint  = {2208.01297},
  archivePrefix = {arXiv}
}

@article{Jinno:2025lfu,
  author  = {Jinno, A. and Murase, K. and Nara, Y. and Haidenbauer, J.},
  title   = {{$\Lambda$} and {$\Sigma$} potentials in dense matter based on chiral {EFT}: bridging heavy-ion collisions, hypernuclei, and neutron stars},
  journal = {EPJ Web Conf.},
  volume  = {364},
  pages   = {14004},
  year    = {2026},
  doi     = {10.1051/epjconf/202636414004},
  eprint  = {2508.19560},
  archivePrefix = {arXiv}
}

@article{Weissenborn:2011ut,
  author={Weissenborn, Simon and Chatterjee, Debarati and Schaffner-Bielich, Juergen},
  title={{Hyperons and massive neutron stars: vector repulsion and SU(3) symmetry}},
  journal={Phys. Rev. C}, volume={85}, pages={065802}, year={2012},
  doi={10.1103/PhysRevC.85.065802}, eprint={1112.0234}, archivePrefix={arXiv}, note={[Erratum: Phys.Rev.C 90, 019904 (2014)]}}

@article{Vidana:2018bdi,
  author={Vida{\~n}a, Isaac},
  title={{Hyperons: the strange ingredients of the nuclear equation of state}},
  journal={Proc. Roy. Soc. Lond. A}, volume={474}, pages={20180145}, year={2018},
  doi={10.1098/rspa.2018.0145}, eprint={1803.00504}, archivePrefix={arXiv}}

@article{Schulze:2011zza,
  author={Schulze, H. -J. and Rijken, T.},
  title={{Maximum mass of hyperon stars with the Nijmegen ES C-08 model}},
  journal={Phys. Rev. C}, volume={84}, pages={035801}, year={2011},
  doi={10.1103/PhysRevC.84.035801}}

@article{Sedrakian:2022ata,
  author={Sedrakian, Armen and Li, Jia-Jie and Weber, Fridolin},
  title={{Heavy baryons in compact stars}},
  journal={Prog. Part. Nucl. Phys.}, volume={131}, pages={104041}, year={2023},
  doi={10.1016/j.ppnp.2023.104041}, eprint={2212.01086}, archivePrefix={arXiv}}

@article{Oertel:2016bki,
  author={Oertel, M. and Hempel, M. and Kl{\"a}hn, T. and Typel, S.},
  title={{Equations of state for supernovae and compact stars}},
  journal={Rev. Mod. Phys.}, volume={89}, pages={015007}, year={2017},
  doi={10.1103/RevModPhys.89.015007}, eprint={1610.03361}, archivePrefix={arXiv}}

@article{Weissenborn:2011kb,
  author={Weissenborn, S. and Chatterjee, D. and Schaffner-Bielich, J.},
  title={{Hyperons and massive neutron stars: the role of hyperon potentials}},
  journal={Nucl. Phys. A}, volume={881}, pages={62--77}, year={2012},
  doi={10.1016/j.nuclphysa.2012.02.012}, eprint={1111.6049}, archivePrefix={arXiv}}

@article{Colucci:2013pya,
  author={Colucci, Giuseppe and Sedrakian, Armen},
  title={{Equation of state of hypernuclear matter: impact of hyperon--scalar-meson couplings}},
  journal={Phys. Rev. C}, volume={87}, pages={055806}, year={2013},
  doi={10.1103/PhysRevC.87.055806}, eprint={1302.6925}, archivePrefix={arXiv}}

@article{Oertel:2014qza,
  author={Oertel, M. and Provid{\^e}ncia, C. and Gulminelli, F. and Raduta, Ad R.},
  title={{Hyperons in neutron star matter within relativistic mean-field models}},
  journal={J. Phys. G}, volume={42}, pages={075202}, year={2015},
  doi={10.1088/0954-3899/42/7/075202}, eprint={1412.4545}, archivePrefix={arXiv}}

@article{Tolos:2016hhl,
  author={Tolos, Laura and Centelles, Mario and Ramos, Angels},
  title={{Equation of State for Nucleonic and Hyperonic Neutron Stars with Mass and Radius Constraints}},
  journal={Astrophys. J.}, volume={834}, pages={3}, year={2017},
  doi={10.3847/1538-4357/834/1/3}, eprint={1610.00919}, archivePrefix={arXiv}}

@article{glendenning1991,
  author  = {Glendenning, N. K. and Moszkowski, S. A.},
  title   = {Reconciliation of neutron-star masses and binding of the {$\Lambda$} in hypernuclei},
  journal = {Phys. Rev. Lett.},
  volume  = {67},
  pages   = {2414},
  year    = {1991},
  doi     = {10.1103/PhysRevLett.67.2414}
}

@article{Millener:1988hp,
  author={Millener, D. J. and Dover, C. B. and Gal, A.},
  title={{Lambda Nucleus Single Particle Potentials}},
  journal={Phys. Rev. C}, volume={38}, pages={2700--2708}, year={1988},
  doi={10.1103/PhysRevC.38.2700}}

@article{Hashimoto:2006aw,
  author={Hashimoto, O. and Tamura, H.},
  title={{Spectroscopy of Lambda hypernuclei}},
  journal={Prog. Part. Nucl. Phys.}, volume={57}, pages={564--653}, year={2006},
  doi={10.1016/j.ppnp.2005.07.001}}

@article{Fortin:2020qin,
  author={Fortin, M. and Raduta, Adriana R. and Avancini, Sidney and Provid{\^e}ncia, Constan{\c{c}}a},
  title={{Relativistic hypernuclear compact stars with calibrated equations of state}},
  journal={Phys. Rev. D}, volume={101}, pages={034017}, year={2020},
  doi={10.1103/PhysRevD.101.034017}, eprint={2001.08036}, archivePrefix={arXiv}}

@article{Fortin:2017cvt,
  author={Fortin, M. and Avancini, S. S. and Provid{\^e}ncia, C. and Vida{\~n}a, I.},
  title={{Hypernuclei and massive neutron stars}},
  journal={Phys. Rev. C}, volume={95}, pages={065803}, year={2017},
  doi={10.1103/PhysRevC.95.065803}, eprint={1701.06373}, archivePrefix={arXiv}}

@article{Batty:1997zp,
  author={Batty, C. J. and Friedman, E. and Gal, A.},
  title={{Strong interaction physics from hadronic atoms}},
  journal={Phys. Rept.}, volume={287}, pages={385--445}, year={1997},
  doi={10.1016/S0370-1573(97)00011-2}}

@article{Friedman:2007zza,
  author={Friedman, E. and Gal, A.},
  title={{In-medium nuclear interactions of low-energy hadrons}},
  journal={Phys. Rept.}, volume={452}, pages={89--153}, year={2007},
  doi={10.1016/j.physrep.2007.08.002}, eprint={0705.3965}, archivePrefix={arXiv}}

@article{Noumi:2001tx,
  author={Noumi, H. and others},
  title={{Sigma nucleus potential in A = 28}},
  journal={Phys. Rev. Lett.}, volume={89}, pages={072301}, year={2002},
  doi={10.1103/PhysRevLett.89.072301}, note={[Erratum: Phys.Rev.Lett. 90, 049902 (2003)]}}

@article{Harada:2005hs,
  author={Harada, T. and Hirabayashi, Y.},
  title={{Is the Sigma-nucleus potential for Sigma- atoms consistent with the Si-28(pi-, K+) data?}},
  journal={Nucl. Phys. A}, volume={759}, pages={143--169}, year={2005},
  doi={10.1016/j.nuclphysa.2005.04.025}}

@article{ALICE:2019eol,
  author={Acharya, Shreyasi and others}, collaboration={ALICE},
  title={{Study of the $\Lambda$-$\Lambda$ interaction with femtoscopy correlations in pp and p-Pb collisions at the LHC}},
  journal={Phys. Lett. B}, volume={797}, pages={134822}, year={2019},
  doi={10.1016/j.physletb.2019.134822}, eprint={1905.07209}, archivePrefix={arXiv}}

@article{Vidana:2010ip,
  author={Vidana, Isaac and Logoteta, Domenico and Providencia, Constanza and Polls, Artur and Bombaci, Ignazio},
  title={{Estimation of the effect of hyperonic three-body forces on the maximum mass of neutron stars}},
  journal={EPL}, volume={94}, pages={11002}, year={2011},
  doi={10.1209/0295-5075/94/11002}, eprint={1006.5660}, archivePrefix={arXiv}}

@article{Yamamoto:2014jga,
  author={Yamamoto, Y. and Furumoto, T. and Yasutake, N. and Rijken, Th. A.},
  title={{Hyperon mixing and universal many-body repulsion in neutron stars}},
  journal={Phys. Rev. C}, volume={90}, pages={045805}, year={2014},
  doi={10.1103/PhysRevC.90.045805}, eprint={1406.4332}, archivePrefix={arXiv}}

@article{Lonardoni:2014bwa,
  author={Lonardoni, Diego and Lovato, Alessandro and Gandolfi, Stefano and Pederiva, Francesco},
  title={{Hyperon Puzzle: Hints from Quantum Monte Carlo Calculations}},
  journal={Phys. Rev. Lett.}, volume={114}, pages={092301}, year={2015},
  doi={10.1103/PhysRevLett.114.092301}, eprint={1407.4448}, archivePrefix={arXiv}}

@article{Haidenbauer:2016vfq,
  author={Haidenbauer, J. and Mei{\ss}ner, U. -G. and Kaiser, N. and Weise, W.},
  title={{Lambda-nuclear interactions and hyperon puzzle in neutron stars}},
  journal={Eur. Phys. J. A}, volume={53}, pages={121}, year={2017},
  doi={10.1140/epja/i2017-12316-4}, eprint={1612.03758}, archivePrefix={arXiv}}

@article{Kohno:2018gby,
  author={Kohno, M.},
  title={{Single-particle potential of the {\ensuremath{\Lambda}} hyperon in nuclear matter with chiral effective field theory NLO interactions including effects of YNN three-baryon interactions}},
  journal={Phys. Rev. C}, volume={97}, pages={035206}, year={2018},
  doi={10.1103/PhysRevC.97.035206}, eprint={1802.05388}, archivePrefix={arXiv}}

@article{Gerstung:2020ktv,
    author = "Gerstung, Dominik and Kaiser, Norbert and Weise, Wolfram",
    title = "{Hyperon{\textendash}nucleon three-body forces and strangeness in neutron stars}",
    eprint = "2001.10563",
    archivePrefix = "arXiv",
    primaryClass = "nucl-th",
    doi = "10.1140/epja/s10050-020-00180-2",
    journal = "Eur. Phys. J. A",
    volume = "56",
    number = "6",
    pages = "175",
    year = "2020"
}

@article{Logoteta:2019utx,
  author={Logoteta, Domenico and Vidana, Isaac and Bombaci, Ignazio},
  title={{Impact of chiral hyperonic three-body forces on neutron stars}},
  journal={Eur. Phys. J. A}, volume={55}, pages={207}, year={2019},
  doi={10.1140/epja/i2019-12909-9}, eprint={1906.11722}, archivePrefix={arXiv}}

@article{Friedman:2022bpw,
  author={Friedman, E. and Gal, A.},
  title={{Constraints from {\ensuremath{\Lambda}} hypernuclei on the {\ensuremath{\Lambda}}NN content of the {\ensuremath{\Lambda}}-nucleus potential}},
  journal={Phys. Lett. B}, volume={837}, pages={137669}, year={2023},
  doi={10.1016/j.physletb.2023.137669}, eprint={2204.02264}, archivePrefix={arXiv}}

@article{miwa2021,
  author  = {Miwa, K. and others},
  collaboration = {J-PARC E40 Collaboration},
  title   = {Measurement of the differential cross sections of the {$\Sigma^- p$}
             elastic scattering in momentum range 470 to 850 {MeV}/$c$},
  journal = {Phys. Rev. C},
  volume  = {104},
  pages   = {045204},
  year    = {2021},
  doi     = {10.1103/PhysRevC.104.045204},
  eprint  = {2104.13608},
  archivePrefix = {arXiv}
}

@article{miwa2022,
  author  = {Miwa, K. and others},
  collaboration = {J-PARC E40 Collaboration},
  title   = {Precise measurement of the differential cross sections of the
             {$\Sigma^- p \to \Lambda n$} reaction},
  journal = {Phys. Rev. Lett.},
  volume  = {128},
  pages   = {072501},
  year    = {2022},
  doi     = {10.1103/PhysRevLett.128.072501},
  eprint = {2111.14277},
  archivePrefix = {arXiv}
}

@article{nanamura2022,
  author  = {Nanamura, T. and others},
  collaboration = {J-PARC E40 Collaboration},
  title   = {Measurement of differential cross sections for {$\Sigma^+ p$}
             elastic scattering in the momentum range 0.44--0.80 {GeV}/$c$},
  journal = {Prog. Theor. Exp. Phys.},
  volume  = {2022},
  pages   = {093D01},
  year    = {2022},
  doi     = {10.1093/ptep/ptac101},
  eprint  = {2203.08393},
  archivePrefix = {arXiv}
}

@article{CLAS:2021gur,
  author={Rowley, J. and others}, collaboration={CLAS},
  title={{Improved $\Lambda p$ Elastic Scattering Cross Sections Between 0.9 and 2.0 GeV/c and Connections to the Neutron Star Equation of State}},
  journal={Phys. Rev. Lett.}, volume={127}, pages={272303}, year={2021},
  doi={10.1103/PhysRevLett.127.272303}, eprint={2108.03134}, archivePrefix={arXiv}}

@article{alice2020nature,
  author  = {Acharya, S. and others},
  collaboration = {ALICE Collaboration},
  title   = {Unveiling the strong interaction among hadrons at the {LHC}},
  journal = {Nature},
  volume  = {588},
  pages   = {232--238},
  year    = {2020},
  doi     = {10.1038/s41586-020-3001-6},
  eprint  = {2005.11495},
  archivePrefix = {arXiv},
  note = {Publisher Correction: Nature 590, E13 (2021)}
}

@article{Fabbietti:2020bfg,
  author={Fabbietti, L. and Mantovani Sarti, V. and Vazquez Doce, O.},
  title={{Study of the Strong Interaction Among Hadrons with Correlations at the LHC}},
  journal={Ann. Rev. Nucl. Part. Sci.}, volume={71}, pages={377--402}, year={2021},
  doi={10.1146/annurev-nucl-102419-034438}, eprint={2012.09806}, archivePrefix={arXiv}}

@article{alice2022coupled,
  author  = {Acharya, S. and others},
  collaboration = {ALICE Collaboration},
  title   = {Exploring the {N$\Lambda$--N$\Sigma$} coupled system with high
             precision correlation techniques at the {LHC}},
  journal = {Phys. Lett. B},
  volume  = {833},
  pages   = {137272},
  year    = {2022},
  doi     = {10.1016/j.physletb.2022.137272},
  eprint  = {2104.04427},
  archivePrefix = {arXiv}
}

@article{ALICE:2022boj,
  author={Acharya, Shreyasi and others}, collaboration={ALICE},
  title={{Towards the understanding of the genuine three-body interaction for p{\textendash}p{\textendash}p and p{\textendash}p{\textendash}$\Lambda $}},
  journal={Eur. Phys. J. A}, volume={59}, pages={145}, year={2023},
  doi={10.1140/epja/s10050-023-00998-6}, eprint={2206.03344}, archivePrefix={arXiv}}

@article{Aoki:2012tk,
  author={Aoki, Sinya and Doi, Takumi and Hatsuda, Tetsuo and Ikeda, Yoichi and Inoue, Takashi and Ishii, Noriyoshi and Murano, Keiko and Nemura, Hidekatsu and Sasaki, Kenji}, collaboration={HAL QCD},
  title={{Lattice QCD approach to Nuclear Physics}},
  journal={PTEP}, volume={2012}, pages={01A105}, year={2012},
  doi={10.1093/ptep/pts010}, eprint={1206.5088}, archivePrefix={arXiv}}

@article{HALQCD:2019wsz,
  author={Sasaki, Kenji and others}, collaboration={HAL QCD},
  title={{$\Lambda\Lambda$ and N$\Xi$ interactions from lattice QCD near the physical point}},
  journal={Nucl. Phys. A}, volume={998}, pages={121737}, year={2020},
  doi={10.1016/j.nuclphysa.2020.121737}, eprint={1912.08630}, archivePrefix={arXiv}}

@article{Inoue:2018axd,
  author={Inoue, Takashi}, collaboration={HAL QCD},
  title={{Strange Nuclear Physics from QCD on Lattice}},
  journal={AIP Conf. Proc.}, volume={2130}, pages={020002}, year={2019},
  doi={10.1063/1.5118370}, eprint={1809.08932}, archivePrefix={arXiv}}

@article{Haidenbauer:2013oca,
  author={Haidenbauer, J. and Petschauer, S. and Kaiser, N. and Meissner, U. -G. and Nogga, A. and Weise, W.},
  title={{Hyperon-nucleon interaction at next-to-leading order in chiral effective field theory}},
  journal={Nucl. Phys. A}, volume={915}, pages={24--58}, year={2013},
  doi={10.1016/j.nuclphysa.2013.06.008}, eprint={1304.5339}, archivePrefix={arXiv}}

@article{Petschauer:2015nea,
  author={Petschauer, S. and Haidenbauer, J. and Kaiser, N. and Mei{\ss}ner, Ulf-G. and Weise, W.},
  title={{Hyperons in nuclear matter from SU(3) chiral effective field theory}},
  journal={Eur. Phys. J. A}, volume={52}, pages={15}, year={2016},
  doi={10.1140/epja/i2016-16015-4}, eprint={1507.08808}, archivePrefix={arXiv}}

@article{petschauer2020,
  author  = {Petschauer, S. and Haidenbauer, J. and Kaiser, N. and Mei{\ss}ner, U.-G. and Weise, W.},
  title   = {Hyperon-nuclear interactions from {SU(3)} chiral effective field theory},
  journal = {Front. Phys.},
  volume  = {8},
  pages   = {12},
  year    = {2020},
  doi     = {10.3389/fphy.2020.00012},
  eprint  = {2002.00424},
  archivePrefix = {arXiv}
}

@article{Haidenbauer:2019boi,
  author={Haidenbauer, J. and Mei{\ss}ner, U. -G. and Nogga, A.},
  title={{Hyperon{\textendash}nucleon interaction within chiral effective field theory revisited}},
  journal={Eur. Phys. J. A}, volume={56}, pages={91}, year={2020},
  doi={10.1140/epja/s10050-020-00100-4}, eprint={1906.11681}, archivePrefix={arXiv}}

@article{Masuda:2012kf,
  author={Masuda, Kota and Hatsuda, Tetsuo and Takatsuka, Tatsuyuki},
  title={{Hadron-Quark Crossover and Massive Hybrid Stars with Strangeness}},
  journal={Astrophys. J.}, volume={764}, pages={12}, year={2013},
  doi={10.1088/0004-637X/764/1/12}, eprint={1205.3621}, archivePrefix={arXiv}}

@article{Baym:2017whm,
  author={Baym, Gordon and Hatsuda, Tetsuo and Kojo, Toru and Powell, Philip D. and Song, Yifan and Takatsuka, Tatsuyuki},
  title={{From hadrons to quarks in neutron stars: a review}},
  journal={Rept. Prog. Phys.}, volume={81}, pages={056902}, year={2018},
  doi={10.1088/1361-6633/aaae14}, eprint={1707.04966}, archivePrefix={arXiv}}

@article{Annala:2019puf,
  author={Annala, Eemeli and Gorda, Tyler and Kurkela, Aleksi and N{\"a}ttil{\"a}, Joonas and Vuorinen, Aleksi},
  title={{Evidence for quark-matter cores in massive neutron stars}},
  journal={Nature Phys.}, volume={16}, pages={907--910}, year={2020},
  doi={10.1038/s41567-020-0914-9}, eprint={1903.09121}, archivePrefix={arXiv}}

@article{McLerran:2018hbz,
    author = "McLerran, Larry and Reddy, Sanjay",
    title = "{Quarkyonic Matter and Neutron Stars}",
    eprint = "1811.12503",
    archivePrefix = "arXiv",
    primaryClass = "nucl-th",
    reportNumber = "INT-PUB-18-060",
    doi = "10.1103/PhysRevLett.122.122701",
    journal = "Phys. Rev. Lett.",
    volume = "122",
    number = "12",
    pages = "122701",
    year = "2019"
}

@article{Fujimoto:2023mzy,
    author = "Fujimoto, Yuki and Kojo, Toru and McLerran, Larry D.",
    title = "{Momentum Shell in Quarkyonic Matter from Explicit Duality: A Dual Model for Cold, Dense QCD}",
    eprint = "2306.04304",
    archivePrefix = "arXiv",
    primaryClass = "nucl-th",
    reportNumber = "INT-PUB-23-018",
    doi = "10.1103/PhysRevLett.132.112701",
    journal = "Phys. Rev. Lett.",
    volume = "132",
    number = "11",
    pages = "112701",
    year = "2024"
}

@article{Margueron:2021dtx,
    author = "Margueron, J. and Hansen, H. and Proust, P. and Chanfray, G.",
    title = "{Quarkyonic stars with isospin-flavor asymmetry}",
    eprint = "2103.10209",
    archivePrefix = "arXiv",
    primaryClass = "nucl-th",
    doi = "10.1103/PhysRevC.104.055803",
    journal = "Phys. Rev. C",
    volume = "104",
    number = "5",
    pages = "055803",
    year = "2021"
}

@article{Fujimoto:2024doc,
    author = "Fujimoto, Yuki and Kojo, Toru and McLerran, Larry",
    title = "{Evolution of strangeness and hyperons in quarkyonic matter}",
    eprint = "2410.22758",
    archivePrefix = "arXiv",
    primaryClass = "nucl-th",
    reportNumber = "INT-PUB-24-056, RIKEN-iTHEMS-Report-24",
    doi = "10.1103/txbp-t8vm",
    journal = "Phys. Rev. C",
    volume = "113",
    number = "3",
    pages = "035206",
    year = "2026"
}

@article{Drago:2013fsa,
  author={Drago, Alessandro and Lavagno, Andrea and Pagliara, Giuseppe},
  title={{Can very compact and very massive neutron stars both exist?}},
  journal={Phys. Rev. D}, volume={89}, pages={043014}, year={2014},
  doi={10.1103/PhysRevD.89.043014}, eprint={1309.7263}, archivePrefix={arXiv}}

@article{Tolos:2017lgv,
  author  = {Tolos, L. and Centelles, M. and Ramos, A.},
  title   = {The equation of state for the nucleonic and hyperonic core of neutron stars},
  journal = {Publ. Astron. Soc. Austral.},
  volume  = {34},
  pages   = {e065},
  year    = {2017},
  doi     = {10.1017/pasa.2017.60},
  eprint  = {1708.08681},
  archivePrefix = {arXiv}
}

@article{Friedman:2023ucs,
  author  = {Friedman, E. and Gal, A.},
  title   = {{$\Lambda$} hypernuclear potentials beyond linear density dependence},
  journal = {Nucl. Phys. A},
  volume  = {1039},
  pages   = {122725},
  year    = {2023},
  doi     = {10.1016/j.nuclphysa.2023.122725},
  eprint  = {2306.06973},
  archivePrefix = {arXiv}
}

@article{benitez2024,
  author  = {Benitez Galan, M. and Alvarez-Ruso, L. and Rafi Alam, M. and Ruiz Simo, I. and Vicente Vacas, M. J.},
  title   = {Cabibbo suppressed hyperon production off nuclei induced by antineutrinos},
  journal = {Phys. Rev. D},
  volume  = {109},
  pages   = {033001},
  year    = {2024},
  doi     = {10.1103/PhysRevD.109.033001},
  eprint  = {2305.17004},
  archivePrefix = {arXiv}
}

@article{microboone2023lambda,
  author  = {{MicroBooNE Collaboration}},
  title   = {First Measurement of Quasielastic {$\Lambda$} Baryon Production in Muon Antineutrino Interactions in the MicroBooNE Detector},
  journal = {Phys. Rev. Lett.},
  volume  = {130},
  pages   = {231802},
  year    = {2023},
  doi     = {10.1103/PhysRevLett.130.231802},
  eprint = {2212.07888},
  archivePrefix = {arXiv}
}

@article{Sobczyk:2019uej,
  author={Sobczyk, J. E. and Rocco, N. and Lovato, A. and Nieves, J.},
  title={{Weak Production of Strange and Charmed Ground-State Baryons in Nuclei}},
  journal={Phys. Rev. C}, volume={99}, pages={065503}, year={2019},
  doi={10.1103/PhysRevC.99.065503}, eprint={1901.10192}, archivePrefix={arXiv}}

@article{Hernandez:2025snr,
  author={Hern{\'a}ndez, E. and Nieves, J. and Sobczyk, J. E.},
  title={{Inclusive quasielastic neutrino and antineutrino scattering off nuclei within the Standard Model and beyond}},
  journal={Phys. Rev. D}, volume={112}, pages={053006}, year={2025},
  doi={10.1103/ll9y-1v4z}, eprint={2506.16774}, archivePrefix={arXiv}}

@article{alberico2002,
  author  = {Alberico, W. M. and Garbarino, G.},
  title   = {Weak decay of {$\Lambda$}-hypernuclei},
  journal = {Phys. Rept.},
  volume  = {369},
  pages   = {1},
  year    = {2002},
  eprint  = {nucl-th/0112036},
  archivePrefix = {arXiv}
}

@article{hinderer2008,
  author  = {Hinderer, Tanja},
  title   = {Tidal Love numbers of neutron stars},
  journal = {Astrophys. J.},
  volume  = {677},
  pages   = {1216},
  year    = {2008},
  doi     = {10.1086/533487},
  eprint  = {0711.2420},
  archivePrefix = {arXiv},
  note = {Erratum: Astrophys. J. 697, 964 (2009)}
}

@article{ligo2017gw170817,
  author  = {{LIGO Scientific Collaboration and Virgo Collaboration}},
  title   = {{GW170817}: Observation of Gravitational Waves from a Binary Neutron Star Inspiral},
  journal = {Phys. Rev. Lett.},
  volume  = {119},
  pages   = {161101},
  year    = {2017},
  doi     = {10.1103/PhysRevLett.119.161101},
  eprint  = {1710.05832},
  archivePrefix = {arXiv}
}

@article{abbott2018eos,
  author  = {{LIGO Scientific Collaboration and Virgo Collaboration}},
  title   = {{GW170817}: Measurements of Neutron Star Radii and Equation of State},
  journal = {Phys. Rev. Lett.},
  volume  = {121},
  pages   = {161101},
  year    = {2018},
  doi     = {10.1103/PhysRevLett.121.161101},
  eprint  = {1805.11581},
  archivePrefix = {arXiv}
}

@article{abbott2019properties,
  author  = {{LIGO Scientific Collaboration and Virgo Collaboration}},
  title   = {Properties of the Binary Neutron Star Merger {GW170817}},
  journal = {Phys. Rev. X},
  volume  = {9},
  pages   = {011001},
  year    = {2019},
  doi     = {10.1103/PhysRevX.9.011001},
  eprint  = {1805.11579},
  archivePrefix = {arXiv}
}

@unpublished{companionPRC,
  author  = {Nowak, J. A.},
  title   = {In-medium hyperon potentials and the quarkyonic hyperon onset:
             charged $\Sigma$'s in $\beta$-equilibrium and the neutrino connection},
  note    = {submitted to Phys. Rev. C},
  year    = {2026},
  eprint = {2607.09280},
  archivePrefix = {arXiv}
}

@misc{Nowak:2026twobody,
  author = {Nowak, J. A.},
  title  = {Strangeness-changing two-body weak currents from {SU(3)} chiral dynamics},
  note   = {submitted to Phys. Rev. C},
  year   = {2026},
  eprint = {2610.32375},
  archivePrefix = {arXiv}
}

@misc{Nowak:2026pheno,
  author = {Nowak, J. A.},
  title  = {Phenomenology of strangeness-changing two-body currents in antineutrino scattering},
  note   = {to be submitted to Phys. Rev. C},
  year   = {2026}
}

@article{thorpe2021,
  author  = {Thorpe, C. and Nowak, J. and Niewczas, K. and Sobczyk, J. T. and Juszczak, C.},
  title   = {Second class currents, axial mass, and nuclear effects in hyperon production},
  journal = {Phys. Rev. C},
  volume  = {104},
  pages   = {035502},
  year    = {2021},
  doi     = {10.1103/PhysRevC.104.035502},
  eprint  = {2010.12361},
  archivePrefix = {arXiv}
}

@article{Sechi-Zorn:1968mao,
  author  = {Sechi-Zorn, B. and Kehoe, B. and Twitty, J. and Burnstein, R. A.},
  title   = {Low-energy {$\Lambda$}-proton elastic scattering},
  journal = {Phys. Rev.},
  volume  = {175},
  pages   = {1735},
  year    = {1968},
  doi     = {10.1103/PhysRev.175.1735}
}

@article{Alexander:1968acu,
  author  = {Alexander, G. and Karshon, U. and Shapira, A. and Yekutieli, G. and Engelmann, R. and Filthuth, H. and Lughofer, W.},
  title   = {Study of the {$\Lambda$}--{$N$} system in low-energy {$\Lambda$}--$p$ elastic scattering},
  journal = {Phys. Rev.},
  volume  = {173},
  pages   = {1452},
  year    = {1968},
  doi     = {10.1103/PhysRev.173.1452}
}

@article{Engelmann:1966npz,
  author  = {Engelmann, R. and Filthuth, H. and Hepp, V. and Kluge, E.},
  title   = {Inelastic {$\Sigma^-p$}-interactions at low momenta},
  journal = {Phys. Lett.},
  volume  = {21},
  pages   = {587},
  year    = {1966},
  doi     = {10.1016/0031-9163(66)91310-2}
}

@article{Buss:2011mx,
  author  = {Buss, O. and Gaitanos, T. and Gallmeister, K. and van Hees, H. and Kaskulov, M. and Lalakulich, O. and Larionov, A. B. and Leitner, T. and Weil, J. and Mosel, U.},
  title   = {Transport-theoretical description of nuclear reactions},
  journal = {Phys. Rept.},
  volume  = {512},
  pages   = {1},
  year    = {2012},
  doi     = {10.1016/j.physrep.2011.12.001},
  eprint  = {1106.1344},
  archivePrefix = {arXiv}
}

@phdthesis{Stephen:1970phd,
  author  = {Stephen, D.},
  school  = {University of Massachusetts},
  year    = {1970},
  note    = {unpublished}
}

@article{Hirtz:2019rqe,
  author={Hirtz, J. and David, J.-C. and Boudard, A. and Cugnon, J. and Leray, S. and Leya, I. and Rodr{\'\i}guez-S{\'a}nchez, J. L. and Schnabel, G.},
  title={{Strangeness production in the new version of the Li{\`e}ge intranuclear cascade model}},
  journal={Phys. Rev. C}, volume={101}, pages={014608}, year={2020},
  doi={10.1103/PhysRevC.101.014608}, eprint={1909.02246}, archivePrefix={arXiv}}

@article{akbar2014,
  author  = {Rafi Alam, M. and Sajjad Athar, M. and Chauhan, S. and Singh, S. K.},
  title   = {Quasielastic hyperon production in $\bar\nu_\mu$--nucleus interactions},
  journal = {J. Phys. G},
  volume  = {42},
  pages   = {055107},
  year    = {2015},
  doi     = {10.1088/0954-3899/42/5/055107},
  eprint  = {1409.2145},
  archivePrefix = {arXiv}
}

@article{fatima2016,
  author  = {Akbar, F. and Rafi Alam, M. and Sajjad Athar, M. and Singh, S. K.},
  title   = {Quasielastic production of polarized hyperons in antineutrino--nucleon reactions},
  journal = {Phys. Rev. D},
  volume  = {94},
  pages   = {114031},
  year    = {2016},
  doi     = {10.1103/PhysRevD.94.114031},
  eprint  = {1608.02103},
  archivePrefix = {arXiv}
}

@article{fatima2021,
  author  = {Fatima, A. and Sajjad Athar, M. and Singh, S. K.},
  title   = {$\bar\nu_\mu$ induced quasielastic production of hyperons leading to pions},
  journal = {Eur. Phys. J. Spec. Top.},
  volume  = {230},
  pages   = {4391},
  year    = {2021},
  doi     = {10.1140/epjs/s11734-021-00272-0},
  eprint  = {2106.14590},
  archivePrefix = {arXiv}
}

@article{Juszczak:2005wk,
  author  = {Juszczak, C. and Nowak, J. A. and Sobczyk, J. T.},
  title   = {Spectrum of recoil nucleons in quasi-elastic neutrino nucleus interactions},
  journal = {Eur. Phys. J. C},
  volume  = {39},
  pages   = {195},
  year    = {2005},
  doi     = {10.1140/epjc/s2004-02086-9}
}

@article{juszczak2005,
  author  = {Juszczak, C. and Nowak, J. A. and Sobczyk, J. T.},
  title   = {Simulations from a new neutrino event generator},
  journal = {Nucl. Phys. B Proc. Suppl.},
  volume  = {159},
  pages   = {211},
  year    = {2006},
  doi     = {10.1016/j.nuclphysbps.2006.08.069},
  eprint  = {hep-ph/0512365},
  archivePrefix = {arXiv}
}

@article{pythia83,
  author  = {Bierlich, C. and others},
  title   = {A comprehensive guide to the physics and usage of {PYTHIA} 8.3},
  journal = {SciPost Phys. Codebases},
  pages   = {8},
  year    = {2022},
  doi     = {10.21468/SciPostPhysCodeb.8},
  eprint  = {2203.11601},
  archivePrefix = {arXiv}
}

@article{golan2012,
  author  = {Golan, T. and Juszczak, C. and Sobczyk, J. T.},
  title   = {Final state interactions effects in neutrino-nucleus interactions},
  journal = {Phys. Rev. C},
  volume  = {86},
  pages   = {015505},
  year    = {2012},
  doi     = {10.1103/PhysRevC.86.015505},
  eprint  = {1202.4197},
  archivePrefix = {arXiv}
}

@article{Douchin:2001sv,
  author  = {Douchin, F. and Haensel, P.},
  title   = {A unified equation of state of dense matter and neutron star structure},
  journal = {Astron. Astrophys.},
  volume  = {380},
  pages   = {151},
  year    = {2001},
  doi     = {10.1051/0004-6361:20011402},
  eprint  = {astro-ph/0111092},
  archivePrefix = {arXiv}
}

@article{Read:2008iy,
  author  = {Read, J. S. and Lackey, B. D. and Owen, B. J. and Friedman, J. L.},
  title   = {Constraints on a phenomenologically parametrized neutron-star equation of state},
  journal = {Phys. Rev. D},
  volume  = {79},
  pages   = {124032},
  year    = {2009},
  doi     = {10.1103/PhysRevD.79.124032},
  eprint  = {0812.2163},
  archivePrefix = {arXiv}
}

@article{Fortin:2016hny,
  author  = {Fortin, M. and Provid{\^e}ncia, C. and Raduta, Ad. R. and Gulminelli, F. and Zdunik, J. L. and Haensel, P. and Bejger, M.},
  title   = {Neutron star radii and crusts: uncertainties and unified equations of state},
  journal = {Phys. Rev. C},
  volume  = {94},
  pages   = {035804},
  year    = {2016},
  doi     = {10.1103/PhysRevC.94.035804},
  eprint  = {1604.01944},
  archivePrefix = {arXiv}
}

@article{rambo,
  author  = {Kleiss, R. and Stirling, W. J. and Ellis, S. D.},
  title   = {A new {Monte Carlo} treatment of multiparticle phase space at high-energies},
  journal = {Comput. Phys. Commun.},
  volume  = {40},
  pages   = {359},
  year    = {1986},
  doi     = {10.1016/0010-4655(86)90119-0}
}

\end{document}